\documentclass[a4paper,10pt,openany]{article}
\usepackage{pst-3d}
\usepackage{pst-3dplot}
\usepackage{etex}
\usepackage[utf8]{inputenc}
\usepackage[english]{babel}
\usepackage{amsmath,amssymb,amsthm,mathrsfs,amsfonts,dsfont}
\usepackage{graphicx}
\usepackage[footnotesize, skip=0pt]{caption}
\usepackage{epigraph}
\usepackage{indentfirst}
\usepackage{booktabs}
\usepackage{fancyhdr}
\usepackage{vmargin}
\usepackage{lscape}
\usepackage{subfig}
\usepackage{textcomp}
\usepackage{pstricks}
\usepackage[metapost]{mfpic}
\usepackage{fancybox}
\usepackage{slashed}
\usepackage[verbose]{wrapfig}
\usepackage{pifont}
\usepackage{keystroke}
\usepackage[title]{appendix}
\usepackage{enumitem}
\usepackage{array}
\usepackage{colortbl}
\usepackage{alltt}
\usepackage{boxedminipage}
\usepackage{tikz}
\usepackage{calc}
\usepackage{subfloat}
\usepackage{multirow}
\usepackage{nicematrix}
\usepackage{cmll}
\usepackage{multicol}
\definecolor{color1}{RGB}{204,0,51}
\definecolor{color2}{RGB}{159,182,205}
\usepackage[open,openlevel=3]{bookmark}

\usepackage{verbatim}
\usepackage{pst-grad} 
\usepackage{pst-plot}
\usepackage{transparent}
\usepackage{color,soul}
\usepackage{cancel}
\usepackage{placeins}
\usepackage{mathtools}
\usepackage{rotating}
\usepackage[refpage]{nomencl}
\usepackage{hhline}
\usepackage{marginnote}
\usepackage{upgreek}
\usepackage{listings}
\usepackage{hyperref}
\usepackage[nameinlink]{cleveref}
\usepackage{afterpage}
\usepackage{cite}
\usepackage{makecell}
\usepackage{accents}
\usepackage{breqn}
\usepackage{environ}
\usepackage{empheq}
\usepackage{bbding}
\usepackage{titlesec}
\usepackage[final]{showlabels}
\usepackage{orcidlink}
\usepackage[outline]{contour}

\newcommand{\dil}{\lambda_{D}}

\newcommand{\coefa}{c_{1}}
\newcommand{\coefb}{c_{2}}
\newcommand{\CA}{\tilde{\alpha}}
\newcommand{\CB}{\tilde{\beta}}
\newcommand{\CC}{\tilde{\gamma}}
\newcommand{\CD}{\tilde{\delta}}
\newcommand{\CE}{\texttt{a}}
\newcommand{\CF}{\texttt{b}}
\newcommand{\CG}{\texttt{c}}

\newcommand{\cc}[1]{\textcolor{ccolor}{c^{#1}}}
\definecolor{ccolor}{RGB}{255, 95, 31}
\definecolor{cdolor}{RGB}{255, 0, 0}

\allowdisplaybreaks
\lstdefinestyle{customc}{
  belowcaptionskip=1\baselineskip,
  breaklines=true,
  frame=L,
  xleftmargin=\parindent,
  language=C,
  showstringspaces=false,
  basicstyle=\footnotesize\ttfamily,
  keywordstyle=\bfseries\color{green!40!black},
  commentstyle=\itshape\color{purple!40!black},
  identifierstyle=\color{blue},
  stringstyle=\color{red217!80!black},
}

\lstdefinestyle{customasm}{
  belowcaptionskip=1\baselineskip,
  frame=L,
  xleftmargin=\parindent,
  language=[x86masm]Assembler,
  basicstyle=\footnotesize\ttfamily,
  commentstyle=\itshape\color{purple!40!black},
}

\sethlcolor{yellow}

\hypersetup{
linkbordercolor={red}
}

\hypersetup{
    bookmarks=false,         
    unicode=false,          
    pdftoolbar=true,        
    pdfmenubar=true,        
    pdffitwindow=false,     
    pdfstartview={FitH},    
    pdftitle={Non-Relativistic Gravity Coupled to Matter},    
    pdfauthor={Eric Bergshoeff, Giacomo Giorgi, Luca Romano},     
    pdfsubject={},   
    pdfnewwindow=true,      
    colorlinks=false,       
    linkcolor=red,          
    citecolor=red,        
    filecolor=red,      
    urlcolor=red           
    linkbordercolor={red},
    citebordercolor={red},
    urlbordercolor={red},
    bookmarksopen={true}
}

\usepackage{listings} 

\makeatletter

\newcommand{\Rmnum}[1]{\expandafter\@slowromancap\romannumeral #1@}
\makeatother

\theoremstyle{definition}

\theoremstyle{remark}

\theoremstyle{proposition}

\usepackage{alphalph}

\makeatletter
\newalphalph{\aalphalph}[mult]{\alphalph@alph}{26}
\newcommand{\alphalphval}[1]{%
  \@ifundefined{c@#1}{
    \aalphalph{#1}
  }{%
    \aalphalph{\value{#1}}
  }
}
\makeatother

\usepackage{color}

\def\chapterautorefname~#1\null{Chap.~(#1)\null}
\def\sectionautorefname~#1\null{Sec.~(#1)\null}
\def\subsectionautorefname~#1\null{sub--Sec.~(#1)\null}
\def\figureautorefname~#1\null{Fig.~(#1)\null}
\def\tableautorefname~#1\null{Tab.~(#1)\null}
\def\equationautorefname~#1\null{~(#1)\null}

\NewEnviron{multieq}[1][2]{

\begin{multicols}{#1}
\begin{subequations}
\setlength{\abovedisplayskip}{-11pt}
\allowdisplaybreaks
\begin{align}
\BODY
\end{align}
\end{subequations}
\end{multicols}
\setlength{\parindent}{0pt}
\noindent
}

\NewEnviron{multieqthree}[1][2]{

\begin{multicols}{#1}
\begin{subequations}
\setlength{\abovedisplayskip}{-16pt}
\allowdisplaybreaks
\begin{align}
\BODY
\end{align}
\end{subequations}
\end{multicols}
\setlength{\parindent}{0pt}
\noindent
}

\NewEnviron{multieqsep}[1][2]{

\begin{multicols}{#1}
\setlength{\columnseprule}{0.4pt}
\begin{subequations}
\setlength{\abovedisplayskip}{-12pt}
\allowdisplaybreaks
\begin{align}
\BODY
\end{align}
\end{subequations}
\end{multicols}
\setlength{\parindent}{0pt}
\noindent
}

\newcommand\multieqreference{}
\NewEnviron{multieqref}[2][2]{
\begin{multicols}{#1}
\def\multieqreference{#2}%
\begin{subequations}\label{\multieqreference}\
\setlength{\abovedisplayskip}{-13pt}
\allowdisplaybreaks
\begin{align}
\BODY
\end{align}
\end{subequations}
\end{multicols}
\setlength{\parindent}{0pt}
\noindent
}

\NewEnviron{multieqrefB}[2][2]{
\begin{subequations}
\begin{multicols}{#1}
\setlength{\abovedisplayskip}{-12pt}
\allowdisplaybreaks
\begin{align}
\BODY
\end{align}
\end{multicols}\label{#2}
\end{subequations}
\setlength{\parindent}{0pt}
\noindent
}

\DeclareMathAlphabet\mathbfcal{OMS}{cmsy}{b}{n}

\usepackage{datetime}
\title{\huge\bf From Relativistic Gravity to the Poisson Equation}

\date{}

\begin{document}

\begin{flushright}
\small
\today
\normalsize
\end{flushright}
{\let\newpage\relax\maketitle}
\maketitle
\def\equationautorefname~#1\null{(#1)\null}
\def\tableautorefname~#1\null{table~#1\null}
\def\sectionautorefname~#1\null{section~#1\null}
\def\appendixautorefname~#1\null{appendix~#1\null}

\vspace{0.8cm}

\begin{center}
\renewcommand{\thefootnote}{\alph{footnote}}
{\bf\large Eric~A.~Bergshoeff\orcidlink{0000-0003-1937-6537}$^{~1}$}\footnote{Email: {\tt e.a.bergshoeff[at]rug.nl}  } {,}
{\bf\large Giacomo~Giorgi\orcidlink{0000-0002-6608-1099}$^{~2}$}\footnote{Email: {\tt giacomo.giorgi[at]um.es}  } {\bf and}
{\bf\large Luca~Romano\orcidlink{0000-0001-9033-1345}$^{~3}$}\footnote{Email: {\tt lucaromano2607[at]gmail.com}}
\setcounter{footnote}{0}
\renewcommand{\thefootnote}{\arabic{footnote}}

\vspace{0.5cm}
${}^1${\it Van Swinderen Institute, University of Groningen\\
Nijenborgh 4, 9747 AG Groningen, The Netherlands}\\
\vskip .2truecm
${}^2${\it Departamento de Física, Universidad de Murcia,\\ Campus de Espinardo, 30100 Murcia, Spain }\\
\vskip .2truecm
${}^3${\it Departamento de Electromagnetismo y Electronica, Universidad de Murcia,\\ Campus de Espinardo, 30100 Murcia, Spain }\\

\vspace{1.8cm}


{\bf Abstract}
\end{center}
\begin{quotation}
{\small

We consider the non-relativistic limit of general relativity coupled to a $(p+1)$-form gauge field and a scalar field in arbitrary dimensions and investigate under which conditions this gives rise to a Poisson equation for a Newton potential describing Newton-Cartan gravity outside a massive $p$-dimensional extended object, a so-called $p$-brane. Given our Ansatz, we show that not all the $p$-branes satisfy the required conditions. We study theories whose dynamics is defined by a Lagrangian as well as systems that are defined by a set of equations of motion not related to a Lagrangian. We show that, within the Lagrangian approach, a Poisson equation can be obtained provided that the coupling of the scalar field is fine-tuned such that the non-relativistic Lagrangian is invariant under an emerging local dilatation symmetry.  On the other hand, we demonstrate that in the absence of a Lagrangian a Poisson equation can be obtained from a set of equations of motion that is not dilatation invariant. We discuss how our Ansatz could be generalized such as to include more $p$-branes giving rise to a Poisson equation.
}
\end{quotation}

\newpage

\tableofcontents

\section*{Introduction}
\addcontentsline{toc}{section}{\protect\numberline{}Introduction}

In recent years much work has been devoted to taking limits and/or expansions of general relativity possibly coupled to matter. One motivation has been to understand the non-relativistic limit of string theory and, in particular, its low-energy limit, with an eye towards possible implications for non-relativistic holography, see, e.g. \cite{Taylor:2008tg, Christensen:2013lma, Christensen:2013rfa,  Oling:2022fft}. Another motivation has been the renewed interest in (post-Newtonian approximations of) gravity in the context of gravitational waves and black hole mergers, see, e.g., \cite{VandenBleeken:2017rij,Hansen:2019svu,Hansen:2020wqw,Elbistan:2022plu,Hartong:2023ckn}. Different non-Lorentzian limits and expansions have been considered as well such as the Carroll limit \cite{Leblond,Gupta}, see, e.g.,\cite{Hansen:2021fxi, Bergshoeff:2022eog, Bidussi:2023rfs}. This limit  plays a relevant role in several different contexts ranging from  the near horizon geometry of black holes and  the BMS group to  Carrollian aspects of flat space holography \cite{Donnay:2019jiz, Duval:2014uva, Bagchi:2022emh,Donnay:2022aba}.\\

In this work we investigate the so-called  Newton-Cartan limits of general relativity \cite{ASENS_1923_3_40__325_0, ASENS_1924_3_41__1_0} since these are the limits that give rise to the well-known non-relativistic gravity that is acting upon the massive particles of the standard model. The Newton-Cartan limits should be distinguished from the more exotic Galilei limits that we will study in an appendix. The difference is that when taking a Galilei limit one ends up with a geometry whose infinitesimal structure group is given by the Galilei algebra whereas when taking a Newton-Cartan limit the structure group gets enhanced with an additional U(1) symmetry that acts as a central extension of the Galilei algebra leading to the so-called Bargmann algebra.\\

Often one defines, instead of a Newton-Cartan limit, a so-called Newtonian limit that is a combination of a non-relativistic limit together with other assumptions like weak gravitation and slowly changing gravitational fields. Upon taking a Newtonian limit,  the Einstein equations reduce to the Poisson equation
\begin{equation}
\triangle \, \Psi(x) =0
\end{equation}
for a gravitational potential $\Psi(x)$ that depends on the spatial coordinates or, equivalently, the coordinates transverse to the particle. The (sourced) Poisson equation describes Newtonian gravity in any frame with constant acceleration. When taking the Newton-Cartan limit one does not make additional assumptions and one ends up with a set of more complicated equations containing more geometric fields that have the advantage of being valid in any frame. Taking the Newton-Cartan limit, however, is slightly more subtle since, to extend the Galilei algebra to a Bargmann algebra one needs to start with general relativity and an auxiliary vector field $A_\mu$ whose field-strength is set to zero by hand in order not to change the physical degrees of freedom described by general relativity.\,\footnote{Such an auxiliary vector field already occurs at the level of a particle sigma model \cite{Gomis:2000bd}. In the field theory approach that we are considering here, the vector field corresponds to the Noether symmetry that leads to the conservation of particles minus anti-particles. Note that general relativity together with this constrained vector field has no Lagrangian formalism. In this work we will also consider the option that the vector field does not satisfy a constraint and there exists a Lagrangian that describes general relativity plus matter. This option occurs for instance when we consider the low-energy limit of string theory.}  One can then show that, after gauge-fixing, the complicated equations reduce to the same Poisson equation where the Newton potential is identified with the time component $A_0$ of the auxiliary vector field $A_\mu$ \cite{Andringa:2013mma,Bergshoeff:2022iyb}. One justification for this identification can be seen from the fact that the Lagrangian describing the coupling of a particle to Newton-Cartan gravity contains, on the one hand, a kinetic term describing the coupling of the particle to the basic Newton-Cartan fields $(\tau_\mu\,, e_\mu{}^a)$ where  $\tau_\mu\, (e_\mu{}^a)$ is the timelike (spacelike) Vierbein.  On the other hand, the vector field $A_\mu$ couples to the particle via a Wess-Zumino term. Upon identification of the Newton potential with the time component of this vector field the Wess-Zumino term precisely reproduces the standard coupling of the Newton potential $\Psi(x)$ to a particle.\,\footnote{When taking the Newtonian limit one usually identifies the Newton potential with the time-time component of the relativistic metric. Combining the kinetic and Wess-Zumino terms in the coupling of a particle to Newton-Cartan gravity one finds the boost-invariant combination $h_{\mu\nu} + A_{(\mu}\tau_{\nu)}$ where $h_{\mu\nu} = e_\mu{}^a e_\nu{}^b \delta_{ab}$ is the spatial metric.  After gauge-fixing  $\tau_\mu = \delta_{\mu,0}$, one then sees that the time-time component of this combination produces the same Newton potential as in the Newtonian limit.}\\

It is the purpose of this work to investigate under which conditions the fact that taking the Newton-Cartan limit of a particle coupled to gravity in general dimensions $D$ leads to a description of  Newton-Cartan gravity in the directions transverse to the particle with, upon gauge-fixing, a corresponding Poisson equation, can be extended to a $p$-brane, i.e.~an object that extends in $p$ spatial directions. In this case we consider general relativity coupled to a $(p+1)$-form gauge field $A_{\mu_1 \cdots \mu_{p+1}}$ where the case $p=0$ corresponds to a particle. We also include the coupling to a scalar field $\Phi$ that will play an important role in obtaining a Poisson equation.
We assume  that the $(p+1)$-form gauge field couples via a Wess-Zumino term to a $p$-brane  such that the Newton-potential $\Psi$, after gauge-fixing,  can be identified with the single component of the   $(p+1)$-form gauge field that is projected onto the longitudinal directions
\begin{equation}
\Psi \equiv A_{01\cdots p+1}\,.
\end{equation}
To describe Newton-Cartan gravity in the $D-p-1$ directions transverse to the $p$-brane this Newton potential should satisfy a Poisson equation in the transverse directions.\\

To better understand the subtleties of the Newton-Cartan limit, it is instructive to compare it with the Galilei limit and see how, in the case of particles, the extra U(1) symmetry gets into the game.  In fact, there exist two kinds of Galilei limits giving rise to what is called electric and magnetic Galilei gravity in the literature.\footnote{This terminology is taken from \cite{Henneaux:2021yzg, Campoleoni:2022ebj}, where it was used in the context of Carroll gravity.} Taking the limit of the Einstein-Hilbert action the electric limit is defined by the leading order expression in $\cc{}$ where $\cc{}$ is a dimensionless contraction parameter that is taken to infinity in the Galilei limit. On the other hand, the magnetic limit is defined by the expression of sub-leading order in $\cc{}$ where first the leading order expression has been eliminated by a so-called Hubbard-Stratonovich transformation introducing a Lagrange multiplier. In both electric and magnetic limits one ends up with a structure group given by the Galilei algebra.\\

In contrast to a Galilei limit, a Newton-Cartan limit is defined by the expression of sub-leading-order in $\cc{}$, like in the case of magnetic Galilei gravity, but where now the expression of leading-order in $\cc{}$ has been cancelled by adding a $(p+1)$-form gauge field to the Einstein-Hilbert term. This cancellation works in general dimensions $D$ provided we divide the tangent space into $p+1$ directions longitudinal to the $p$-brane and $D-p-1$  directions transverse to the same $p$-brane.\,\footnote{This choice of foliation implies that we only consider fundamental $p$-branes. This should be distinguished from, for instance, the many branes in string theory where, when taking the non-relativistic limit of the fundamental string, one should use the same string foliation for all the other branes. In the case of strings only the 2-form gauge field kinetic term plays a crucial role in cancelling the divergence from the Einstein-Hilbert term.} The absence of a leading divergence requires a specific expansion of the $(p+1)$-form gauge field together with a fine-tuning of coefficients in the Lagrangian. The corresponding limit is therefore sometimes called a critical limit. In the case of particles ($p=0$) the Abelian gauge transformation of the vector gauge field is identified with the central extension of the Bargmann algebra.  The algebraic interpretation of the $(p+1)$-form gauge transformation in terms of an algebra is less straightforward.\\

Although not essential for the cancellation of the leading-order expression the inclusion of a scalar field $\Phi$ will play a crucial role in realizing an emergent dilatation symmetry of the non-relativistic theory and obtaining a Poisson equation. One consequence of this emerging dilatation symmetry is that going on-shell and then taking the Newton-Cartan limit is not the same as taking the Newton-Cartan limit and then going on-shell. In the first case one obtains one more equation of motion that cannot be obtained by varying the non-relativistic action. It turns out that this `missing' equation of motion for all cases that we consider is precisely the Poisson equation defining Newton-Cartan gravity. This makes the construction of a Lagrangian describing Newton-Cartan gravity a non-trivial matter.\,\footnote{For efforts in constructing an action for Newton-Cartan gravity based upon extended symmetries beyond the Bargman algebra, see \cite{Hansen:2019pkl}.}\\

In this work we will consider two approaches to taking  Newton-Cartan limits which we will call the {\it Lagrangian approach} and the {\it constrained on-shell approach}. In the Lagrangian approach our starting point is a relativistic Lagrangian describing general relativity coupled to a dynamical $(p+1)$-form gauge field and a scalar field. Taking the Newton-Cartan limit in the Lagrangian, the contribution to the leading order divergence in $\cc{}$ coming from the Einstein-Hilbert (EH) term is cancelled by a similar contribution from the gauge field kinetic term.  By contrast,
in the constrained on-shell approach we start from the same set of relativistic equations of motion that we find in the Lagrangian approach, but we also impose by hand an additional zero-field-strength constraint on the $(p+1)$-form gauge field and/or the scalar field, \footnote{There are a few exceptions where the zero-field-strength constraint on the $(p+1)$-form gauge field can be considered an equation of motion that follows from a Lagrangian provided one introduces another gauge field whose field-strength is also zero on-shell. Examples are three-dimensional extended Bargmann gravity with two gauge fields corresponding to two central extensions \cite{Bergshoeff:2016lwr,Hartong:2016yrf} and the theories of \cite{Bergshoeff:2018vfn}.} which allows us at the on-shell level to control the leading divergence in $\cc{}$ in a different way.
In this way we can extract a Poisson equation out of a Newton-Cartan limit of general relativity without the occurrence of an emerging dilatation symmetry. This does not solve of course constructing an action principle for Newton-Cartan gravity but it makes the need of an emerging dilatation symmetry less essential.\\

\begin{figure}
\resizebox{\textwidth}{!}{%

\tikzset {_av4v3k2wi/.code = {\pgfsetadditionalshadetransform{ \pgftransformshift{\pgfpoint{0 bp } { 0 bp }  }  \pgftransformrotate{0 }  \pgftransformscale{2 }  }}}
\pgfdeclarehorizontalshading{_r10ld2a7p}{150bp}{rgb(0bp)=(0.25,0.46,0.02);
rgb(37.5bp)=(0.25,0.46,0.02);
rgb(62.5bp)=(0.82,0.97,0);
rgb(100bp)=(0.82,0.97,0)}


\tikzset {_vw500b468/.code = {\pgfsetadditionalshadetransform{ \pgftransformshift{\pgfpoint{0 bp } { 0 bp }  }  \pgftransformrotate{0 }  \pgftransformscale{2 }  }}}
\pgfdeclarehorizontalshading{_1zovbnu7y}{150bp}{rgb(0bp)=(0.25,0.46,0.02);
rgb(37.5bp)=(0.25,0.46,0.02);
rgb(62.5bp)=(0.82,0.97,0);
rgb(100bp)=(0.82,0.97,0)}
\tikzset{_lr8jn5m3q/.code = {\pgfsetadditionalshadetransform{\pgftransformshift{\pgfpoint{0 bp } { 0 bp }  }  \pgftransformrotate{0 }  \pgftransformscale{2 } }}}
\pgfdeclarehorizontalshading{_gdkewmbvm} {150bp} {color(0bp)=(transparent!0);
color(37.5bp)=(transparent!0);
color(62.5bp)=(transparent!10);
color(100bp)=(transparent!10) }
\pgfdeclarefading{_8raobx5ax}{\tikz \fill[shading=_gdkewmbvm,_lr8jn5m3q] (0,0) rectangle (50bp,50bp); }


\tikzset {_psj5tz9w2/.code = {\pgfsetadditionalshadetransform{ \pgftransformshift{\pgfpoint{0 bp } { 0 bp }  }  \pgftransformrotate{0 }  \pgftransformscale{2 }  }}}
\pgfdeclarehorizontalshading{_lm4jp2zbs}{150bp}{rgb(0bp)=(0.54,0,0.06);
rgb(37.5bp)=(0.54,0,0.06);
rgb(62.5bp)=(0.93,0,0.11);
rgb(100bp)=(0.93,0,0.11)}


\tikzset {_e51gm6v5w/.code = {\pgfsetadditionalshadetransform{ \pgftransformshift{\pgfpoint{0 bp } { 0 bp }  }  \pgftransformrotate{0 }  \pgftransformscale{2 }  }}}
\pgfdeclarehorizontalshading{_wku2pdggv}{150bp}{rgb(0bp)=(0.31,0.29,0);
rgb(37.5bp)=(0.31,0.29,0);
rgb(62.5bp)=(0.97,0.91,0.11);
rgb(100bp)=(0.97,0.91,0.11)}
\tikzset{_k0vc5cqlx/.code = {\pgfsetadditionalshadetransform{\pgftransformshift{\pgfpoint{0 bp } { 0 bp }  }  \pgftransformrotate{0 }  \pgftransformscale{2 } }}}
\pgfdeclarehorizontalshading{_fklou1555} {150bp} {color(0bp)=(transparent!0);
color(37.5bp)=(transparent!0);
color(62.5bp)=(transparent!10);
color(100bp)=(transparent!10) }
\pgfdeclarefading{_s4tsbuw9a}{\tikz \fill[shading=_fklou1555,_k0vc5cqlx] (0,0) rectangle (50bp,50bp); }
\tikzset{every picture/.style={line width=0.75pt}} 

\begin{tikzpicture}[x=0.75pt,y=0.75pt,yscale=-1,xscale=1]

\path  [shading=_r10ld2a7p,_av4v3k2wi] (80,370) -- (80,239.85) .. controls (80,239.85) and (80,239.85) .. (80,239.85) -- (184.2,239.85) -- (184.2,230) -- (220,250.45) -- (184.2,270.9) -- (184.2,261.05) -- (101.2,261.05) .. controls (101.2,261.05) and (101.2,261.05) .. (101.2,261.05) -- (101.2,370) -- cycle ; 
 \draw  [line width=0.75]  (80,370) -- (80,239.85) .. controls (80,239.85) and (80,239.85) .. (80,239.85) -- (184.2,239.85) -- (184.2,230) -- (220,250.45) -- (184.2,270.9) -- (184.2,261.05) -- (101.2,261.05) .. controls (101.2,261.05) and (101.2,261.05) .. (101.2,261.05) -- (101.2,370) -- cycle ; 

\path  [shading=_1zovbnu7y,_vw500b468,path fading= _8raobx5ax ,fading transform={xshift=2}] (760,240) -- (837.57,240) -- (837.57,230) -- (860,250) -- (837.57,270) -- (837.57,260) -- (760,260) -- cycle ; 
 \draw  [line width=0.75]  (760,240) -- (837.57,240) -- (837.57,230) -- (860,250) -- (837.57,270) -- (837.57,260) -- (760,260) -- cycle ; 

\path  [shading=_lm4jp2zbs,_psj5tz9w2] (80,455) -- (80,585.15) .. controls (80,585.15) and (80,585.15) .. (80,585.15) -- (184.2,585.15) -- (184.2,595) -- (220,574.55) -- (184.2,554.1) -- (184.2,563.95) -- (101.2,563.95) .. controls (101.2,563.95) and (101.2,563.95) .. (101.2,563.95) -- (101.2,455) -- cycle ; 
 \draw  [line width=0.75]  (80,455) -- (80,585.15) .. controls (80,585.15) and (80,585.15) .. (80,585.15) -- (184.2,585.15) -- (184.2,595) -- (220,574.55) -- (184.2,554.1) -- (184.2,563.95) -- (101.2,563.95) .. controls (101.2,563.95) and (101.2,563.95) .. (101.2,563.95) -- (101.2,455) -- cycle ; 

\path  [shading=_wku2pdggv,_e51gm6v5w,path fading= _s4tsbuw9a ,fading transform={xshift=2}] (390,240) -- (490.84,240) -- (490.84,230) -- (520,250) -- (490.84,270) -- (490.84,260) -- (390,260) -- cycle ; 
 \draw  [line width=0.75]  (390,240) -- (490.84,240) -- (490.84,230) -- (520,250) -- (490.84,270) -- (490.84,260) -- (390,260) -- cycle ; 

\draw  [line width=0.75]   (2,380) .. controls (2,378.9) and (2.9,378) .. (4,378) -- (165,378) .. controls (166.1,378) and (167,378.9) .. (167,380) -- (167,440) .. controls (167,441.1) and (166.1,442) .. (165,442) -- (4,442) .. controls (2.9,442) and (2,441.1) .. (2,440) -- cycle  ;
\draw (84.5,410) node  [font=\Large] [align=left] {\begin{minipage}[lt]{109.65pt}\setlength\topsep{0pt}
\begin{center}
\textbf{No Divergence }\\\textbf{Condition}
\end{center}

\end{minipage}};
\draw    (868,220) .. controls (868,218.9) and (868.9,218) .. (870,218) -- (966,218) .. controls (967.1,218) and (968,218.9) .. (968,220) -- (968,280) .. controls (968,281.1) and (967.1,282) .. (966,282) -- (870,282) .. controls (868.9,282) and (868,281.1) .. (868,280) -- cycle  ;
\draw (918,250) node  [font=\Large] [align=left] {\begin{minipage}[lt]{65.55pt}\setlength\topsep{0pt}
\begin{center}
\textbf{Poisson }\\\textbf{Equation}
\end{center}

\end{minipage}};
\draw    (228,555) .. controls (228,553.9) and (228.9,553) .. (230,553) -- (400,553) .. controls (401.1,553) and (402,553.9) .. (402,555) -- (402,595) .. controls (402,596.1) and (401.1,597) .. (400,597) -- (230,597) .. controls (228.9,597) and (228,596.1) .. (228,595) -- cycle  ;
\draw (315,575) node   [align=left] {\begin{minipage}[lt]{115.6pt}\setlength\topsep{0pt}
\begin{center}
\textbf{{\Large Not Explored}}
\end{center}

\end{minipage}};
\draw    (228,210) .. controls (228,208.9) and (228.9,208) .. (230,208) -- (380,208) .. controls (381.1,208) and (382,208.9) .. (382,210) -- (382,290) .. controls (382,291.1) and (381.1,292) .. (380,292) -- (230,292) .. controls (228.9,292) and (228,291.1) .. (228,290) -- cycle  ;
\draw (305,250) node   [align=left] {\begin{minipage}[lt]{102pt}\setlength\topsep{0pt}
\begin{center}
\textbf{{\Large + Constraints}}\\
\vspace{2mm}
\large $F=0$,\\ \large $d\Phi =0$
\end{center}

\end{minipage}};
\draw    (528,216) .. controls (528,214.9) and (528.9,214) .. (530,214) -- (750,214) .. controls (751.1,214) and (752,214.9) .. (752,216) -- (752,286) .. controls (752,287.1) and (751.1,288) .. (750,288) -- (530,288) .. controls (528.9,288) and (528,287.1) .. (528,286) -- cycle  ;
\draw (640,251) node   [align=left] {\begin{minipage}[lt]{149.6pt}\setlength\topsep{0pt}
\begin{center}
\textbf{{\Large NO Local \ Dilatation}}\\\textbf{{\Large Symmetry}}
\end{center}

\end{minipage}};
\draw (115,250) node  [font=\large,color={rgb, 255:red, 255; green, 255; blue, 255 }  ,opacity=1 ] [align=left] {\begin{minipage}[lt]{34pt}\setlength\topsep{0pt}
{fulfilled}
\end{minipage}};
\draw (135,574.35) node  [font=\large,color={rgb, 255:red, 255; green, 255; blue, 255 }  ,opacity=1 ] [align=left] {\begin{minipage}[lt]{67.6pt}\setlength\topsep{0pt}
{not fulfilled}
\end{minipage}};
\draw (442,250) node  [font=\large,color={rgb, 255:red, 255; green, 255; blue, 255 }  ,opacity=1 ] [align=left] {\begin{minipage}[lt]{72.17pt}\setlength\topsep{0pt}
\begin{center}
{no fine-tuning}
\end{center}

\end{minipage}};
\draw (645,129) node  [font=\LARGE] [align=left] {\begin{minipage}[lt]{591.6pt}\setlength\topsep{0pt}
{\Huge \textbf{The Constrained On-Shell Approach}}
\end{minipage}};

\end{tikzpicture}
}
\vspace{4mm}
\caption{This Figure summarizes the constrained on-shell approach that is discussed in \autoref{sec:constrained}. In this approach it is possible to obtain the Poisson equation from general relativity without further dynamical degrees of freedom and without an emerging local dilatation symmetry. }\label{fig:constrainedonshell}
\end{figure}
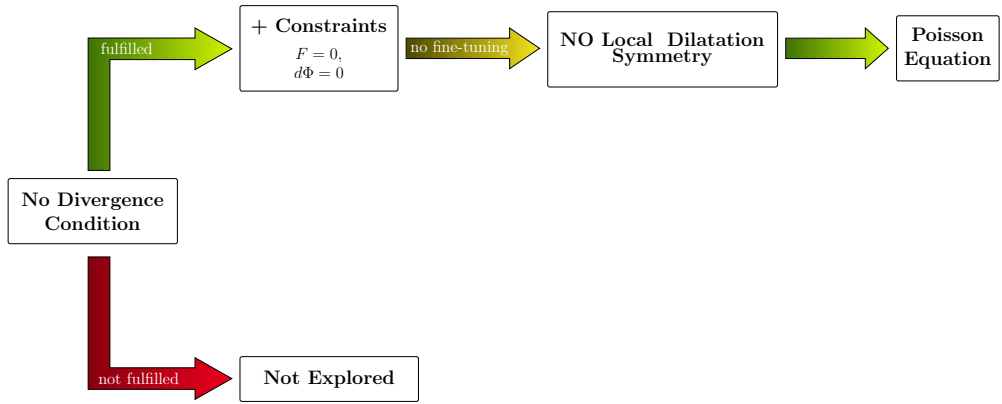

This paper is organized as visualized in \autoref{fig:lagrangian} and \autoref{fig:constrainedonshell}. We first discuss the Lagrangian approach, see \autoref{fig:lagrangian}. Starting from a Lagrangian in general dimensions $D$ for general relativity coupled to a $(p+1)$-form gauge field and a scalar field, we calculate in \autoref{sec:DivCcancellation} the so-called {\it no-divergence condition} under which there is a cancellation of the leading order term in the Lagrangian. Furthermore, we calculate the finite non-relativistic action in the sub-leading order. In \autoref{sec:localdil} we calculate the conditions under which the non-relativistic action is invariant under an emergent local dilatation symmetry. It turns out that besides satisfying the no-divergence condition, this requires a fine-tuning of the scalar field coupling. We are then ready to consider in \autoref{sec:Poisson} the non-relativistic equations of motion and search for a  Poisson equation. We find that such a Poisson equation can be identified for precisely the same fine-tuned scalar field coupling that led to the emergent local dilatation symmetry.  Furthermore, we find that the Poisson equation is precisely the single `missing' equation of motion that does not follow from the non-relativistic action.
Next, in \autoref{sec:constrained} we consider the constrained on-shell approach, see \autoref{fig:constrainedonshell}.  In this approach, we start from the same relativistic equations of motion as in the Lagrangian approach but also impose an additional zero-field-strength constraint by hand. We show that in this way one obtains a field theory without dilatation symmetry that, upon gauge fixing nevertheless gives the Poisson equation. We assume that in this approach the same no-divergence condition that we derived in the Lagrangian approach is satisfied, thereby pointing out how this condition also here plays a crucial role in taking the limit. We will not explore here the possible options if this condition is not satisfied.  Finally, in \autoref{sec:multiplet} we give, for the Lagrangian approach, the multiplet structure of the equations of motion under boost symmetry in the two cases with and without emerging dilatation symmetry. In particular, we show that in the case with local dilatation symmetry, the Poisson equation together with the other equations of motion that follow from the non-relativistic action form a so-called reducible but indecomposable representation under boosts. We have included three appendices. \hyperref[sec:Notation]{appendix~\ref*{sec:Notation}} gives our notation and conventions while \autoref{sec:appB} gives some technical details of the expansion in $\cc{}$ of the equations of motion that have been used in the main text. Finally, \cref{sec:ElectricGalilei2} gives an analysis of a matter-coupled electric Galilei gravity theory.
\begin{figure}
\resizebox{\textwidth}{!}{%


\tikzset {_npouaie1t/.code = {\pgfsetadditionalshadetransform{ \pgftransformshift{\pgfpoint{0 bp } { 0 bp }  }  \pgftransformrotate{0 }  \pgftransformscale{2 }  }}}
\pgfdeclarehorizontalshading{_lzjzoubkk}{150bp}{rgb(0bp)=(0.25,0.46,0.02);
rgb(37.5bp)=(0.25,0.46,0.02);
rgb(62.5bp)=(0.82,0.97,0);
rgb(100bp)=(0.82,0.97,0)}


\tikzset {_qj202mptw/.code = {\pgfsetadditionalshadetransform{ \pgftransformshift{\pgfpoint{0 bp } { 0 bp }  }  \pgftransformrotate{0 }  \pgftransformscale{2 }  }}}
\pgfdeclarehorizontalshading{_m7zewfdnq}{150bp}{rgb(0bp)=(0,0.03,0.5);
rgb(37.5bp)=(0,0.03,0.5);
rgb(62.5bp)=(0.23,0.45,1);
rgb(100bp)=(0.23,0.45,1)}


\tikzset {_3cxn68sez/.code = {\pgfsetadditionalshadetransform{ \pgftransformshift{\pgfpoint{0 bp } { 0 bp }  }  \pgftransformrotate{0 }  \pgftransformscale{2 }  }}}
\pgfdeclarehorizontalshading{_uf5wv9y5s}{150bp}{rgb(0bp)=(0.31,0.29,0);
rgb(37.5bp)=(0.31,0.29,0);
rgb(62.5bp)=(0.97,0.91,0.11);
rgb(100bp)=(0.97,0.91,0.11)}


\tikzset {_o9mxs8o7h/.code = {\pgfsetadditionalshadetransform{ \pgftransformshift{\pgfpoint{0 bp } { 0 bp }  }  \pgftransformrotate{0 }  \pgftransformscale{2 }  }}}
\pgfdeclarehorizontalshading{_bg26jzwv5}{150bp}{rgb(0bp)=(0.25,0.46,0.02);
rgb(37.5bp)=(0.25,0.46,0.02);
rgb(62.5bp)=(0.82,0.97,0);
rgb(100bp)=(0.82,0.97,0)}
\tikzset{_xy45gigrv/.code = {\pgfsetadditionalshadetransform{\pgftransformshift{\pgfpoint{0 bp } { 0 bp }  }  \pgftransformrotate{0 }  \pgftransformscale{2 } }}}
\pgfdeclarehorizontalshading{_oqrvc0ts3} {150bp} {color(0bp)=(transparent!0);
color(37.5bp)=(transparent!0);
color(62.5bp)=(transparent!10);
color(100bp)=(transparent!10) }
\pgfdeclarefading{_2xhd63u5f}{\tikz \fill[shading=_oqrvc0ts3,_xy45gigrv] (0,0) rectangle (50bp,50bp); }


\tikzset {_88sywchkl/.code = {\pgfsetadditionalshadetransform{ \pgftransformshift{\pgfpoint{0 bp } { 0 bp }  }  \pgftransformrotate{0 }  \pgftransformscale{2 }  }}}
\pgfdeclarehorizontalshading{_avd0jd64u}{150bp}{rgb(0bp)=(0.54,0,0.06);
rgb(37.5bp)=(0.54,0,0.06);
rgb(62.5bp)=(0.93,0,0.11);
rgb(100bp)=(0.93,0,0.11)}


\tikzset {_g3gavopfd/.code = {\pgfsetadditionalshadetransform{ \pgftransformshift{\pgfpoint{0 bp } { 0 bp }  }  \pgftransformrotate{0 }  \pgftransformscale{2 }  }}}
\pgfdeclarehorizontalshading{_lpbsfkbc0}{150bp}{rgb(0bp)=(0,0,0);
rgb(37.5bp)=(0,0,0);
rgb(62.5bp)=(0.57,0.57,0.57);
rgb(100bp)=(0.57,0.57,0.57)}
\tikzset{every picture/.style={line width=0.75pt}} 

\begin{tikzpicture}[x=0.75pt,y=0.75pt,yscale=-1,xscale=1]

\path  [shading=_lzjzoubkk,_npouaie1t] (80,390) -- (80,239.85) .. controls (80,239.85) and (80,239.85) .. (80,239.85) -- (184.2,239.85) -- (184.2,230) -- (220,250.45) -- (184.2,270.9) -- (184.2,261.05) -- (101.2,261.05) .. controls (101.2,261.05) and (101.2,261.05) .. (101.2,261.05) -- (101.2,390) -- cycle ; 
 \draw  [line width=0.75]  (80,390) -- (80,239.85) .. controls (80,239.85) and (80,239.85) .. (80,239.85) -- (184.2,239.85) -- (184.2,230) -- (220,250.45) -- (184.2,270.9) -- (184.2,261.05) -- (101.2,261.05) .. controls (101.2,261.05) and (101.2,261.05) .. (101.2,261.05) -- (101.2,390) -- cycle ; 

\path  [shading=_m7zewfdnq,_qj202mptw] (290,225) -- (290,139.78) .. controls (290,139.78) and (290,139.78) .. (290,139.78) -- (399.9,139.78) -- (399.9,130) -- (430,149.98) -- (399.9,169.95) -- (399.9,160.18) -- (310.4,160.18) .. controls (310.4,160.18) and (310.4,160.18) .. (310.4,160.18) -- (310.4,225) -- cycle ; 
 \draw  [line width=0.75]  (290,225) -- (290,139.78) .. controls (290,139.78) and (290,139.78) .. (290,139.78) -- (399.9,139.78) -- (399.9,130) -- (430,149.98) -- (399.9,169.95) -- (399.9,160.18) -- (310.4,160.18) .. controls (310.4,160.18) and (310.4,160.18) .. (310.4,160.18) -- (310.4,225) -- cycle ; 

\path  [shading=_uf5wv9y5s,_3cxn68sez] (289.6,275) -- (289.6,360.22) .. controls (289.6,360.22) and (289.6,360.22) .. (289.6,360.22) -- (399.5,360.22) -- (399.5,370) -- (429.6,350.02) -- (399.5,330.05) -- (399.5,339.82) -- (310,339.82) .. controls (310,339.82) and (310,339.82) .. (310,339.82) -- (310,275) -- cycle ; 
 \draw  [line width=0.75]  (289.6,275) -- (289.6,360.22) .. controls (289.6,360.22) and (289.6,360.22) .. (289.6,360.22) -- (399.5,360.22) -- (399.5,370) -- (429.6,350.02) -- (399.5,330.05) -- (399.5,339.82) -- (310,339.82) .. controls (310,339.82) and (310,339.82) .. (310,339.82) -- (310,275) -- cycle ; 

\path  [shading=_bg26jzwv5,_o9mxs8o7h,path fading= _2xhd63u5f ,fading transform={xshift=2}] (670,140) -- (770.84,140) -- (770.84,130) -- (800,150) -- (770.84,170) -- (770.84,160) -- (670,160) -- cycle ; 
 \draw  [line width=0.75]  (670,140) -- (770.84,140) -- (770.84,130) -- (800,150) -- (770.84,170) -- (770.84,160) -- (670,160) -- cycle ; 

\path  [shading=_avd0jd64u,_88sywchkl] (80,480) -- (80,630.15) .. controls (80,630.15) and (80,630.15) .. (80,630.15) -- (184.2,630.15) -- (184.2,640) -- (220,619.55) -- (184.2,599.1) -- (184.2,608.95) -- (101.2,608.95) .. controls (101.2,608.95) and (101.2,608.95) .. (101.2,608.95) -- (101.2,480) -- cycle ; 
 \draw  [line width=0.75]  (80,480) -- (80,630.15) .. controls (80,630.15) and (80,630.15) .. (80,630.15) -- (184.2,630.15) -- (184.2,640) -- (220,619.55) -- (184.2,599.1) -- (184.2,608.95) -- (101.2,608.95) .. controls (101.2,608.95) and (101.2,608.95) .. (101.2,608.95) -- (101.2,480) -- cycle ; 

\path  [shading=_lpbsfkbc0,_g3gavopfd] (670,350) -- (770.84,350) -- (770.84,340) -- (800,360) -- (770.84,380) -- (770.84,370) -- (670,370) -- cycle ; 
 \draw  [line width=0.75]  (670,350) -- (770.84,350) -- (770.84,340) -- (800,360) -- (770.84,380) -- (770.84,370) -- (670,370) -- cycle ; 

\draw  [line width=0.75]   (2,405) .. controls (2,403.9) and (2.9,403) .. (4,403) -- (165,403) .. controls (166.1,403) and (167,403.9) .. (167,405) -- (167,465) .. controls (167,466.1) and (166.1,467) .. (165,467) -- (4,467) .. controls (2.9,467) and (2,466.1) .. (2,465) -- cycle  ;
\draw (84.5,435) node  [font=\Large] [align=left] {\begin{minipage}[lt]{109.65pt}\setlength\topsep{0pt}
\begin{center}
\textbf{No Divergence }\\\textbf{Condition}
\end{center}

\end{minipage}};
\draw    (810,120) .. controls (810,118.9) and (810.9,118) .. (812,118) -- (908,118) .. controls (909.1,118) and (910,118.9) .. (910,120) -- (910,180) .. controls (910,181.1) and (909.1,182) .. (908,182) -- (812,182) .. controls (810.9,182) and (810,181.1) .. (810,180) -- cycle  ;
\draw (860,150) node  [font=\Large] [align=left] {\begin{minipage}[lt]{65.55pt}\setlength\topsep{0pt}
\begin{center}
\textbf{Poisson }\\\textbf{Equation}
\end{center}

\end{minipage}};
\draw    (810,310) .. controls (810,308.9) and (810.9,308) .. (812,308) -- (908,308) .. controls (909.1,308) and (910,308.9) .. (910,310) -- (910,400) .. controls (910,401.1) and (909.1,402) .. (908,402) -- (812,402) .. controls (810.9,402) and (810,401.1) .. (810,400) -- cycle  ;
\draw (860,355) node  [font=\Large] [align=left] {\begin{minipage}[lt]{65.55pt}\setlength\topsep{0pt}
\begin{center}
\textbf{NO }\\\textbf{Poisson }\\\textbf{Equation}
\end{center}

\end{minipage}};
\draw (340,150) node  [font=\large,color={rgb, 255:red, 255; green, 255; blue, 255 }  ,opacity=1 ] [align=left] {\begin{minipage}[lt]{60.83pt}\setlength\topsep{0pt}
\begin{center}
{fine-tuning}
\end{center}

\end{minipage}};
\draw (350,350) node  [font=\large,color={rgb, 255:red, 255; green, 255; blue, 255 }  ,opacity=1 ] [align=left] {\begin{minipage}[lt]{72.17pt}\setlength\topsep{0pt}
\begin{center}
{no fine-tuning}
\end{center}

\end{minipage}};
\draw    (228,600) .. controls (228,598.9) and (228.9,598) .. (230,598) -- (400,598) .. controls (401.1,598) and (402,598.9) .. (402,600) -- (402,640) .. controls (402,641.1) and (401.1,642) .. (400,642) -- (230,642) .. controls (228.9,642) and (228,641.1) .. (228,640) -- cycle  ;
\draw (315,620) node   [align=left] {\begin{minipage}[lt]{115.6pt}\setlength\topsep{0pt}
\begin{center}
\textbf{{\Large NO Finite Limit}}
\end{center}

\end{minipage}};
\draw    (228,230) .. controls (228,228.9) and (228.9,228) .. (230,228) -- (370,228) .. controls (371.1,228) and (372,228.9) .. (372,230) -- (372,270) .. controls (372,271.1) and (371.1,272) .. (370,272) -- (230,272) .. controls (228.9,272) and (228,271.1) .. (228,270) -- cycle  ;
\draw (300,250) node   [align=left] {\begin{minipage}[lt]{95.2pt}\setlength\topsep{0pt}
\begin{center}
\textbf{{\Large Finite Limit}}
\end{center}

\end{minipage}};
\draw    (438,320) .. controls (438,318.9) and (438.9,318) .. (440,318) -- (660,318) .. controls (661.1,318) and (662,318.9) .. (662,320) -- (662,390) .. controls (662,391.1) and (661.1,392) .. (660,392) -- (440,392) .. controls (438.9,392) and (438,391.1) .. (438,390) -- cycle  ;
\draw (550,355) node   [align=left] {\begin{minipage}[lt]{149.6pt}\setlength\topsep{0pt}
\begin{center}
\textbf{{\Large NO Local \ Dilatation}}\\\textbf{{\Large Symmetry}}
\end{center}

\end{minipage}};
\draw    (448,115) .. controls (448,113.9) and (448.9,113) .. (450,113) -- (640,113) .. controls (641.1,113) and (642,113.9) .. (642,115) -- (642,185) .. controls (642,186.1) and (641.1,187) .. (640,187) -- (450,187) .. controls (448.9,187) and (448,186.1) .. (448,185) -- cycle  ;
\draw (545,150) node   [align=left] {\begin{minipage}[lt]{129.2pt}\setlength\topsep{0pt}
\begin{center}
\textbf{{\Large Local \ Dilatation}}\\\textbf{{\Large Symmetry}}
\end{center}

\end{minipage}};
\draw (115,250) node  [font=\large,color={rgb, 255:red, 255; green, 255; blue, 255 }  ,opacity=1 ] [align=left] {\begin{minipage}[lt]{34pt}\setlength\topsep{0pt}
{fulfilled}
\end{minipage}};
\draw (135,619.35) node  [font=\large,color={rgb, 255:red, 255; green, 255; blue, 255 }  ,opacity=1 ] [align=left] {\begin{minipage}[lt]{67.6pt}\setlength\topsep{0pt}
{not fulfilled}
\end{minipage}};
\draw (730,358.4) node  [font=\small,color={rgb, 255:red, 255; green, 255; blue, 255 }  ,opacity=1 ] [align=left] {\begin{minipage}[lt]{68pt}\setlength\topsep{0pt}
\end{minipage}};
\draw (300,-2) node [anchor=north west][inner sep=0.75pt]  [font=\LARGE] [align=left] {{\Huge \textbf{The Lagrangian Approach}}};

\end{tikzpicture}
}
\vspace{4mm}
\caption{This Figure summarizes the Lagrangian approach discussed in sections 1-3. Starting from the left, the no-divergence condition is derived in \autoref{sec:DivCcancellation}. Assuming that this condition is satisfied, we show in \autoref{sec:localdil} that there is an emergent local dilatation symmetry provided the scalar field coupling is fine-tuned. In \autoref{sec:Poisson} we take the non-relativistic limit of the equations of motion and show that, only in the case with local dilatation symmetry, the Poisson equation for a Newton potential in the directions transverse to the $p$-brane is obtained. Furthermore, we show that this Poisson equation is precisely the single equation of motion that does not follow from the non-relativistic action.}\label{fig:lagrangian}
\end{figure}
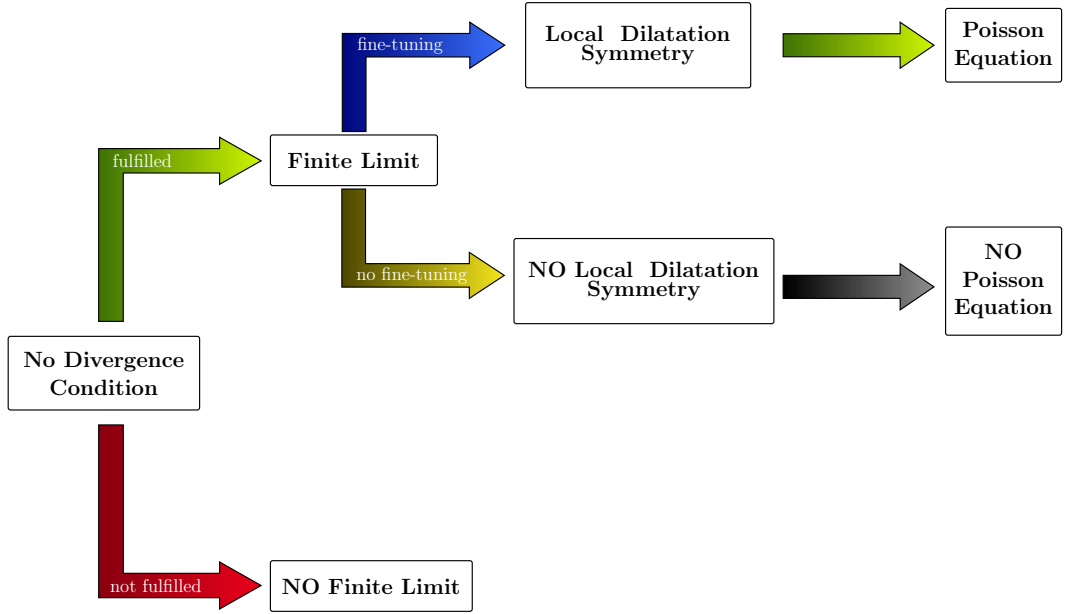

\FloatBarrier
\section{Cancellation of Divergences and Finite Action}\label{sec:DivCcancellation}

In the first three sections we will consider the Lagrangian approach where we assume the existence of a relativistic Lagrangian. In the introduction we already discussed three different ways of taking the non-relativistic limit of an action: the electric Galilei limit, the magnetic Galilei limit and the Newton-Cartan limit. Here we will focus on the Newton-Cartan limit where the leading divergence in the action coming from the Einstein-Hilbert term is canceled by a similar divergence originating from a $(p+1)$-form gauge field kinetic term for different dimensions $D$. Taking the limit we will distinguish between the $p+1$ directions longitudinal to a $p$-brane extended object and the remaining $D-p-1$ transverse directions. We will comment about a result that we obtained when taking an electric Galilei limit in \autoref{sec:ElectricGalilei2}.\\

In this section, we consider a generic p-brane foliation, i.e.~the $D$-dimensional flat index decomposes as $\hat{A}=\{A,a\}$ with $A=0,...,p$. We refer the reader to \autoref{sec:Notation} for details about our notation.

\subsection{The Relativistic Theory}

We consider a relativistic theory containing a metric $g_{\mu\nu}$ (with corresponding Vielbein $E_\mu{}^{\hat A}$), a $(p+1)$-form gauge potential $A_{\mu_{1}...\mu_{p+1}}$ and a scalar $\Phi$,  whose dynamics is described by the following action:
\begin{align}
\mathcal{S}=\int\,d^{D}x\, E\, {\rm e}^{-\coefa\Phi}\bigg[ {\rm R} -\frac{1}{2(p+2)!} F_{\mu_{1}...\mu_{p+2}}F^{\mu_{1}...\mu_{p+2}}+\coefb\,g^{\mu\nu}\partial_{\mu}\Phi\partial_{\nu}\Phi\bigg]\,,
\label{eq:RelAction}
\end{align}
with
\begin{subequations}
\begin{align}
{\rm R}&=g^{\mu\nu}(\partial_{\rho}\Gamma^{\rho}_{\mu\nu}-\partial_{\nu}\Gamma^{\rho}_{\rho \mu}+\Gamma^{\rho}_{\rho\sigma}\Gamma^{\sigma}_{\nu\mu}-\Gamma^{\rho}_{\nu\sigma}\Gamma^{\sigma}_{\rho\mu})\,,\\
\Gamma^{\rho}_{\mu\nu}&=\frac{1}{2}g^{\rho\sigma}(\partial_{\mu}g_{\nu\sigma}+\partial_{\nu}g_{\sigma\mu}-\partial_{\sigma}g_{\mu\nu})\,,\\
F_{\mu_{1}...\mu_{p+2}}&= (p+2) \partial_{[\mu_{1}}A_{\mu_{2}...\mu_{p+2}]}\,,\\
E&={\rm det}\, E_{\mu}{}^{\hat{A}}
\end{align}
\end{subequations}
and $\coefa,\coefb$ constants. If $\coefa\neq 0$ it can be removed by a field redefinition of the scalar field, thereby simultaneously modifying $\coefb$ to $\coefb /\coefa^2$. We prefer to work with both constants since this allows us to easily study the special case in which there is no scalar field, obtained by setting both $\coefa$ and $\coefb$ equal to zero or the special cases with no exponential factor in front of the square bracket or without a kinetic term for the scalar field, that can be obtained by setting only one of the two constants to zero. In the case of a domain-wall foliation with only one transverse direction, i.e. $p=D-2$, the gauge field term in the action corresponds to the presence of a cosmological constant. The action \autoref{eq:RelAction} is invariant under diffeomorphisms, Lorentz transformations and $(p+1)$-form gauge transformations.\\

We note that the way the scalar field $\Phi$ occurs in the action \eqref{eq:RelAction} is reminiscent of the dilaton in string theory. However, to keep the calculations manageable, the scalar field, unlike the dilaton, does not couple to the $(p+1)$-form gauge field for any value of $p$. For this reason, the branes considered in this work can not all be identified with the branes of string theory and therefore the scalar field cannot be identified with the dilaton, see also the discussion in the 
\hyperref[sec:conclusions]{Conclusions}.\\

The relativistic equations of motion $[G]^{\mu\nu}=0, [\Phi]=0$ and $[A]^{\mu_{1}...\mu_{p+1}}=0$
corresponding to  the action \autoref{eq:RelAction} are defined by
\begin{align}
\delta \mathcal{S}&= \int\,d^{D}x\, E\, {\rm e}^{-\coefa\Phi}\bigg[-\coefb[\Phi]\delta\Phi + \bigg(-[G]^{\mu\nu}+\frac{\coefb}{2\coefa}g^{\mu\nu}[\Phi]\bigg)\delta g_{\mu\nu}+\frac{1}{(p+1)!}[A]^{\mu_{1}...\mu_{p+1}}\delta A_{\mu_{1}...\mu_{p+1}}\bigg]\,.
\end{align}
We find the following expressions for $[G]^{\mu\nu}, [\Phi]$ and $[A]^{\mu_{1}...\mu_{p+1}}$:
\begin{subequations}
\begin{align}
[G]^{\mu\nu}&=R^{\mu\nu}-\frac{1}{2(p+1)!}F^{\mu}{}_{\rho_{1}...\rho_{p+1}}F^{\nu\rho_{1}...\rho_{p+1}}+\coefa \nabla^{\mu}\partial^{\nu}\Phi+\nonumber\\
&-(\coefa^{2}-\coefb)\bigg[\partial^{\mu}\Phi\partial^{\nu}\Phi-g^{\mu\nu}\partial_{\rho}\Phi\partial^{\rho}\Phi+\frac{1}{\coefa}g^{\mu\nu}\nabla_{\rho}\partial^{\rho}\Phi\bigg]\,,\\
\nonumber\\
[\Phi]&=\frac{\coefa}{\coefb}{\rm R}-\frac{\coefa}{2\coefb (p+2)!}F_{\mu_{1}...\mu_{p+2}}F^{\mu_{1}...\mu_{p+2}}+2\nabla_{\mu}\partial^{\mu}\Phi-\coefa \partial_{\mu}\Phi\partial^{\mu}\Phi\,,\\
\nonumber\\
[A]^{\mu_{1}...\mu_{p+1}}&=\nabla_{\nu}F^{\nu\mu_{1}...\mu_{p+1}}-\coefa\partial_{\nu}\Phi\, F^{\nu\mu_{1}...\mu_{p+1}}\,.
\end{align}\label{eq:RelEOM}
\end{subequations}
  In the next subsection, we are going to discuss how and under which conditions  the non-relativistic limit can be taken in the Lagrangian and in the equations of motion and how the results will depend on the dimensions $D$ and the foliation defined by $p$.\\
In the following, we also use the equations motion with flat indices, defined by
\begin{align}
[G]_{\hat{A}\hat{B}}&=[G]^{\mu\nu}E_{\mu\hat{A}}E_{\nu\hat{B}}=0\,,&
[A]^{\hat{A}_{1}...\hat{A}_{p+1}}&=[A]^{\mu_{1}...\mu_{p+1}}E_{\mu_{1}}{}^{\hat{A}_{1}}...E_{\mu_{p+1}}{}^{\hat{A}_{p+1}}=0
\,.\label{eq:relEOMflat}
\end{align}

\subsection{No-Divergence Condition}\label{sec:NoDivergenceCondition}

To define the limit we consider the following  expansion in terms of $\cc{}$:
\begin{subequations}
\begin{align}
E_{\mu}{}^{A}&=\cc{\alpha}\tau_{\mu}{}^{A}\,,\label{eq:ansatztau}\\
E_{\mu}{}^{a}&=\cc{\beta}e_{\mu}{}^{a}\,,\\
E^{\mu}{}_{A}&=\cc{-\alpha}\tau^{\mu}{}_{A}\,,\\
E^{\mu}{}_{a}&=\cc{-\beta}e^{\mu}{}_{a}\,,\\
\Phi&=\phi +\frac{p+1}{\coefa}\ln \cc{}\,,\\
A_{\mu_{1}...\mu_{p+1}}&=\cc{\xi}\tau_{\mu_{1}}{}^{A_{1}}...\tau_{\mu_{p+1}}{}^{A_{p+1}}\epsilon_{A_{1}...A_{p+1}}+ \cc{\gamma} a_{\mu_{1}...\mu_{p+1}}\,.\label{eq:ansatzA}
\end{align}
\end{subequations}
Here,  $\alpha,\beta, \gamma$, and $\xi$ and are arbitrary parameters. We will see how these parameters together with the dimension $D$, the rank $p+1$ of the gauge field and the constants  $\coefa$ and $\coefb$ are constrained by requiring the absence of divergences in the action. The power of $\cc{}$ in the first term of \autoref{eq:ansatzA} needs to be chosen such to achieve a cancellation between the leading divergences when taking the limit that $\cc{} \to \infty$ coming from the Ricci scalar and the Maxwell term. After taking the limit, in the non-relativistic theory, we convert curved indices into flat ones using the non-relativistic inverse Vielbeine $\tau^\mu{}_A$ and $e^\mu{}_a$  via
\begin{multieq}[2]
\tau^{\mu}{}_{A}T_{\mu}&=T_{A}\,,\\
e^{\mu}{}_{a}T_{\mu}&=T_{a}\,
\end{multieq}
for any vector  $T_{\mu}$. \\

As a first step, we consider the limit of the boost transformations and redefine the boost parameter as follows:
\begin{align}
\Lambda_{Aa}&=\cc{\delta}\lambda_{Aa}\,.
\end{align}
We now require that the limit of the boost transformation of the Vielbein reproduces the usual Galilean boost. This imposes the following restrictions:
\begin{multieqref}[2]{eq:nodivboost}
\delta&=\beta-\alpha\,,\\
\alpha-\beta&>0\,.\
\end{multieqref}
With this result acquired, the expansion of the Ricci scalar is given by:
\begin{align}
{\rm R} &=\cc{2\alpha-4\beta}\quad \accentset{(2\alpha-4\beta)}{\rm R}\, +\, \cc{-2\beta}\,\,\,\accentset{(-2\beta)}{\rm R}\, +\, \cc{-2\alpha}\,\,\, \accentset{(-2\alpha)}{\rm R}\, +\cc{-4\alpha-2\beta}\quad\, \accentset{(-4\alpha+2\beta)}{\rm R}\,,\label{eq:Rexpansion}
\end{align}
where we have denoted with $\accentset{(n)}{\rm R}$ the coefficient of the $\cc{n}$ term in the expansion of ${\rm R}$. This is a general notation that we adopt in the present work, with some slight variations, in some cases, about which we will detail when necessary.
Note that  the jumps in power of $\cc{}$ are at steps of $\cc{2(\alpha-\beta)}$. From \autoref{eq:nodivboost} it follows that this is a positive power of $\cc{}$.
Therefore,  the leading and sub-leading order terms in $\cc{}$ appear in  \autoref{eq:Rexpansion} with decreasing power of $\cc{}$  from the left to the right. The leading power term is given by
\begin{align}
\accentset{(2\alpha-4\beta)}{\rm R}&=-\frac{1}{4}t_{abA}t^{abA}\,,\label{eq:EHleading}
\end{align}
where
\begin{equation}
t_{ab}{}^{A}=e^{\mu}{}_{a}e^{\nu}{}_{b}\Big(\partial_{\mu}\tau_{\nu}{}^{A} - \partial_{\nu}\tau_{\mu}{}^{A}\Big)
\end{equation}
 is an intrinsic torsion tensor written with flat indices.
Taking a Newton-Cartan limit,  we regard this as a divergent part and try to cancel it against a similar term coming from the the gauge field kinetic term in the action. The sub-leading order  $\cc{-2\beta}$ term will play the role of ``finite action''. To write the leading divergence term \autoref{eq:EHleading} we have implicitly assumed that the transverse space is at least two-dimensional, namely
\begin{align}
D\geqslant p+3\,,\label{eq:existencecondition}
\end{align}
i.e.~we assume p-branes for general $0\le p \le D-4$  and defect branes with $p=D-3$   only.  The case of domain-walls with $p=D-2$ is special and will be discussed separately below.\\

To obtain another inequality on the parameters we also require that the boost transformation of the gauge field remains finite. Before taking the limit this boost transformation is given by
\begin{align}
\delta a_{\mu_{1}...\mu_{p+1}}&=-(p+1)\cc{\xi-2\alpha+2\beta-\gamma}\lambda^{A_{1}}{}_{b}e_{[\mu_{1}}{}^{b}\tau_{\mu_{2}}{}^{A_{2}}...\tau_{\mu_{p+1}]}{}^{A_{p+1}}\epsilon_{A_{1}...A_{p+1}}\,.\label{eq:aboost}
\end{align}
Thus we deduce that to avoid divergences in the limit we must satisfy the following inequality:
\begin{align}
\gamma\geqslant \xi + 2(\beta -\alpha)\,.\label{eq:gammaboost}
\end{align}
We now require that a divergence arising from the kinetic term of the gauge field cancels against the ``divergent" part coming from the Ricci scalar. With the Ansatz \autoref{eq:ansatzA} we have
\begin{align}
F_{\mu_{1}...\mu_{p+2}}&=\cc{\xi}\,\accentset{(\xi)}{F}_{\mu_{1}...\mu_{p+2}}+\cc{\gamma}\,\accentset{(\gamma)}{F}_{\mu_{1}...\mu_{p+2}}=\nonumber\\
&=\frac{1}{2}\cc{\xi}(p+2)(p+1)\, t_{[\mu_{1}\mu_{2}}{}^{A_{1}}\tau_{\mu_{3}}{}^{A_{2}}...\tau_{\mu_{p+2}]}{}^{A_{p+1}}\epsilon_{A_{1}...A_{p+1}}+
\cc{\gamma}f_{\mu_{1}...\mu_{p+2}}\,,\label{eq:FExpansion}
\end{align}
where
\begin{align}
f_{\mu_{1}...\mu_{p+2}}&=(p+2)\partial_{[\mu_1}a_{\mu_2 ... \mu_{p+2}]}\,.
\end{align}
For the field strength with flat indices we use the following convention:
\begin{align}
\accentset{(\gamma)}{F}_{A_{1}...A_{k}a_{1}...a_{p+2-k}}&=\tau^{\mu_{1}}{}_{A_{1}}...
\tau^{\mu_{k}}{}_{A_{k}}e^{\mu_{k+1}}{}_{a_{1}}...e^{\mu_{p+2}}{}_{a_{p+2-k}}\accentset{(\gamma)}{F}_{\mu_{1}...\mu_{p+2}}\,,
\end{align}
i.e.~the super-index refers to the power of $\cc{}$ in the curved index basis. Using this notation,
we get the following three types of terms in the expansion of the Maxwell Lagrangian term:
\begin{align}\label{three}
&F_{\mu_{1}...\mu_{p+2}}F^{\mu_{1}...\mu_{p+2}}=\sum_{k=\max\{0,2p+3-D\}}^{p+1}\, \binom{p+2}{k}\, F_{A_{1}...A_{k}a_{1}...a_{p+2-k}}F^{A_{1}...A_{k}a_{1}...a_{p+2-k}}=\nonumber\\
&=\sum_{k=\max\{0,2p+3-D\}}^{p+1}\, \binom{p+2}{k}\, \cc{2\xi-2k\alpha-2\beta(p+2-k)} \accentset{(\xi)}{F}_{A_{1}...A_{k}a_{1}...a_{p+2-k}}\accentset{(\xi)}{F}^{A_{1}...A_{k}a_{1}...a_{p+2-k}}+\nonumber\\
&+2\sum_{k=\max\{0,2p+3-D\}}^{p+1}\, \binom{p+2}{k}\, \cc{\xi+\gamma-2k\alpha-2\beta(p+2-k)} \accentset{(\gamma)}{F}_{A_{1}...A_{k}a_{1}...a_{p+2-k}}\accentset{(\xi)}{F}^{A_{1}...A_{k}a_{1}...a_{p+2-k}}+\nonumber\\
&+\sum_{k=\max\{0,2p+3-D\}}^{p+1}\, \binom{p+2}{k}\, \cc{2\gamma-2k\alpha-2\beta(p+2-k)} \accentset{(\gamma)}{F}_{A_{1}...A_{k}a_{1}...a_{p+2-k}}\accentset{(\gamma)}{F}^{A_{1}...A_{k}a_{1}...a_{p+2-k}}\,.
\end{align}
The lower bound in the sum, $\max\{0,2p+3-D\}$, is due to the fact it is in general not possible to write the  $(p+2)$-form with all flat transverse indices only. For example, in the case of a string foliation in three spacetime dimensions, i.e.~$p=1$ and $D=3$, one can only write down a 3-form with 2 longitudinal and 1 transverse index. The quantity $2p+3-D$ is the difference between the rank of the form and the number of transverse indices. We should not only require a cancellation of one of the types of divergences coming from the gauge field kinetic term against the leading part of the Einstein-Hilbert term, but we should also require that this gauge field kinetic term does not develop other divergences as well.  We call the three terms in eq.~\eqref{three} that are quadratic in the field strength as follows:
\begin{subequations}
\begin{align}
I_{1}(k)&=\accentset{(\xi)}{F}_{A_{1}...A_{k}a_{1}...a_{p+2-k}}\accentset{(\xi)}{F}^{A_{1}...A_{k}a_{1}...a_{p+2-k}}\,,\\
I_{2}(k)&=\accentset{(\xi)}{F}_{A_{1}...A_{k}a_{1}...a_{p+2-k}}\accentset{(\gamma)}{F}^{A_{1}...A_{k}a_{1}...a_{p+2-k}}\,,\\
I_{3}(k)&=\accentset{(\gamma)}{F}_{A_{1}...A_{k}a_{1}...a_{p+2-k}}\accentset{(\gamma)}{F}^{A_{1}...A_{k}a_{1}...a_{p+2-k}}\,.
\end{align}\label{eq:F2terms}
\end{subequations}
To evaluate $I_{1}(k)$ we note that from \autoref{eq:FExpansion} it follows that this term exists only for $k=p$ and $k=p+1$ at orders $\cc{2\xi-2p\alpha-4\beta}$ and $\cc{2\xi-2\beta-2(p+1)\alpha}$ respectively:
\begin{subequations}
\begin{align}
\accentset{(\xi)}{F}_{abA_{1}...A_{p}}&=t_{ab}{}^{B}\epsilon_{BA_{1}...A_{p}}\,,\label{eq:tabcancel}\\
\accentset{(\xi)}{F}_{aA_{1}...A_{p+1}}&=(-)^{p}\, (p+1)\, t_{a[ A_{1}}{}^{B}\epsilon_{A_{2}...A_{p+1}]B}=\epsilon_{A_{1}...A_{p+1}}t_{aB}{}^{B}\,.
\end{align}
\end{subequations}
The term in \autoref{eq:tabcancel} is needed to cancel the leading part coming from the Einstein-Hilbert term.  For this to happen we have to fix $\xi$ in such a way the powers of $\cc{}$ of the two terms match, i.e.
\begin{align}
\xi&=(p+1)\alpha\,. \label{eq:1}
\end{align}
Having fixed the parameter $\xi$ to this value, we give the order in $\cc{}$ and the allowed values of the parameter $k$ for the three different terms $I_{1}(k), I_{2}(k)$ and $I_{3}(k)$ in the table below.
\begin{table}[!ht]
\centering
\renewcommand{\arraystretch}{2.5}
\begin{center}
\resizebox{\textwidth}{!}{
\begin{tabular}{|c|c|c|}
\hline
\bf Term & \bf Order in $\mathbf \cc{}$ &\bf Existence\\
\hline
$I_{1}(k)=\accentset{(p\alpha + \alpha)}{F}_{A_{1}...A_{k}a_{1}...a_{p+2-k}}\accentset{(p\alpha+\alpha)}{F}^{A_{1}...A_{k}a_{1}...a_{p+2-k}}$&$2(\alpha-\beta)(p+1-k)-2\beta$&$k=p,p+1$\\
$I_{2}(k)=\accentset{(p\alpha+\alpha)}{F}_{A_{1}...A_{k}a_{1}...a_{p+2-k}}\accentset{(\gamma)}{F}^{A_{1}...A_{k}a_{1}..._{p+2-k}}$&$\gamma+\alpha(p+1-2k)-2\beta(p+2-k)$&$k=p,p+1$\\
$I_{3}(k)=\accentset{(\gamma)}{F}_{A_{1}...A_{k}a_{1}...a_{p+2-k}}\accentset{(\gamma)}{F}^{A_{1}...A_{k}a_{1}...a_{p+2-k}}$&$2\gamma-2k\alpha-2\beta(p+2-k)$&$k=\max\{0,2p+3-D\},...,p+1$\\
\hline
\end{tabular}
}
\end{center}
\caption{This table describes the three different terms coming from the Maxwell term in the action, the order in the power of $\cc{}$ and the allowed values of the number $k$ of longitudinal flat indices.}\label{Tab:IExtistence}
\end{table}
\FloatBarrier
The requirement that  the term $I_{2}(k)$ does not lead to a divergence  reads
\begin{align}
\gamma+\alpha(p+1-2k)-2\beta(p+2-k)\leqslant -2\beta\,,
\end{align}
for $k=p,\, p+1$. This  leads to the inequality
\begin{align}
\gamma\leqslant 2\beta+\alpha(p-1)\,.
\end{align}
Together with  eqs.~\autoref{eq:gammaboost} and \eqref{eq:1} this  fixes the parameter $\gamma$ as follows:
\begin{align}
\gamma=2\beta+\alpha(p-1).\label{eq:3}
\end{align}
Finally, the requirement that the term  $I_{3}(k)$ does not lead to a divergence leads to the inequality
\begin{align}
2\gamma-2k\alpha-2\beta(p+2-k)\leqslant-2\beta
\end{align}
for $k=\max\{0,2p+3-D\},...,p+1$. Plugging in the value just found for $\gamma$ we obtain
\begin{align}
(\alpha-\beta)(p-1-k)\leqslant 0\,.
\end{align}
Since we already found that $\alpha-\beta>0$, see \eqref{eq:nodivboost} this implies that
\begin{align}
p-1-k\leqslant 0\,.
\end{align}
Since this is an expression decreasing in $k$ it implies the following inequality:
\begin{align}
\min\{p-1,D-p-4\}\leqslant 0\,.\label{eq:4}
\end{align}

We have shown that the requirement that taking the non-relativistic limit gives both finite boost transformations and a finite action with a non-trivial term in the Ricci scalar expansion provided that the conditions \autoref{eq:nodivboost}, \autoref{eq:1}, \autoref{eq:3} and \autoref{eq:4} are satisfied. These conditions impose the inequalities 
\begin{subequations}
\begin{align}
\alpha-\beta&>0\,,\\
D&\geqslant p+3\,,\\
\min&\{D-p-4,p-1\}\leqslant 0
\end{align}
and fix the expansion parameters $\delta,\gamma$ and $\xi$ in terms of $\alpha, \beta$ and $p$ as follows:
\begin{align}
\delta&=\beta-\alpha\,,\\
\gamma&=(p+1)\alpha-2(\alpha-\beta)\,,\\
\xi&=(p+1)\alpha\,.
\end{align}\label{eq:conditions1}
\end{subequations}
We note that the inequalities given above only constrain the quantity $\alpha-\beta$. Therefore,  without loss of  generality we can set
\begin{align}
\alpha&=1\,,&\beta&=0\,.\label{eq:expansion1}
\end{align}
Then the inequalities and conditions \autoref{eq:conditions1} reduce to 
\begin{multieqref}[2]{eq:expansion2}
\delta&=-1\,,\\
\xi&=p+1\,,\\
\gamma&=p-1\,,\\
D-p-3&\geqslant 0\,,\\
\min\{D-p-4,p-1\}&\leqslant 0\,.
\end{multieqref}

As described above, the first equation in the second line is needed to match the order of power of $\cc{}$ between the leading terms coming from the Ricci scalar and the Maxwell term, the first equation in the first line is required to avoid divergences in the boost transformations, the inequality in the first line requires the transverse space to be at least 2-dimensional, while the purpose of the inequality in the second line is to avoid divergences in the action with the given Ansatz. \\

We note that the condition\,\footnote{Among all possibilities,  we do not discuss the extremal cases $p=-1$ corresponding to instantons with no longitudinal directions and a fully transverse Euclidean space and $p=D-1$ corresponding to space-filling branes with no transverse directions.}
\begin{align}
D\geqslant & p+3\hskip 1truecm {\rm or}\hskip 1truecm  p\leqslant D-3
\end{align}
 followed by the requirement that the leading term of the Einstein-Hilbert term is non-zero which requires that we have at least two transverse directions. This means that the Einstein-Hilbert term has a non-trivial leading term for  $p\in \{0,..., D-3\}$. Therefore, the only brane that we excluded by this requirement is the domain wall with $p=D-2$ that has a one-dimensional transverse space. In this case, no divergences are coming from the Ricci scalar. One may verify that in this case neither divergences are coming from the Maxwell term. Thus the theory is finite and we will include this special case although the absence of divergences for this special case does not require the gauge field Ansatz considered above.

Summarizing, we have shown that  the set of theories satisfying the conditions
\begin{multieqref}[2]{eq:NoDivCondition}
\xi&=p+1\,,\\
\gamma&=p-1\,,\\
0\leqslant &p\leqslant D-2\,,\\
\min\{D-p-4,p-1\}&\leqslant 0\label{eq:NoDivCondition4}\,,
\end{multieqref}
do not develop divergences neither in the action nor in the boost transformation when $\cc{}$ is sent to infinity, performing the non-relativistic limit. In the rest of this paper, we will refer to the condition  \autoref{eq:NoDivCondition4},
\begin{align}
\min\{D-p-4,p-1\}&\leqslant 0
\end{align}
as the {\bf no-divergence condition}.
\\

We have visualized the allowed brane spectrum in \autoref{tab:pDCases}.
\begin{table}[!h]
\centering
\renewcommand{\arraystretch}{1.5}
\begin{center}
\begin{tabular}{|c|c||*{9}{p{0.3cm}|}}
\cline{3-11}
\multicolumn{2}{c|}{}&\multicolumn{9}{c|}{\bf Dimensions}\\
\cline{3-11}
\multicolumn{2}{c|}{}&\bf 3&\bf 4&\bf 5&\bf 6&\bf 7&\bf 8&\bf 9&\bf 10&\bf 11\\
\hhline{--::=========}
&\bf 11&&&&&&&&&\\
\hhline{~-||---------}
&\bf 10&&&&&&&&&{\cellcolor{black}}\\
\hhline{~-||---------}

&\bf 9&&&&&&&&{\cellcolor{black}}&{\cellcolor{yellow}}\\
\hhline{~-||---------}
&\bf 8&&&&&&&{\cellcolor{black}}&{\cellcolor{yellow}}&{\cellcolor{green}}\\
\hhline{~-||---------}
&\bf 7&&&&&&{\cellcolor{black}}&{\cellcolor{yellow}}&{\cellcolor{green}}&{\cellcolor{green}}\\
\hhline{~-||---------}
&\bf 6&&&&&{\cellcolor{black}}&{\cellcolor{yellow}}&{\cellcolor{green}}&{\cellcolor{green}}&{\cellcolor{red}}\\
\hhline{~-||---------}
&\bf 5&&&&{\cellcolor{black}}&{\cellcolor{yellow}}&{\cellcolor{green}}&{\cellcolor{green}}&{\cellcolor{red}}&{\cellcolor{red}}\\
\hhline{~-||---------}
&\bf 4&&&{\cellcolor{black}}&{\cellcolor{yellow}}&{\cellcolor{green}}&{\cellcolor{green}}&{\cellcolor{red}}&{\cellcolor{red}}&{\cellcolor{red}}\\
\hhline{~-||---------}

&\bf 3&&{\cellcolor{black}}&{\cellcolor{yellow}}&{\cellcolor{green}}&{\cellcolor{green}}&{\cellcolor{red}}&{\cellcolor{red}}&{\cellcolor{red}}&{\cellcolor{red}}\\
\hhline{~-||---------}

&\bf 2&{\cellcolor{black}}&{\cellcolor{yellow}}&{\cellcolor{green}}&{\cellcolor{green}}&{\cellcolor{red}}&{\cellcolor{red}}&{\cellcolor{red}}&{\cellcolor{red}}&{\cellcolor{red}}\\
\hhline{~-||---------}

&\bf 1&{\cellcolor{yellow}}&{\cellcolor{green}}&{\cellcolor{green}}&{\cellcolor{green}}&{\cellcolor{green}}&{\cellcolor{green}}&{\cellcolor{green}}&{\cellcolor{green}}&{\cellcolor{green}}\\
\hhline{~-||---------}

 \parbox[t]{2mm}{\multirow{-11}{*}{\rotatebox[origin=c]{90}{\bf p-Brane}}}&\bf 0&{\cellcolor{green}}&{\cellcolor{green}}&{\cellcolor{green}}&{\cellcolor{green}}&{\cellcolor{green}}&{\cellcolor{green}}&{\cellcolor{green}}&{\cellcolor{green}}&{\cellcolor{green}}\\

\hhline{--||---------}

\end{tabular}
\end{center}
\caption{This table shows for which cases for dimensions $3 \le D \le 11$ the conditions \autoref{eq:NoDivCondition} are satisfied. A white cell corresponds to a theory/foliation not admitted. Green cells signal branes satisfying the no-divergence condition \autoref{eq:NoDivCondition4} and red cells correspond to branes that do not satisfy this condition. Black cells denote the space-filling branes that we do not treat, while yellow cells correspond to domain wall branes for which the Ricci scalar is not divergent.\label{tab:pDCases}}
\end{table}
\\

We notice a few regularities in the table. Starting from $D=11$, all domain walls, defect branes and branes with three transverse directions are related to each other via a double dimensional reduction. This means that if for these cases the no-divergence condition is satisfied for a given value of $(D,p)$ then the same no-divergence condition is also satisfied for the double-dimensionally reduced brane with values $(D-1,p-1)$.
The case of domain walls, i.e.~$p=D-2$ is rather special. In that case the $(D-1)$-form potential is equal to an integration constant that can be identified as a cosmological constant. The domain wall can interpolate between two different values of this cosmological constant.
The 10D-case with a 9-form potential is part of the  IIA massive supergravity theory of Romans \cite{Romans:1985tz}.

Another regularity is that there are no restrictions for $p=0$, i.e.~particles, neither on their `dual' $(D-4)$-branes.\,\footnote{We use the word `dual' here in a loose way ignoring the specific values of the scalar field coupling to the gauge field and dual gauge field used in our Ansatz.}
We should distinguish between the non-relativistic duality between fundamental branes versus non-fundamental or solitonic branes. We will call two fundamental branes dual if they are the limits of two relativistic branes that are dual to each other. Due to the different foliations we are using in the limits we cannot define a duality directly between the two non-relativistic fundamental branes. For solitonic branes one can use the same foliation as for the corresponding fundamental brane and define a non-relativistic duality both before as well as after taking the limit.
This has, for instance, been done for the $\mathcal{N}=1$  solitonic NS-NS 5-brane solution which allows a non-relativistic limit with a string foliation and a Newton potential that is harmonic in the 4 transverse directions \cite{Bergshoeff:2022pzk}. We note that there is a second diagonal with $(D-3)$-branes which can be considered to be `dual' to $(p=-1)$-branes, i.e.~instantons.\\

In the case of strings, i.e.~$p=1$, there are no restrictions either.
Remarkably,  the fine-tuning of the scalar field coupling in $10D$ given by $\coefb/\coefa^2=1$ precisely coincides with the dilaton coupling of the $\mathcal{N}=1\ 10D$ supergravity theory. Concerning duality, we see that for $D>6$ none of the strings has a corresponding `dual' $(D-5)$-brane.\,\footnote{Note that for $D=6$ the dual of a string is again a string.}

Substituting the conditions  \autoref{eq:NoDivCondition}  back into the expansion \autoref{eq:FExpansion} we find the simplified expression
\begin{align}
F_{\mu_{1}...\mu_{p+2}}&=
\cc{p+1}\,\accentset{(p+1)}{F}_{\mu_{1}...\mu_{p+2}}+
\cc{p-1}\,\accentset{(p-1)}{F}_{\mu_{1}...\mu_{p+2}}=\nonumber\\
&=\frac{1}{2}\cc{p+1}(p+2)(p+1)\, t_{[\mu_{1}\mu_{2}}{}^{A_{1}}\tau_{\mu_{3}}{}^{A_{2}}...\tau_{\mu_{p+2}]}{}^{A_{p+1}}\epsilon_{A_{1}...A_{p+1}}+\cc{p-1}f_{\mu_{1}...\mu_{p+2}}\,.\label{eq:FExpansion2}
\end{align}
implying
\begin{align}
F_{\mu_{1}...\mu_{p+2}}F^{\mu_{1}...\mu_{p+2}}&=-\frac{1}{2}\cc{2}(p+2)! t_{abA}t^{abA} +\mathcal{O}(\cc{0})\,.
\end{align}
This ensures the cancellation of the divergent term in the Lagrangian,\autoref{eq:EHleading}.
\\

Instead of taking a Newton-Cartan limit, which is the principle focus of this work, our calculations can also be used to take an electric Galilei limit in the presence of the gauge field and the scalar field. In that case, we do not cancel leading divergences and consider the sub-leading terms but, instead, we use these leading terms to construct an electric Galilei action. In the presence of the Maxwell term this only works if this term does not produce terms that are of higher order in $\cc{}$ than the leading term in the Einstein-Hilbert action. We give an example of such an electric Galilei limit in  \autoref{sec:ElectricGalilei2}.

\subsection{The Non-Relativistic Action}
We now consider the resulting finite action. We assume that the indices $A$ and $a$ take at least one value. We first define a non-relativistic connection as
\begin{align}
\accentset{(nr)}{\Gamma}_{\mu\nu}^{\rho}&=\frac{1}{2}h^{\rho\xi}(\partial_{\mu}h_{\nu\xi}+\partial_{\nu}h_{\mu\xi}-\partial_{\xi}h_{\mu\nu})+\tau^{\rho}{}_{A}\partial_{\mu}\tau_{\nu}{}^{A}\,,
\end{align}
such that
\begin{multieq}[2]
\accentset{(nr)}{\nabla}_{\mu}\tau_{\nu\rho}&=0\,,\\
\accentset{(nr)}{\nabla}_{\mu}h^{\nu\rho}&=0\,,
\end{multieq}
where $\accentset{(nr)}{\nabla}_{\mu}$ denotes the covariant derivative containing only the non-relativistic connection defined above.  This induces the definition of a non-relativistic Ricci tensor
\begin{align}
\accentset{(nr)}{\rm R}_{\mu\nu}&=\partial_{\rho}\accentset{(nr)}{\Gamma}_{\nu\mu}^{\rho}-\partial_{\nu}\accentset{(nr)}{\Gamma}_{\rho\mu}^{\rho}+\accentset{(ur)}{\Gamma}_{\rho\xi}^{\rho}\accentset{(nr)}{\Gamma}_{\nu\mu}^{\xi}-\accentset{(nr)}{\Gamma}_{\nu\xi}^{\rho}\accentset{(nr)}{\Gamma}_{\rho\mu}^{\xi}\,.
\end{align}
We adopt the following convention
\begin{align}
\nabla_{\mu}T_{\nu}{}^{\rho}\equiv \partial_{\mu}T_{\nu}{}^{\rho}-\Gamma_{\mu\nu}^{\xi}T_{\xi}{}^{\rho}+\Gamma_{\mu\xi}^{\rho}T_{\nu}{}^{\xi}\,,
\end{align}
for a generic tensor $T_{\mu}{}^{\nu}$ and $\nabla_{\mu}$ a generic covariant derivatice. The connection we just introduced also satisfies
\begin{align}
\accentset{(nr)}{\nabla}_{\mu}\tau_{\nu}{}^{A}&=0\,.
\end{align}
Expanding all the fields, we find at zero order in $\cc{}$ the following finite action:
\begin{align}
\mathcal{S}_{NR}&=\int\,d^{D}x\, E\, {\rm e}^{-\coefa\phi}\mathcal{L}_{NR}=\nonumber\\
&=\int\,d^{D}x\, E\, {\rm e}^{-\coefa\phi}\bigg[e^{\nu}\,_{a} e^{\mu a}\accentset{(nr)}{{\rm R}}_{\mu\nu}-2h^{\mu\nu}\tau^{\rho}{}_{A}\accentset{(nr)}{\nabla}_{\mu}t_{\nu \rho}{}^{A} -\frac{1}{2}t_{aA}{}^{A} t^{a B}\,_{B} - t_{a A B}t^{a (A B)} +\nonumber\\
&-\frac{1}{12(p-1)!}f_{a_{1}a_{2}a_{3}A_{1}...A_{p-1}}f^{a_{1}a_{2}a_{3}A_{1}...A_{p-1}}-\frac{1}{2\, p!}f^{abA_{1}...A_{p}}t_{ab}{}^{B}\epsilon_{BA_{1}...A_{p}}+\coefb\partial_{a}\phi\partial^{a}\phi\bigg]\,,\label{eq:NRAction}
\end{align}
where the first term in the last line only exists for $p\geqslant 1$ and $D\geqslant p+4$ while the second term in the same line exists only for $D\geqslant p+3$. This implies they both terms vanish in the case of a domain-wall foliation.\\

As a check,  it is interesting to verify if the action that we have just obtained is invariant under boost transformations.  Denoting  the different terms of the Lagrangian in \autoref{eq:NRAction} by
\begin{subequations}
\begin{align}
\mathcal{L}_{\rm R}&=e^{\nu}\,_{a} e^{\mu a}\accentset{(nr)}{{\rm R}}_{\mu\nu}-2h^{\mu\nu}\tau^{\rho}{}_{A}\accentset{(nr)}{\nabla}_{\mu}t_{\nu \rho}{}^{A}-\frac{1}{2}t_{aA}{}^{A} t^{a B}\,_{B} - t_{a A B}t^{a (A B)}\,,\\
\mathcal{L}_{F}&=-\frac{1}{12(p-1)!}f_{a_{1}a_{2}a_{3}A_{1}...A_{p-1}}f^{a_{1}a_{2}a_{3}A_{1}...A_{p-1}}-\frac{1}{2\, p!}f^{abA_{1}...A_{p}}t_{ab}{}^{B}\epsilon_{BA_{1}...A_{p}}\,,\\
\mathcal{L}_{\phi}&=\coefb\partial_{a}\phi\partial^{a}\phi\,,
\end{align}
\end{subequations}
and using that under boost transformations we have
\begin{subequations}
\begin{align}
\delta (h^{\mu\nu}\accentset{(nr)}{\rm R}_{\mu\nu})=&
-2\lambda^{Aa}h^{\mu\rho}e^{\nu}{}_{a}\accentset{(nr)}{\nabla}_{\mu}t_{\nu\rho A}-3\partial_{\mu}\lambda^{Aa}e^{\mu b}t_{abA}+\frac{3}{2}\lambda^{Aa}t^{bc}{}_{A}z_{bca}+\lambda^{Aa}t_{abA}t_{B}{}^{bB}\,,\\
\delta (h^{\mu\nu}\tau^{\rho}{}_{A}\accentset{(nr)}{\nabla}_{\mu}t_{\nu\rho}{}^{A})=&-\lambda^{Aa}h^{\mu\rho}e^{\nu}{}_{a}\accentset{(nr)}{\nabla}_{\mu}t_{\nu\rho A}-\partial_{\mu}\lambda^{Aa}e^{\mu b}t_{abA}+\frac{1}{2}\lambda^{Aa}t^{bc}{}_{A}z_{bca}-\lambda^{Aa}t_{abA}t_{B}{}^{bB}\,,\\
\delta t_{\mu\nu}{}^{A}=&0\,,\\
\delta a_{\mu_1 ...\mu_{p+1}}=&-(p+1)\lambda^{A_{1}}{}_{b}e_{[\mu_{1}}{}^{b}\tau_{\mu_{2}}{}^{A_{2}}...\tau_{\mu_{p+1}]}{}^{A_{p+1}}\epsilon_{A_{1}...A_{p+1}}\,,
\end{align}
\end{subequations}
we find the following boost transformations of the different terms in the Lagrangian:
\begin{align}
\delta \mathcal{L}_{\rm R}&=\lambda^{A a} t_{a}\,^{b}\,_{A} t^{B}\,_{b B}-\lambda^{A a} t_{a}\,^{b B} t_{A b B}+\frac{1}{2}\lambda^{A a} t^{b c}\,_{A} z_{b c a}-\partial^{b}{\lambda_{A}\,^{a}} t_{a b}\,^{A}\,,\\
\delta \mathcal{L}_{F}&=-\delta \mathcal{L}_{\rm R}-\frac{1}{6}\lambda_{A_{1}}{}^{d}f_{abcdA_{2}...A_{p-1}}f^{abcA_{1}...A_{p-1}}\,,\\
\delta \mathcal{L}_{\phi}&=0\,,
\end{align}
where $z_{\mu\nu}{}^{a} = 2\partial_{[\mu}  e_{\nu]}{}^a$. This implies the  following boost transformation of the action:
\begin{align}
\delta \mathcal{S}&= -\frac{1}{6}\int d^{D}x\, E{\rm e}^{-\coefa \phi}\, \lambda_{B}{}^{d}f_{abcd A_{1}...A_{p-2}}f^{abcBA_{1}...A_{p-2}}\,.
\end{align}
This term is non-zero/exists only for
\begin{align}
D-p-1&\geqslant 4\,,&p\geqslant 2\,,
\end{align}
but these conditions are inconsistent with the no-divergence condition , \autoref{eq:NoDivCondition}, that we report here for convenience,
\begin{align}
\min \{D-p-1,p+2\}\leqslant 3\,.
\end{align}
Therefore, for all the green branes in \autoref{tab:pDCases}, the action is invariant under boost transformations.

\section{An Emerging Local Dilatation Symmetry}\label{sec:localdil}

As discussed in the introduction, an interesting feature of taking a Newton-Cartan limit is the emergence of a local dilatation symmetry in some cases. In this section we investigate, under which conditions the theories that already satisfy the no-divergence condition \autoref{eq:NoDivCondition} exhibit such an emerging local dilatation symmetry. For this purpose,  we consider the non-relativistic Lagrangian \autoref{eq:NRAction}, perform a general local rescaling of the fields with parameter $\lambda_D$, and investigate under which condition this is indeed a local symmetry of the non-relativistic action.\\

We consider the following general local dilatation transformations
\begin{subequations}\label{localD}
\begin{align}
\delta_{D}\tau_{\mu}{}^{A}&=\alpha \dil \tau_{\mu}{}^{A}\,,\\
\delta_{D}e_{\mu}{}^{a}&=\beta \dil e_{\mu}{}^{a}\,,\\
\delta_{D}\tau^{\mu}{}_{A}&=-\alpha \dil \tau^{\mu}{}_{A}\,,\\
\delta_{D}e^{\mu}{}_{a}&=-\beta \dil e^{\mu}{}_{a}\,,\\
\delta_{D} a_{\mu_{1}...\mu_{p+1}}&=\gamma \dil a_{\mu_{1}...\mu_{p+1}} \,,\\
\delta_{D}\phi&=\delta \dil\,,
\end{align}
\end{subequations}
where $\alpha,\beta ,\gamma$ and $\delta$ are constant parameters, while $\dil$ is local. We note that the scalar field undergoes a shift under a local dilatation. The
variation of the non-relativistic action under these local dilatations is given by
\begin{align}
\delta_{D} \mathcal{S}_{NR}&=\int d^{D}x\,
\, 2[\beta(2-D+p)-(p+1)\alpha]\partial_{\mu}(E{\rm e}^{-\coefa\phi}h^{\mu\nu}\partial_{\nu}\dil)+\nonumber\\
&+E{\rm e}^{-\coefa\phi}\Bigg\{
\dil [\alpha(p+1)+\beta (D-p-3)-\coefa\delta] \mathcal{L}_{NR}+\nonumber\\
&-\frac{1}{6(p-1)!}f^{abcA_{1}...A_{p-1}}\Big[\dil[\gamma-2\beta-\alpha(p-1)]f_{abcA_{1}...A_{p-1}}+\nonumber\\
&+\gamma(p+2) e^{\mu_{1}}{}_{a}e^{\mu_{2}}{}_{b}e^{\mu_{3}}{}_{c}\tau^{\mu_{4}}{}_{A_{1}}...\tau^{\mu_{p+2}}{}_{A_{p-1}}\partial_{[\mu_{1}}\dil a_{\mu_{2}...\mu_{p+2}]}\Big]+\nonumber\\
&\textcolor{black}{-\frac{1}{2\, p!}\dil [\gamma-2\beta-(p-1)\alpha]f^{abA_{1}...A_{p}}t_{ab}{}^{B}\epsilon_{BA_{1}...A_{p}}+}\nonumber\\
&\textcolor{black}{-\frac{1}{2\, p!}\gamma e^{\mu_{1}}{}_{a}e^{\mu_{2}}{}_{b}\tau^{\mu_{3}}{}_{A_{1}}...\tau^{\mu_{p+2}}{}_{A_{p}}\partial_{[\mu_{1}}\dil a_{\mu_{2}...\mu_{p+2}]}t^{ab}{}_{B}\epsilon^{BA_{1}...A_{p}}}+\nonumber\\
&+2\Big[\coefb\delta+\coefa\beta(p+2-D)-\coefa\alpha(p+1)\Big]h^{\mu\nu}\partial_{\mu}\dil\partial_{\nu}\phi+\nonumber\\
&+\Big[-2\beta+\alpha(1-p)\Big]e^{\mu a}\partial_{\mu}\dil t_{Aa}{}^{A}\Bigg\}\,,\label{eq:SDil}
\end{align}
where the terms in the third and fourth lines are non-zero only for $p\geqslant 1$. We assume that  $p\leqslant D-3$, such that all the terms in the action can exist. Invariance under local dilatation requires all the coefficients in \autoref{eq:SDil}, except the one in the first line multiplying the total derivative term, to vanish. Note that the terms in the third and fourth lines of \autoref{eq:SDil} do not impose any condition that has not yet already been imposed by another term. The resulting system of equations admits the following solutions
\begin{align}
\beta&=\frac{1-p}{2}\alpha\,,&\gamma&=0\,,&\delta&=
\frac{\alpha}{2\coefa}\Big[p^2+p(4-D)+D-1\Big]\,,&\frac{\coefb}{\coefa^2}&=\frac{p^2+p(3-D)+D}{p^2+p(4-D)+D-1}\,. \label{eq:DilCond}
\end{align}
One can show that for theories fulfilling the no-divergence condition the denominator in $\coefb/\coefa^2$  is non-zero. The solutions we have found imply that all theories that satisfy the no-divergence condition also exhibit an emerging local dilatation symmetry provided that the scalar field coupling corresponding to $\coefa$ and $\coefb$  is fine-tuned according to \eqref{eq:DilCond}.\\

Among the solutions above we find for the case of strings:
\begin{align}
p&=1\,,&\beta&=0\,,&\gamma&=0\,,&\delta&=
\frac{2\alpha}{\coefa}\,,&\frac{\coefb}{\coefa^2}&=1\,.
\end{align}
The interesting aspect of this case is that the scalings are independent of the dimension and that the ratio $\coefb/\coefa^2$ corresponds to that of ten-dimensional supergravity.
We can easily include the case of domain walls, i.e.~$p=D-2$. In that case the only difference is that the parameter $\gamma$ is arbitrary, while for the other parameters the same solutions hold. \\

We stress that the scalar field plays a pivotal role in the emergence of the local dilatation symmetry. For all cases where one of the coefficients $\coefa$ or $\coefb$ is zero, we cannot establish a local dilatation symmetry. Note, however, that the theories without fine-tuning that do not satisfy the conditions \autoref{eq:DilCond} and therefore do not have a local dilatation symmetry, are still valid theories as long as the no-divergence condition is satisfied.

\section{The Poisson Equation}\label{sec:Poisson}
In this section, we will study under which conditions the non-relativistic limit of the equations of motion that follows from the relativistic action \autoref{eq:RelAction} contains the Poisson equation. Although our starting point is given by the equations of motion coming from an action, the procedures that we apply in this section neither rely on the existence of an action nor imply it.

\subsection{Non-Relativistic Limit of the Equations of Motion}
Taking the non-relativistic limit of an equation of motion amounts to select its leading order term in the corresponding $\cc{}$  expansion.

However, in doing so,  it is important to organize the equations of motion in linear combinations such as to minimize the amount of $\cc{}$ powers in their expansion. For instance, let us suppose that there are two equations, say $[X]=0$ and $[Y]=0$ with leading order terms in the $\cc{}$ expansion proportional to each other, for example
\begin{align}
[X]&=\cc{n}\accentset{(n)}{[X]}+\cc{n-2}\accentset{(n-2)}{[X]}+\mathcal{O}(\cc{n-4})\,,&[Y]&=\cc{n}\accentset{(n)}{[X]}+\cc{n-2}\accentset{(n-2)}{[Y]}+\mathcal{O}(\cc{n-4})\,.
\end{align}
In that case it is important to combine them, before taking the limit, in such a way that in one of the two linearly independent combinations the term at order $\cc{n}$ cancels. Following the example this would lead to considering the combinations
\begin{subequations}
\begin{align}
[X]+[Y]&=2\cc{n}\accentset{(n)}{[X]}+\cc{n-2}\Big(\accentset{(n-2)}{[X]}+\accentset{(n-2)}{[Y]}\Big)+\mathcal{O}(\cc{n-4})\,,\\
[X]-[Y]&=\cc{n-2}\Big(\accentset{(n-2)}{[X]}-\accentset{(n-2)}{[Y]}\Big)+\mathcal{O}(\cc{n-4})\,.
\end{align}
\end{subequations}
It is clear that taking the limit of the two equations $[X]$ and $[Y]$ before or after combining them as above leads to two different results. In particular, the procedure described above prevents us from losing an equation of motion in the limit.   Studying the possible combinations and the different leading order terms is a fundamental step to setting up the limit properly.\\

The equations of motion expand as follows
\begin{subequations}
\begin{alignat}{3}
[G]_{Aa}&=\cc{}&&\accentset{(1)}{[G]}_{Aa}+\mathcal{O}(\cc{-3})\,,\\
[G]_{\{AB\}}&=&&\accentset{(0)}{[G]}_{\{AB\}}+\mathcal{O}(\cc{-2})\,,\\
[G]_{A}{}^{A}&=\cc{2}&&\accentset{(2)}{[G]}_{A}{}^{A}+\mathcal{O}(\cc{0})\,,\\
[G]_{ab}&=&&\accentset{(0)}{[G]}_{ab}+\mathcal{O}(\cc{-2})\,,\\
[\Phi]&=&&\accentset{(0)}{[\Phi]}+\mathcal{O}(\cc{-2})\,,\\
[A]_{A_{1}...A_{p+1}}&=\cc{2}&&\accentset{(2)}{[A]}_{A_{1}...A_{p+1}}+\mathcal{O}(\cc{0})\\
[A]_{A_{1}...A_{p}a}&=\cc{}&&\accentset{(1)}{[A]}_{A_{1}...A_{p}a}+\mathcal{O}(\cc{-1})\\
[A]_{A_{1}...A_{p-1}a_{1}a_{2}}&=&&\accentset{(0)}{[A]}_{A_{1}...A_{p-1}a_{1}a_{2}}+\mathcal{O}(\cc{-2})\,,\\
[A]_{A_{1}...A_{p-2}a_{1}a_{2}a_{3}}&=\cc{}&&\accentset{(1)}{[A]}_{A_{1}...A_{p-2}a_{1}a_{2}a_{3}}+\mathcal{O}(\cc{-1})\,,
\end{alignat}\label{eq:EOMExpansion}
\end{subequations}
where the explicit expressions of the different leading order terms in the expansion can be found in \autoref{sec:appB} and the next section. In \autoref{sec:appB} we also show how the no-divergence condition removes some terms in the expansion that otherwise would have been of leading order. This is crucial to ensure the closure of the multiplet of non-relativistic equations of motion under the symmetries of the theory. This shows that the no-divergence condition also can play a crucial role in the constrained on-shell approach.\\

To properly perform the limit we define the following combinations
\begin{subequations}
\begin{align}
[V_{\pm}]_{Aa}&=[G]_{Aa}\mp\frac{1}{2p!}[A]_{a}{}^{B_{1}...B_{p}}\epsilon_{AB_{1}...B_{p}}\,,\label{eq:Vpm}\\
[P_{\pm}]&=\CA [G]_{A}{}^{A}+\CB [G]_{a}{}^{a}\pm \CC [A]^{A_{1}...A_{p+1}}\epsilon_{A_{1}...A_{p+1}}+\CD [\Phi]\,,\label{eq:Ppm}
\end{align}
\end{subequations}
where  $\CA,\CB,\CC$ and $\CD$ are arbitrary coeffcients.
The two combinations given in eq.~\autoref{eq:Vpm} expand as
\begin{align}
[V_{+}]_{Aa}&=\cc{-1}\accentset{(-1)}{[V_{+}]}_{Aa}+\mathcal{O}(\cc{-3})\,,\\
[V_{-}]_{Aa}&=\cc{+1}\accentset{(1)}{[V_{-}]}_{Aa}+\mathcal{O}(\cc{-1})\,,
\end{align}
reducing the number of terms occurring at order $\cc{}$ by one. The second two combinations require a more refined analysis. For one of them, say $[P_{+}]$, the emergence of the local dilatation symmetry \eqref{localD}, by the second Noether's theorem, could help us in reducing the power of $\cc{}$ of the leading term even further, since the combination corresponding to the Noether identity will ensure that $\accentset{(0)}{[P_{+}]}=0$.\\

Physically, for the limit to describe Newton-Cartan gravity and its foliated generalizations, we expect one of the combinations $[P_{\pm}]$ to provide, after taking the limit, the analogue of the Poisson equation, that is the marker of Newton-Cartan gravity. In the next subsection we will study the appearance of this Poisson equation in detail.

\subsection{Poisson Equation and Dilatation Symmetry}\label{sec:PoissonDilatation}

As we will see the presence of the Poisson equation in the limit of the equations of motion is strictly related to the emergence of a local dilatation symmetry. This was already observed in the case of $(D,p) = (10,1)$ \cite{Bergshoeff:2021bmc}. Since we know that the Poisson equation is a scalar equation we consider the most general scalar combinations of \autoref{eq:Ppm} focusing on one of them, say $[P_{+}]$\,:\,\footnote{For $D=2p+2$ another term could be added that is $[A]^{a_{1}...a_{p+1}}\epsilon_{a_{1}...a_{p+1}}$. Due to the no-divergence condition the transverse space could be at most three-dimensional for $p>1$ Thus, the only two cases with $D\geqslant 3$ where this term could exist are $p=1, D=4$ and $p=2, D=6$. In these cases we have
\begin{subequations}
\begin{align}
D&=4\,, p=1\,,&\accentset{(0)}{[A]}_{ab}&=e^{\mu}{}_{a}e^{\nu}{}_{b}h^{\alpha\beta}\,\accentset{(nr)}{\nabla}_{\alpha}f_{\beta\mu\nu}+\coefa \partial^{A}\phi t_{ab}{}^{B}\epsilon_{AB}+t_{A}{}^{BA}\epsilon_{BC}t_{ab}{}^{C}\,,\\
D&=6\,, p=2\,,&\accentset{(1)}{[A]}_{abc}&=e^{\mu}{}_{a}e^{\nu}{}_{b}e^{\rho}{}_{c}h^{\alpha\beta}\,\accentset{(nr)}{\nabla}_{\alpha}f_{\beta\mu\nu\rho}\,.
\end{align}
\end{subequations}
These terms will not play any roles in the cancellation, since they contain terms not appearing in the expansion of the other contributions, so we prefer to stick directly to \eqref{eq:Poissoneq}.}
\begin{align}
[P_{+}]=\CA [G]_{A}{}^{A}+\CB [G]_{a}{}^{a}+\CC [A]^{A_{1}...A_{p+1}}\epsilon_{A_{1}...A_{p+1}}+\CD [\Phi]\,,\label{eq:Poissoneq}
\end{align}
where the relativistic equations of motions with flat indices are defined in \autoref{eq:relEOMflat}. The expansion of  \eqref{eq:Poissoneq} reads
\begin{align}
[P_{+}]=\cc{2}\,\accentset{(2)}{[P_{+}]}+\,\accentset{(0)}{[P_{+}]}+\cc{-2}\,\accentset{(-2)}{[P_{+}]}+...\,,
\end{align}
where the dots denote terms of lower order in $\cc{}$. Explicitly, we find the following expressions:
\begin{subequations}
\begin{align}
\accentset{(2)}{[P_{+}]}&=\frac{p+1}{4}t_{abA}t^{abA}\Big(\CA +2\CC p!\Big)\,,\\
\nonumber\\
\accentset{(0)}{[P_{+}]}&=-\bigg[\CB+\CA+(p+1)!\CC+\frac{2\coefa}{\coefb}\CD\bigg]\, e^{\mu a}e^{\nu}{}_{a}\tau^{\rho}{}_{A}\accentset{(nr)}{\nabla}_{\mu}t_{\nu\rho}{}^{A}+\nonumber\\
&-\frac{1}{2p!}\bigg[2\CB+p\CA+\frac{\coefa }{\coefb}\CD+(p+1)!\CC\bigg]\epsilon^{A_{1}...A_{p+1}}t^{ab}{}_{A_{1}}f_{abA_{2}...A_{p+1}}+\nonumber\\
&-\frac{1}{12(p-1)!}\bigg[3\CB+(p-1)\CA+\frac{\coefa }{\coefb}\CD\bigg]f_{abcA_{1}...A_{p-1}}f^{abcA_{1}...A_{p-1}}+\nonumber\\
&+\frac{1}{2}t_{aA}{}^{A}t^{aB}{}_{B}\bigg[\CB+(p-1)\CA-\frac{\coefa }{\coefb}\CD\bigg]+\nonumber\\
&-\bigg[\coefa\CA-\frac{1}{\coefa}(\coefa^2-\coefb)(\delta_{B}^{B}\CA+\CB\delta_{b}^{b})+2\CD+\coefa (p+1)!\CC\bigg]t_{A}{}^{aA}\partial_{a}\phi+
\nonumber\\
&+\bigg[\CB\coefa-\frac{1}{\coefa}(\coefa^2-\coefb)(\delta_{B}^{B}\CA+\CB\delta_{b}^{b})+2\CD\bigg]e^{\mu a}e^{\nu}{}_{a}\accentset{(nr)}{\nabla}_{\mu}\partial_{\nu}\phi+\nonumber\\
&-\Big[(\coefa^2-\coefb)(\CB-\delta_{B}^{B}\CA-\CB\delta_{b}^{b})+\coefa \CD\Big]e^{\mu a}e^{\nu}{}_{a}\partial_{\mu}\phi\partial_{\nu}\phi+\nonumber\\
&+\bigg(\CB+\frac{\coefa}{\coefb}\CD\bigg)\Big[\,\accentset{(nr)}{\rm R}_{a}{}^{a}-t_{aAB}t^{a(AB)}\Big]\,,\label{eq:P0}\\
\nonumber\\
\accentset{(-2)}{[P_{+}]}&=\bigg(\CB+\frac{\coefa}{\coefb}\CD\bigg)\Big(\tau^{\rho A}\tau^{\mu}{}_{A}e^{\nu}{}_{a}\accentset{(nr)}{\nabla}_{\mu}z_{\nu\rho}{}^{a}-t^{AB}{}_{B}z_{Aa}{}^{a}-z_{Aa}{}^{a}z^{Ab}{}_{b}-z^{Aab}z_{A[ab]}\Big)+\nonumber\\
&+\bigg(\CA+\frac{\coefa}{\coefb}\CD\bigg)\bigg[\,\accentset{(nr)}{\rm R}_{A}{}^{A}+2\tau^{\mu}{}_{B}\tau^{\rho B}\tau^{\nu}{}_{A}\accentset{(nr)}{\nabla}_{\mu}t_{\nu\rho}{}^{A}-t^{AB}{}_{B}(t_{AC}{}^{C}+z_{Aa}{}^{a})+\nonumber\\
&-\frac{1}{4}t^{ABC}(t_{ABC}+2t_{ACB})+\frac{1}{2}t^{AaB}z_{ABa}\bigg]+\nonumber\\
&+\CC\tau^{\mu_{1}}{}_{A_{1}}...\tau^{\mu_{p+1}}{}_{A_{p+1}}\epsilon^{A_{1}...A_{p+1}}h^{\alpha\nu}\accentset{(nr)}{\nabla}_{\alpha}f_{\nu\mu_{1}...\mu_{p+1}}-\frac{1}{4p!}
\bigg(2\CB+p\CA+\frac{\coefa}{\coefb}\CD\bigg)f_{A_{1}...A_{p}ab}f^{A_{1}...A_{p}ab}+\nonumber\\
&-\frac{1}{p!}\bigg(\CB+(p+1)\CA+\frac{\coefa}{\coefb}\CD\bigg)f^{aA_{1}...A_{p+1}}t_{aA_{1}}{}^{B}\epsilon_{BA_{2}...A_{p+1}}
-\coefa B\epsilon^{A_{1}...A_{p+1}}\partial^{b}\phi\, f_{bA_{1}...A_{p+1}}+\nonumber\\
&+\bigg[\coefa\CA-\frac{1}{\coefa}(\coefa^{2}-\coefb)(\delta_{C}^{C}\CA+\CB\delta_{b}^{b})+2\CD\bigg]\Big(\tau^{\mu}{}_{A}\tau^{\nu A}\accentset{(nr)}{\nabla}_{\mu}\partial_{\nu}\phi+t^{BA}{}_{A}\partial_{B}\phi\Big)+\nonumber\\
&+\bigg[\CB\coefa-\frac{1}{\coefa}(\coefa^{2}-\coefb)(\delta_{C}^{C}\CA+\CB\delta_{b}^{b})+2\CD\bigg]z^{Aa}{}_{a}\partial_{A}\phi+\nonumber\\
&-\bigg[(\coefa^{2}-\coefb)(\CA-\delta_{C}^{C}\CA-\CB\delta_{b}^{b})+\coefa \CD\bigg]\partial^{A}\phi\partial_{A}\phi\,,
\end{align}\label{eq:PoissonExpansion}
\end{subequations}
where $z_{\mu\nu}{}^{a}$ and $t_{\mu\nu}{}^{A}$ are defined in \autoref{sec:Notation}. We see a Poisson equation arising at order $\cc{-2}$ due to the following term in  $\accentset{(-2)}{[P_{+}]}$ :
\begin{align}
\tau^{\mu_{1}}{}_{A_{1}}...\tau^{\mu_{p+1}}{}_{A_{p+1}}\epsilon^{A_{1}...A_{p+1}}h^{\alpha\nu}
\accentset{(nr)}{\nabla}_{\alpha}f_{\nu\mu_{1}...\mu_{p+1}}\,.
\end{align}
This term  produces a contribution of the form
\begin{align}
h^{\mu\nu}\partial_{\mu}\partial_{\nu}a_{A_{1}...A_{p+1}}
\end{align}
that can be identified with the characteristic Laplacian term acting on the Newton's potential. For further details we refer to \cite{Andringa:2012uz, Bergshoeff:2021bmc}. In order to obtain a Poisson equation at order $\cc{-2}$ all higher-order terms, i.e. all terms of  order $\cc{2}$ and $\cc{0}$ should vanish so that we are left with $\accentset{(-2)}{[P_{+}]}$ in the limit. \\

If we require all the coefficients of the order $\cc{2}$ terms and of the order $\cc{0}$ terms to vanish we obtain the following system of equations
\begin{subequations}
\begin{align}
\CA +2\CC p!&=0\,,\\
\CB+\CA+(p+1)!\CC+\frac{2\coefa}{\coefb}\CD&=0\,,\\
2\CB+p\CA+\frac{\coefa }{\coefb}\CD+(p+1)!\CC&=0\,,\\
3\CB+(p-1)\CA+\frac{\coefa }{\coefb}\CD&=0\,,\\
\CB+(p-1)\CA-\frac{\coefa }{\coefb}\CD&=0\,,\\
\coefa\CA-\frac{1}{\coefa}(\coefa^2-\coefb)\Big[(p+1)\CA+(D-p-1)\CB\Big]+2\CD+\coefa (p+1)!\CC&=0\,,\\
\CB\coefa-\frac{1}{\coefa}(\coefa^2-\coefb)\Big[(p+1)\CA+(D-p-1)\CB\Big]+2\CD&=0\,,\\
(\coefa^2-\coefb)\Big[(p+2-D)\CB-(p+1)\CA\Big]+\coefa \CD&=0\,,\\
\CB+\frac{\coefa}{\coefb}\CD&=0\,.
\end{align}\label{eq:systemP}
\end{subequations}
These equations admit the following  solutions for the theories already respecting the no-divergence condition:
\begin{align}
\CB&=-\frac{p-1}{2}\CA\,,&\CC&=-\dfrac{1}{2p!}\CA\,,&\CD&=\dfrac{\coefb(p-1)}{2\coefa}\CA\,,&\frac{\coefb}{\coefa^2}&=\frac{p^2+p(3-D)+D}{p^2+p(4-D)+D-1}\equiv \xi_{Dil}\,.\label{eq:PoissonYes}
\end{align}
Since these solutions do not impose any further condition on $D$ and $p$ this implies that for all theories that satisfy the no-divergence condition, it is possible, by fine-tuning the scalar field coupling corresponding to   $\coefa$ and $\coefb$, to obtain the Poisson the equation from the limit of the equations of motion. Note that in deriving the equations of motion we have implicitly assumed that $c_{1}$ and $c_{2}$ are different from zero. Remarkably, the fine-tuning of the scalar field coupling that is needed to obtain the Poisson equation is identical to the fine-tuning that leads to an emerging dilatation symmetry. In particular, the fact that $\accentset{(0)}{[P_{+}]}=0$ follows from the fact that this is the Noether identity for local dilatations.
 The relation between the coefficients $\alpha,\beta,\gamma,\delta$ determining the scaling weights of the different non-relativistic fields and the coefficients $\tilde \alpha, \tilde \beta, \tilde \gamma, \tilde \delta$ determining the Noether identity is,  up to an overall rescaling, given by
\begin{align}
\CA&=\alpha\,,&\CB&=\beta\,,&\CC&=-\frac{1}{2p!}\alpha\,,&\CD&=
-\frac{\coefb}{2\coefa}[(p+1)\alpha+\beta(D-p-1)-\coefa \delta] \,.
\end{align}

The fact that the scalar field plays a crucial role in obtaining the Poisson equation can be understood by inspecting the case where it is truncated from the beginning, which amounts to choosing $\coefa=\coefb=0$  in the action. Setting them to zero at the beginning now is equivalent to set
\begin{align}
\CD&=-\frac{\coefb}{2\coefa}[(p+1)\CA+(D-p-1)\CB]\,,&\phi&=0\,.
\end{align}
Substituting this back into the equations \autoref{eq:systemP}, the resulting system of equations does not admit solutions for theories without divergences. This is consistent with the result of \cite{3823ef0bb56b497c9abff0dec317478b} where it was shown that for an Einstein-Maxwell system, cancellation of divergences could be established but no local dilatation symmetry and Poisson equation could be obtained.

\section{The Constrained On-Shell Approach}\label{sec:constrained}

At this point it is natural to ask how it is possible to reproduce the textbook result of obtaining a Poisson equation for $p=0$  from general relativity without using a scalar field or, more generally, a scalar field without fine-tuned coupling. Such a limit, leading to a covariant formulation of Newtonian gravity, called Newton-Cartan gravity, has been discussed in \cite{Bergshoeff:2015uaa} and shown to lead to a Poisson equation upon gauge-fixing.
The starting point of \cite{Bergshoeff:2015uaa} was the same set of equations of motion used so far supplemented with an additional constraint such that the total number of equations did not follow from an action. This approach, which we denominated in the introduction as the {\sl constrained on-shell approach}, will be investigated in this section.

Therefore, the starting point in this section will be an extended set of equations of motion, containing the ones we already considered before, such that the full set of equations (equations of motion plus constraints)  do not follow from an action. Strictly speaking, without an action our starting theory should only satisfy the conditions that follow from requiring a finite boost transformation. Nevertheless, we will assume that the full no-divergence condition is satisfied and we will use the same expansion defined in \autoref{eq:expansion1} and \autoref{eq:expansion2} as before.\\

The analysis of  \cite{Bergshoeff:2015uaa} only considered the $p=0$ case. The starting point was the set of equations of motion that follows from an  Einstein-Hilbert-Maxwell action  supplemented by hand with the constraint that the 2-form field-strength of the vector $A_\mu$ is zero such that this vector field does not add any degree of freedom to the ones described by general relativity:
\begin{align}
F_{\mu\nu}&=0\,.\label{eq:constrainttot}
\end{align}
The vector field is purely an auxiliary field that is needed to define the limit.
After expansion, the relativistic condition \eqref{eq:constrainttot}
results in the following equation\,\footnote{The factor 1/2 is due to the fact that in \cite{Bergshoeff:2015uaa} the expansion Ansatz for the longitudinal Vielbein is slightly different with respect to ours. }
\begin{align}
t_{\mu\nu}&=\frac{1}{2\cc{2}}f_{\mu\nu}
\end{align}
with the intrinsic torsion tensor $t_{\mu\nu}$ given by $t_{\mu\nu} = \partial_\mu\tau_\nu - \partial_\nu\tau_\mu$ and $f_{\mu\nu} = \partial_\mu a_\nu - \partial_\nu a_\mu$.
This equation allows to shift the power of $\cc{}$ of all the terms proportional to the intrinsic torsion, and combined with an opportune choice of the coefficients $\CA,\CB,\CC$ and $\CD$, defining the specific combination of equations of motion like in the previous approach, enables one to find, after taking the limit, the Poisson equation among the equations of motion.\\

We wish to generalize this approach for general $p$ and $D$. Pragmatically, what we are looking for is a way to eliminate one or more of the conditions appearing in \autoref{eq:systemP} occurring in the Lagrangian approach in such a way to be able to find solutions even when the scalar field coupling corresponding to $\coefa$ and $\coefb$ is not fine-tuned as in eq.~\autoref{eq:PoissonYes}. One way of achieving this is by imposing one or more constraints by hand. However, these constraints should not over-constrain the non-relativistic theory or remove the Poisson equation itself. Below we will discuss two possible constraints that do the job. \\

\paragraph*{Field Strength Constraint.} The first option is to generalize directly  \autoref{eq:constrainttot} to
\begin{align}
&F_{\mu_{1}...\mu_{p+2}}=\frac{1}{2}\cc{p+1}(p+2)(p+1)\, t_{[\mu_{1}\mu_{2}}{}^{A_{1}}\tau_{\mu_{3}}{}^{A_{2}}...\tau_{\mu_{p+2}]}{}^{A_{p+1}}\epsilon_{A_{1}...A_{p+1}}+
\cc{p-1}f_{\mu_{1}...\mu_{p+2}}=0\,.\label{eq:relconstraint}
\end{align}
Projecting this $(p+2)$-form on the different flat directions we obtain the following conditions:
\begin{subequations}
\begin{align}
t_{aA}{}^{A}&=\frac{\cc{-2}}{(p+1)!}f_{aB_{1}...B_{p+1}}\epsilon^{B_{1}...B_{p+1}}\,,&&\\
t_{ab}{}^{A}&=\frac{\cc{-2}}{p!}f_{abB_{1}...B_{p}}\epsilon^{AB_{1}...B_{p}}\,,&&\\
f_{A_{1}...A_{k}a_{1}...a_{p+2-k}}&=0\,,\quad&& \text{for}\quad k\leqslant p-1\,.
\end{align}
\end{subequations}
By the no-divergence condition \autoref{eq:NoDivCondition4} the transverse space could be at most three dimensional if $p>1$ so the conditions above can be rewritten as
\begin{subequations}
\begin{align}t_{aA}{}^{A}&=\frac{\cc{-2}}{(p+1)!}f_{aB_{1}...B_{p+1}}\epsilon^{B_{1}...B_{p+1}}\,,\label{eq:constraints1}\\
t_{ab}{}^{A}&=\frac{\cc{-2}}{p!}f_{abB_{1}...B_{p}}\epsilon^{AB_{1}...B_{p}}\,,\label{eq:constraints2}\\
f_{A_{1}...A_{p-1}a_{1}a_{2}a_{3}}&=0\,,\label{eq:constraints3}
\end{align}\label{eq:constraints}
\end{subequations}
where it is understood that each condition appears once it is allowed by the dimension and foliation. These conditions can be used in \autoref{eq:PoissonExpansion} to lower the power of $\cc{}$ of the terms that occur at the left hand side of eq.~\autoref{eq:constraints}. Combined with taking the right combination of equations of motion, defined by the same coefficients  $\CA,\CB$ and $\CD$\,\footnote{Note the absence of the coefficient $\CC$ due to the constraint we imposed.} as used in the Lagrangian approach, one can promote the correct order $\cc{-2}$ term to be the leading order term in the expansion of the equations of motion, thereby unveiling the Poisson equations in the limit. \\

Using \autoref{eq:constraints} and requiring that all the coefficients of the order $\cc{0}$ terms that are  not affected by the constraints above in $\accentset{(0)}{[P_{+}]}$ vanish, leads to the following set of conditions:
\begin{subequations}
\begin{align}
\CB\coefa-\frac{1}{\coefa}(\coefa^2-\coefb)\Big[(p+1)\CA+(D-p-1)\CB\Big]+2\CD&=0\,,\\
(\coefa^2-\coefb)\Big[(p+2-D)\CB-(p+1)\CA\Big]+\coefa \CD&=0\,,\\
\CB+\frac{\coefa}{\coefb}\CD&=0\,.\label{eq:system2}
\end{align}
\end{subequations}
These conditions should be supplemented together with the request that the coefficient of the Laplacian term, characteristic of the Poisson equation, is not zero. The condition \autoref{eq:relconstraint} eliminates the contribution of the equations of motion of the gauge field from $[P_{+}]$, however by \autoref{eq:constraints1} the first line of \autoref{eq:P0} provides us with a Laplacian term at order $\cc{-2}$, since
\begin{align}
h^{\mu\nu}\tau^{\rho}{}_{A}\accentset{(nr)}{\nabla}_{\mu}t_{\nu\rho}{}^{A}&=\accentset{(nr)}{\nabla}_{\mu}(e^{\mu a}t_{aA}{}^{A})
\,,
\end{align}
using
\begin{align}
\accentset{(nr)}{\nabla}_{\mu}\tau^{\rho}{}_{A}&=\frac{1}{2}e^{\rho a}\tau_{\mu}{}^{B}z_{ABa}+\frac{1}{2}e^{\rho a}e_{\mu}{}^{b}(z_{Aab}+z_{Aba})\,.
\end{align}
Thus the request for the coefficient of the Laplacian term to be different from zero reads
\begin{align}
\CA+\CB+\frac{2\coefa}{\coefb}\CD\neq 0\,. \label{eq:nozero}
\end{align}

The set of conditions \autoref{eq:system2} together with the condition \autoref{eq:nozero} admits the following solutions:
\begin{subequations}
\begin{enumerate}[label=\bf (\alph*)]
\item \begin{align}
\CA&\neq 0\,,&\CB&=\frac{(\coefb/\coefa^2-1)(p+1)}{(\coefb/\coefa^2-1)(D-p-3)-1}\CA\,,&\CD&=-\frac{\coefb}{\coefa}\CB\,,&\frac{\coefb}{\coefa^2}&\neq \bigg\{1,\frac{D-p-2}{D-p-3},\xi_{Dil}\bigg\}\,,
\end{align}
\item \begin{align}
\CA&\neq 0\,,&\CB&=0\,,&\CD&=0\,,&\frac{\coefb}{\coefa^2}&=1\,,&p&\neq 1\,,
\end{align}
\item \begin{align}
\CA&=0\,,&\CB&\neq 0\,,&\CD&=-\frac{\coefb}{\coefa}\CB\,,& \frac{\coefb}{\coefa^2}&= \frac{D-p-2}{D-p-3}\,.
\end{align}
\end{enumerate}
\end{subequations}

With these prescriptions the leading order term in the expansion of $[P_{+}]$ is of order $\cc{-2}$ which makes it possible to get the Poisson equation from the limit. \\

After taking the limit the relativistic constraints \autoref{eq:constraints}  become the non-relativistic constraints
\begin{subequations}
\begin{align}
t_{ab}{}^{A}&=0\,,\\
t_{aA}{}^{A}&=0\,,\\
f_{A_{1}...A_{p-1}a_{1}a_{2}a_{3}}&=0\,,
\end{align}
\end{subequations}
on the intrinsic torsion and the field strength of the non-relativistic gauge field. These constraints form a closed set under  non-relativistic boost transformations as follows
\begin{subequations}
\begin{align}
\delta t_{ab}{}^{A}&=0\,,\\
\delta t_{aA}{}^{A}&=\lambda_{A}{}^{b}t_{ab}{}^{A}\,,\\
\delta f_{abcA_{1}...A_{p-1}}&=(p-1)!\lambda^{B_{1}}{}_{[a}t_{bc]}{}^{B_{2}}\epsilon_{B_{1}B_{2}A_{1}...A_{p-1}}\,,
\end{align}
\end{subequations}
where we have used again the fact that for theories with no divergences,  $p>1$, the transverse space is at most three-dimensional. \\

We have not investigated whether there is a milder version of the constraint \autoref{eq:constraints} that can be imposed leading to a Poisson equation upon taking the limit.\\

\paragraph*{Scalar Field Constraint.} Another option is to constrain   the scalar field by setting
\begin{align}
\partial_{\mu}\Phi&=0\,.
\end{align}
This corresponds to setting $\partial_{\mu}\phi=0$ in the expression for $[P_{+}]$.  We stress that there is an important difference between this constraint and the one discussed at the end of \autoref{sec:PoissonDilatation} since in that case the constraint was imposed at the level of the action while here we impose the constraint at the level of the equations of motions. The system of equations \autoref{eq:systemP} that followed by requiring the cancellation of all terms at order $\cc{2}$ and $\cc{0}$ in \autoref{eq:PoissonExpansion} in this case reads
\begin{subequations}
\begin{align}
\CA +2\CC p!&=0\,,\\
\CB+\CA+(p+1)!\CC+\frac{2\coefa}{\coefb}\CD&=0\,,\\
2\CB+p\CA+\frac{\coefa }{\coefb}\CD+(p+1)!\CC&=0\,,\\
3\CB+(p-1)\CA+\frac{\coefa }{\coefb}\CD&=0\,,\\
\CB+(p-1)\CA-\frac{\coefa }{\coefb}\CD&=0\,,\\
\CB+\frac{\coefa}{\coefb}\CD&=0\,.
\end{align}
\end{subequations}
It admits the following unique solution.
\begin{align}
\CA&\neq0\,,&\CB&=\frac{1-p}{2}\CA\,,&\CC&=-\frac{1}{2p!}\CA\,,&\CD&=\frac{\coefb}{2\coefa}(p-1)\CA\,.
\end{align}
\\

This finishes our discussion of the constrained on-shell approach. We have shown in two cases how by imposing a constraint by hand one can obtain via this approach the Poisson equation as a non-relativistic limit of the equations of motion. It would be interesting to extend this analysis and classify the possible constraints which have this property.\\

\section{Multiplet Structure of the Non-Relativistic Equations of Motion}\label{sec:multiplet}
In this section we describe how the non-relativistic equations of motion transform under the Galilean boost transformations. We limit our attention to the two different cases in the Lagrangian approach, i.e.
\begin{enumerate}[start=1,label={\bfseries (\roman*)}]
\item  the case with emerging dilatation symmetry and 
\item the case without emerging dilatation symmetry, i.e.~with no Poisson equation.
\end{enumerate}

We remind that the non-relativistic boost transformation acting on the fundamental fields is given by
\begin{subequations}
\begin{align}
\delta \tau_{\mu}{}^{A}&=0\,,\\
\delta e_{\mu}{}^{a}&=\lambda^{a}{}_{A}\tau_{\mu}{}^{A}\,,\\
\delta \tau^{\mu}{}_{A}&=\lambda_{A}{}^{a}e^{\mu}{}_{a}\,,\\
\delta e^{\mu}{}_{a}&=0\,,\\
\delta a_{\mu_{1}...\mu_{p+1}}&=-(p+1)\lambda^{A_{1}}{}_{b}e_{[\mu_{1}}{}^{b}\tau_{\mu_{2}}{}^{A_{2}}...
\tau_{\mu_{p+1}]}{}^{A_{p+1}}\epsilon_{A_{1}...A_{p+1}}\,.
\end{align}
\end{subequations}

We can find the boost transformation of the non-relativistic equations of motion by just taking the $\cc{}\rightarrow \infty$ limit of the boost transformations of the relativistic equations of motion.  These are given by
\begin{subequations}
\begin{align}
\delta [G]_{AB}&=\Lambda_{A}{}^{c}[G]_{cB}+\Lambda_{B}{}^{c}[G]_{Ac}\,,\\
\delta [G]_{Ab}&=\Lambda_{A}{}^{c}[G]_{cb}+\Lambda_{b}{}^{C}[G]_{AC}\,,\\
\delta [G]_{ab}&=\Lambda_{a}{}^{C}[G]_{Cb}+\Lambda_{b}{}^{C}[G]_{aC}\,,\\
\delta [\Phi]&=0\,,\\
\delta [A]_{A_{1}...A_{k}a_{1}...a_{p+1-k}}&=
(-)^{p}k\Lambda_{[A_{1}}{}^{b}[A]_{A_{2}...A_{k}]a_{1}...a_{p+1-k}b}+\nonumber\\
&+(-)^{p(k+1)}(p+1-k)\Lambda_{[a_{1}}{}^{B}[A]_{a_{2}...a_{p+1-k}]A_{1}...A_{k}B}\,.
\end{align}\label{eq:BoostEOMRelativistic}
\end{subequations}
We recall that the expansion of the equations of motion is given by eq.~\autoref{eq:EOMExpansion}.\\

Considering the combinations
\begin{align}
[P_{\pm}]=\CA [G]_{A}{}^{A}+\CB [G]_{a}{}^{a}\pm \CC [A]^{A_{1}...A_{p+1}}\epsilon_{A_{1}...A_{p+1}}+\CD [\Phi]\,,
\end{align}
with coefficients $\CA,\CB$ and $\CC$ given by eq.~\autoref{eq:PoissonYes} for both cases with and without local dilatation symmetry. The coefficient $\CD$ will be as in \autoref{eq:PoissonYes} when we are in the case with local dilatation symmetry, otherwise it will be zero because the scalar field equation of motion is boost invariant by itself, so when it does not contribute to cancellations in the $[P_{\pm}]$ expansion it is easier to treat it separately. In summary
\begin{align}
\CA&\neq 0\,,&\CB&=-\frac{p-1}{2}\CA\,,&\CC&=-\frac{1}{2p!}\CA\,,&\CD&=
\left\{
\begin{array}{ccl}
\coefa \xi_{Dil}\dfrac{p-1}{2}\CA&&\text{for theories with Local}\\
&&\text{Dilatation Symmetry}\,,\\
\\
0&&\text{for theories without Local}\\
&&\text{Dilatation Symmetry}\,.\\
\end{array}
\right.\label{eq:Combination}
\end{align}

Under boost transformations  $[P_{\pm}]$ and $[V_{\pm}]_{Aa}$ transform as

\begin{subequations}
\begin{align}
\delta [P_{\pm}]&=(p+1)\CA\,\Lambda^{Aa}[V_{\pm}]_{Aa}\,,\\
\delta [V_{\pm}]_{Aa}&=\Lambda_{A}{}^{b}\bigg([G]_{ba}+\frac{p-1}{2(p+1)}\delta_{ab}[G]_{d}{}^{d}\bigg)+\Lambda_{a}{}^{B}[G]_{\{AB\}}+\nonumber\\
&\mp \frac{1}{2(p-1)!}\Lambda^{B_{1}b}[A]_{ab}{}^{B_{2}...B_{p}}\epsilon_{AB_{1}...B_{p}}+\frac{1}{(p+1)\CA}\Lambda_{aA}[P_{\pm}]-\frac{\CD}{(p+1)\CA}\Lambda_{aA}[\Phi]\,.
\end{align}
\end{subequations}

We recall that the leading order expansion of these combinations is given by
\begin{multieq}[2]
\text{\bf With } &\text{\bf Dilatation Symmetry}\nonumber\\
\nonumber\\
[P_{+}]&=\cc{-2}\accentset{(-2)}{[P_{+}]}+\mathcal{O}(\cc{-4})\,,\\
[P_{-}]&=\cc{+2}\accentset{(2)}{[P_{-}]}+\mathcal{O}(\cc{0})\,,\\
[V_{+}]_{Aa}&=\cc{-1}\accentset{(-1)}{[V_{+}]}_{Aa}+\mathcal{O}(\cc{-3})\,,\\
[V_{-}]_{Aa}&=\cc{+1}\accentset{(1)}{[V_{-}]}_{Aa}+\mathcal{O}(\cc{-1})\,,\\
\text{\bf Without } &\text{\bf Dilatation Symmetry}\nonumber\\
\nonumber\\
[P_{+}]&=\phantom{\cc{+1}}\accentset{(0)}{[P_{+}]}+\mathcal{O}(\cc{-2})\,,\\
[P_{-}]&=\cc{+2}\accentset{(2)}{[P_{-}]}+\mathcal{O}(\cc{0})\,,\\
[V_{+}]_{Aa}&=\cc{-1}\accentset{(-1)}{[V_{+}]}_{Aa}+\mathcal{O}(\cc{-3})\,,\\
[V_{-}]_{Aa}&=\cc{+1}\accentset{(1)}{[V_{-}]}_{Aa}+\mathcal{O}(\cc{-1})\,.
\end{multieq}

We have now all the information to determine how the boost transformations act on the non-relativistic equations of motion. We discuss the two cases separately below.
\vskip .3truecm

\noindent {\bf With Local Dilatation Symmetry}\ \
\vskip .1truecm

In theories with local dilatation symmetry the non-relativistic EOMs are
\begin{align}
\Big\{\accentset{(-2)}{[P_{+}]}\,, \accentset{(+2)}{[P_{-}]}\,, \accentset{(-1)}{[V_{+}]}_{Aa}\,, \accentset{(+1)}{[V_{-}]}_{Aa}\,,
\accentset{(0)}{[G]}_{ab}\,,
\accentset{(0)}{[G]}_{\{AB\}}\,,\accentset{(0)}{[\Phi]}\,,
\accentset{(0)}{[A]}_{A_{1}...A_{p-1}a_{1}a_{2}}\,,\accentset{(+1)}{[A]}_{A_{1}...A_{p-2}a_{1}a_{2}a_{3}}\Big\}\,,
\end{align}
where each equation should be understood to appear if allowed by the theory/foliation ( for example in the particle case the last two are absent).
The boost transformations of these equations of motion   are given by

\begin{subequations}
\begin{align}
\delta\accentset{(-2)}{[P_{+}]}&=(p+1)\CA\, \lambda^{Aa}\accentset{(-1)}{[V_{+}]}_{Aa}\,,\\
\delta \accentset{(+2)}{[P_{-}]}&=0\,,\\
\delta \accentset{(-1)}{[V_{+}]}_{Aa}&=\lambda_{A}{}^{b}\bigg(\accentset{(0)}{[G]}_{ba}+\frac{p-1}{2(p+1)}\delta_{ab}\accentset{(0)}{[G]}_{d}{}^{d}\bigg)+\lambda_{a}{}^{B}\accentset{(0)}{[G]}_{\{AB\}}+\nonumber\\
&\mp \frac{1}{2(p-1)!}\lambda^{B_{1}b}\accentset{(0)}{[A]}_{ab}{}^{B_{2}...B_{p}}\epsilon_{AB_{1}...B_{p}}-\frac{\CD}{(p+1)\CA}\lambda_{aA}\accentset{(0)}{[\Phi]}\,,\label{eq:BoostVp}\\
\delta \accentset{(+1)}{[V_{-}]}_{Aa}&=\frac{1}{(p+1)\CA}\lambda_{Aa}\accentset{(+2)}{[P_{-}]}\,,\\
\delta \accentset{(0)}{[G]}_{ab}&=\accentset{(+1)}{[V_{-}]_{A(a}}\lambda_{b)}{}^{A}\,,\\
\delta \accentset{(0)}{[G]}_{\{AB\}}&=\lambda_{\{A}{}^{a}\accentset{(+1)}{[V_{-}]}_{B\}a}\,,\\
\delta \accentset{(0)}{[\Phi]}&=0\,,\\
\delta \accentset{(0)}{[A]}_{A_{1}...A_{p-1}a_{1}a_{2}}&=
(-)^{p+1}\epsilon^{ABC_{1}...C_{p-1}}\accentset{(+1)}{[V_{-}]}_{A[b}\,\lambda_{a]B}-(p-1)\lambda^{c[C_{1}}\,\,\accentset{(+1)}{[A]}_{abc}{}^{C_{2}...C_{p-1}]}\,,\\
\delta \accentset{(+1)}{[A]}_{A_{1}...A_{p-2}a_{1}a_{2}a_{3}}&=0\,.
\end{align}
\end{subequations}

\vskip .1truecm

\noindent {\bf Without Local Dilatation Symmetry}\ \
\vskip .1truecm

In theories without local dilatation symmetry  the non-relativistic equations of motion are given by
\begin{align}
\Big\{\accentset{(0)}{[P_{+}]}\,, \accentset{(+2)}{[P_{-}]}, \accentset{(-1)}{[V_{+}]}_{Aa}\,, \accentset{(+1)}{[V_{-}]}_{Aa}\,,
\accentset{(0)}{[G]}_{ab}\,,
\accentset{(0)}{[G]}_{\{AB\}}\,,\accentset{(0)}{[\Phi]}\,,
\accentset{(0)}{[A]}_{A_{1}...A_{p-1}a_{1}a_{2}}\,,\accentset{(+1)}{[A]}_{A_{1}...A_{p-2}a_{1}a_{2}a_{3}}\Big\}\,,
\end{align}
where each equation should be understood to appear if allowed by the theory/foliation. The boost transformation of the non-relativistic equations of motion for this case read:
\begin{subequations}
\begin{align}
\delta\accentset{(0)}{[P_{+}]}&=0\,,\\
\delta \accentset{(+2)}{[P_{-}]}&=0\,,\\
\delta \accentset{(-1)}{[V_{+}]}_{Aa}&=\lambda_{A}{}^{b}\bigg(\accentset{(0)}{[G]}_{ba}+\frac{p-1}{2(p+1)}\delta_{ab}\accentset{(0)}{[G]}_{d}{}^{d}\bigg)+\lambda_{a}{}^{B}\accentset{(0)}{[G]}_{\{AB\}}+\frac{1}{(p+1)\CA}\accentset{(0)}{[P_{+}]}+\nonumber\\
&\mp \frac{1}{2(p-1)!}\lambda^{B_{1}b}\accentset{(0)}{[A]}_{ab}{}^{B_{2}...B_{p}}\epsilon_{AB_{1}...B_{p}}\,,\\
\delta \accentset{(+1)}{[V_{-}]}_{Aa}&=\frac{1}{(p+1)\CA}\lambda_{Aa}\accentset{(+2)}{[P_{-}]}\,,\\
\delta \accentset{(0)}{[G]}_{ab}&=\accentset{(+1)}{[V_{-}]_{A(a}}\lambda_{b)}{}^{A}\,,\\
\delta \accentset{(0)}{[G]}_{\{AB\}}&=\lambda_{\{A}{}^{a}\accentset{(+1)}{[V_{-}]}_{B\}a}\,,\\
\delta \accentset{(0)}{[\Phi]}&=0\,,\\
\delta \accentset{(0)}{[A]}_{A_{1}...A_{p-1}a_{1}a_{2}}&=
(-)^{p+1}\epsilon^{ABC_{1}...C_{p-1}}\accentset{(+1)}{[V_{-}]}_{A[b}\,\lambda_{a]B}-(p-1)\lambda^{c[C_{1}}\,\,\accentset{(+1)}{[A]}_{abc}{}^{C_{2}...C_{p-1}]}\,,\\
\delta \accentset{(+1)}{[A]}_{A_{1}...A_{p-2}a_{1}a_{2}a_{3}}&=0\,.
\end{align}
\end{subequations}

\begin{landscape}
\begin{figure}

\resizebox{1.5\textwidth}{!}{%

\tikzset{every picture/.style={line width=0.25pt}} 
\hspace{2cm}
\centering
\begin{tikzpicture}[x=0.75pt,y=0.75pt,yscale=-1,xscale=1]

\draw  [fill={rgb, 255:red, 179; green, 179; blue, 255 }  ]  (894,673.5) .. controls (894,671.84) and (895.34,670.5) .. (897,670.5) -- (963,670.5) .. controls (964.66,670.5) and (966,671.84) .. (966,673.5) -- (966,742.5) .. controls (966,744.16) and (964.66,745.5) .. (963,745.5) -- (897,745.5) .. controls (895.34,745.5) and (894,744.16) .. (894,742.5) -- cycle  ;
\draw (930,708) node  [font=\huge]  {$\accentset{( 0)}{[ P_{+}]}$};
\draw  [fill={rgb, 255:red, 186; green, 224; blue, 140 }  ]  (875.5,497.5) .. controls (875.5,495.84) and (876.84,494.5) .. (878.5,494.5) -- (981.5,494.5) .. controls (983.16,494.5) and (984.5,495.84) .. (984.5,497.5) -- (984.5,578.5) .. controls (984.5,580.16) and (983.16,581.5) .. (981.5,581.5) -- (878.5,581.5) .. controls (876.84,581.5) and (875.5,580.16) .. (875.5,578.5) -- cycle  ;
\draw (930,538) node  [font=\huge]  {$\accentset{( -1)}{[ V_{+}]}_{Aa}$};
\draw  [fill={rgb, 255:red, 165; green, 200; blue, 241 }]  (1014,669.5) .. controls (1014,667.84) and (1015.34,666.5) .. (1017,666.5) -- (1093,666.5) .. controls (1094.66,666.5) and (1096,667.84) .. (1096,669.5) -- (1096,746.5) .. controls (1096,748.16) and (1094.66,749.5) .. (1093,749.5) -- (1017,749.5) .. controls (1015.34,749.5) and (1014,748.16) .. (1014,746.5) -- cycle  ;
\draw (1055,708) node  [font=\huge]  {$\accentset{( 0)}{[ G]}_{ab}$};
\draw  [fill={rgb, 255:red, 168; green, 241; blue, 225 }]  (749,669.5) .. controls (749,667.84) and (750.34,666.5) .. (752,666.5) -- (858,666.5) .. controls (859.66,666.5) and (861,667.84) .. (861,669.5) -- (861,746.5) .. controls (861,748.16) and (859.66,749.5) .. (858,749.5) -- (752,749.5) .. controls (750.34,749.5) and (749,748.16) .. (749,746.5) -- cycle  ;
\draw (805,708) node  [font=\huge]  {$\accentset{( 0)}{[ G]}_{\{AB\}}$};
\draw  [fill={rgb, 255:red, 255; green, 247; blue, 82 }]  (1112,669.5) .. controls (1112,667.84) and (1113.34,666.5) .. (1115,666.5) -- (1245,666.5) .. controls (1246.66,666.5) and (1248,667.84) .. (1248,669.5) -- (1248,746.5) .. controls (1248,748.16) and (1246.66,749.5) .. (1245,749.5) -- (1115,749.5) .. controls (1113.34,749.5) and (1112,748.16) .. (1112,746.5) -- cycle  ;
\draw (1180,708) node  [font=\huge]  {$\accentset{( 0)}{[ A]}_{( p-1\vert 2)}$};
\draw  [fill={rgb, 255:red, 205; green, 205; blue, 205 } ]  (876.5,837.5) .. controls (876.5,835.84) and (877.84,834.5) .. (879.5,834.5) -- (980.5,834.5) .. controls (982.16,834.5) and (983.5,835.84) .. (983.5,837.5) -- (983.5,918.5) .. controls (983.5,920.16) and (982.16,921.5) .. (980.5,921.5) -- (879.5,921.5) .. controls (877.84,921.5) and (876.5,920.16) .. (876.5,918.5) -- cycle  ;
\draw (930,878) node  [font=\huge]  {$\accentset{( +1)}{[ V_{-}]}_{Aa}$};
\draw  [fill={rgb, 255:red, 241; green, 128; blue, 128 }]  (894.5,1013.5) .. controls (894.5,1011.84) and (895.84,1010.5) .. (897.5,1010.5) -- (962.5,1010.5) .. controls (964.16,1010.5) and (965.5,1011.84) .. (965.5,1013.5) -- (965.5,1082.5) .. controls (965.5,1084.16) and (964.16,1085.5) .. (962.5,1085.5) -- (897.5,1085.5) .. controls (895.84,1085.5) and (894.5,1084.16) .. (894.5,1082.5) -- cycle  ;
\draw (930,1048) node  [font=\huge]  {$\accentset{( +2)}{[ P_{-}]}$};
\draw  [fill={rgb, 255:red, 255; green, 255; blue, 255 }]  (1237,839.5) .. controls (1237,837.84) and (1238.34,836.5) .. (1240,836.5) -- (1370,836.5) .. controls (1371.66,836.5) and (1373,837.84) .. (1373,839.5) -- (1373,916.5) .. controls (1373,918.16) and (1371.66,919.5) .. (1370,919.5) -- (1240,919.5) .. controls (1238.34,919.5) and (1237,918.16) .. (1237,916.5) -- cycle  ;
\draw (1305,878) node  [font=\huge]  {$\accentset{( +1)}{[ A]}_{( p-2\vert 3)}$};
\draw  [fill={rgb, 255:red, 179; green, 179; blue, 255 }]  (102,333.5) .. controls (102,331.84) and (103.34,330.5) .. (105,330.5) -- (171,330.5) .. controls (172.66,330.5) and (174,331.84) .. (174,333.5) -- (174,402.5) .. controls (174,404.16) and (172.66,405.5) .. (171,405.5) -- (105,405.5) .. controls (103.34,405.5) and (102,404.16) .. (102,402.5) -- cycle  ;
\draw (138,368) node  [font=\huge]  {$\accentset{( -2)}{[ P_{+}]}$};
\draw  [fill={rgb, 255:red, 186; green, 224; blue, 140 }]  (83.5,497.5) .. controls (83.5,495.84) and (84.84,494.5) .. (86.5,494.5) -- (189.5,494.5) .. controls (191.16,494.5) and (192.5,495.84) .. (192.5,497.5) -- (192.5,578.5) .. controls (192.5,580.16) and (191.16,581.5) .. (189.5,581.5) -- (86.5,581.5) .. controls (84.84,581.5) and (83.5,580.16) .. (83.5,578.5) -- cycle  ;
\draw (138,538) node  [font=\huge]  {$\accentset{( -1)}{[ V_{+}]}_{Aa}$};
\draw  [fill={rgb, 255:red, 165; green, 200; blue, 241 } ]  (222,669.5) .. controls (222,667.84) and (223.34,666.5) .. (225,666.5) -- (301,666.5) .. controls (302.66,666.5) and (304,667.84) .. (304,669.5) -- (304,746.5) .. controls (304,748.16) and (302.66,749.5) .. (301,749.5) -- (225,749.5) .. controls (223.34,749.5) and (222,748.16) .. (222,746.5) -- cycle  ;
\draw (263,708) node  [font=\huge]  {$\accentset{( 0)}{[ G]}_{ab}$};
\draw  [fill={rgb, 255:red, 168; green, 241; blue, 225 } ]  (-43,669.5) .. controls (-43,667.84) and (-41.66,666.5) .. (-40,666.5) -- (66,666.5) .. controls (67.66,666.5) and (69,667.84) .. (69,669.5) -- (69,746.5) .. controls (69,748.16) and (67.66,749.5) .. (66,749.5) -- (-40,749.5) .. controls (-41.66,749.5) and (-43,748.16) .. (-43,746.5) -- cycle  ;
\draw (13,708) node  [font=\huge]  {$\accentset{( 0)}{[ G]}_{\{AB\}}$};
\draw  [fill={rgb, 255:red, 255; green, 247; blue, 82 }]  (320,669.5) .. controls (320,667.84) and (321.34,666.5) .. (323,666.5) -- (453,666.5) .. controls (454.66,666.5) and (456,667.84) .. (456,669.5) -- (456,746.5) .. controls (456,748.16) and (454.66,749.5) .. (453,749.5) -- (323,749.5) .. controls (321.34,749.5) and (320,748.16) .. (320,746.5) -- cycle  ;
\draw (388,708) node  [font=\huge]  {$\accentset{( 0)}{[ A]}_{( p-1\vert 2)}$};
\draw  [fill={rgb, 255:red, 205; green, 205; blue, 205 }]  (84.5,837.5) .. controls (84.5,835.84) and (85.84,834.5) .. (87.5,834.5) -- (188.5,834.5) .. controls (190.16,834.5) and (191.5,835.84) .. (191.5,837.5) -- (191.5,918.5) .. controls (191.5,920.16) and (190.16,921.5) .. (188.5,921.5) -- (87.5,921.5) .. controls (85.84,921.5) and (84.5,920.16) .. (84.5,918.5) -- cycle  ;
\draw (138,878) node  [font=\huge]  {$\accentset{( +1)}{[ V_{-}]}_{Aa}$};
\draw  [fill={rgb, 255:red, 241; green, 128; blue, 128 }]  (102.5,1013.5) .. controls (102.5,1011.84) and (103.84,1010.5) .. (105.5,1010.5) -- (170.5,1010.5) .. controls (172.16,1010.5) and (173.5,1011.84) .. (173.5,1013.5) -- (173.5,1082.5) .. controls (173.5,1084.16) and (172.16,1085.5) .. (170.5,1085.5) -- (105.5,1085.5) .. controls (103.84,1085.5) and (102.5,1084.16) .. (102.5,1082.5) -- cycle  ;
\draw (138,1048) node  [font=\huge]  {$\accentset{( +2)}{[ P_{-}]}$};
\draw  [fill={rgb, 255:red, 255; green, 255; blue, 255 } ]  (445,839.5) .. controls (445,837.84) and (446.34,836.5) .. (448,836.5) -- (578,836.5) .. controls (579.66,836.5) and (581,837.84) .. (581,839.5) -- (581,916.5) .. controls (581,918.16) and (579.66,919.5) .. (578,919.5) -- (448,919.5) .. controls (446.34,919.5) and (445,918.16) .. (445,916.5) -- cycle  ;
\draw (513,878) node  [font=\huge]  {$\accentset{( +1)}{[ A]}_{( p-2\vert 3)}$};
\draw (141.01,200) node  [font=\huge] [align=left] {\begin{minipage}[lt]{370.57pt}\setlength\topsep{0pt}
\centering
{\Huge \textbf{With }}\\{\Huge \textbf{Local Dilatation Symmetry}}

\end{minipage}};
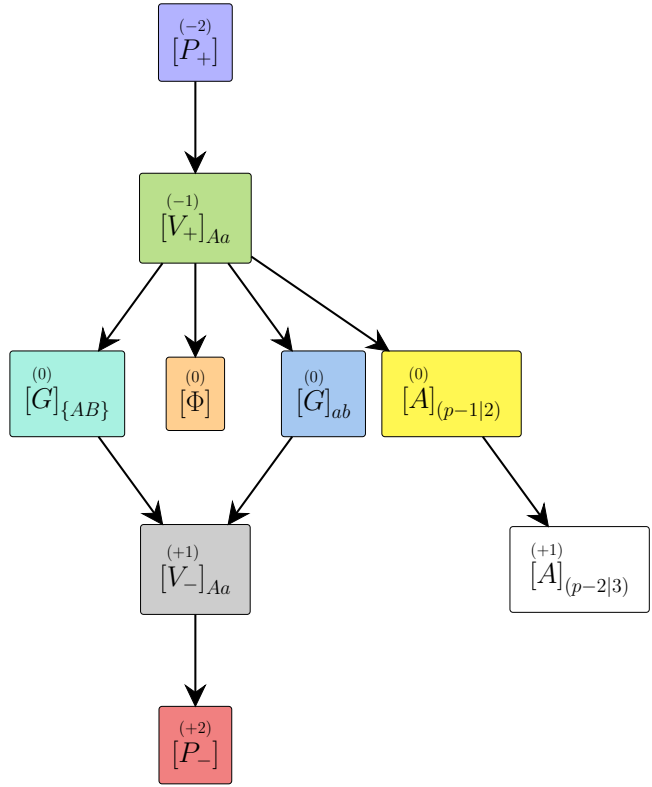
\draw (1050.97,200) node  [font=\huge] [align=left] {\begin{minipage}[lt]{469.95pt}\setlength\topsep{0pt}
\centering
\textbf{{\Huge Without}\\ {\Huge Local Dilatation Symmetry}}

\end{minipage}};
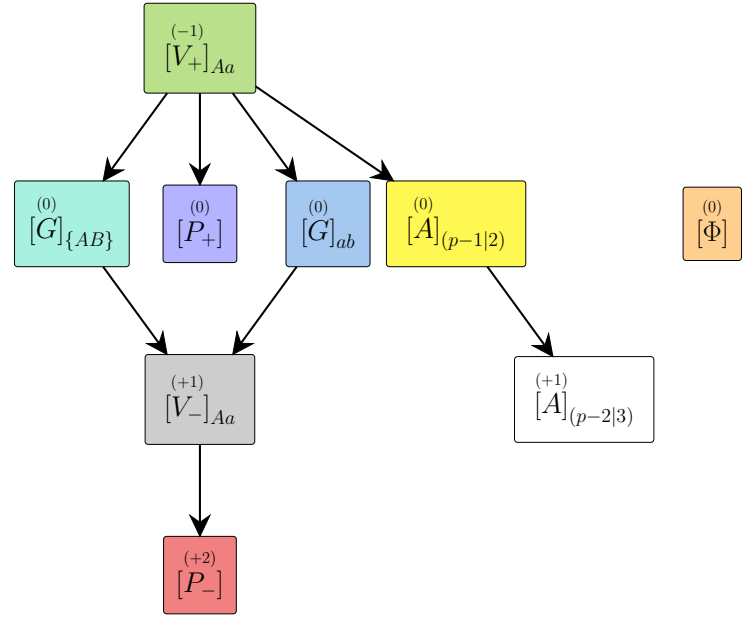
\draw  [fill={rgb, 255:red, 255; green, 207; blue, 144 } ]  (109.5,675.5) .. controls (109.5,673.84) and (110.84,672.5) .. (112.5,672.5) -- (163.5,672.5) .. controls (165.16,672.5) and (166.5,673.84) .. (166.5,675.5) -- (166.5,740.5) .. controls (166.5,742.16) and (165.16,743.5) .. (163.5,743.5) -- (112.5,743.5) .. controls (110.84,743.5) and (109.5,742.16) .. (109.5,740.5) -- cycle  ;
\draw (138,708) node  [font=\huge]  {$\accentset{( 0)}{[ \Phi ]}$};
\draw  [fill={rgb, 255:red, 255; green, 207; blue, 144 }]  (1401.5,675.5) .. controls (1401.5,673.84) and (1402.84,672.5) .. (1404.5,672.5) -- (1455.5,672.5) .. controls (1457.16,672.5) and (1458.5,673.84) .. (1458.5,675.5) -- (1458.5,740.5) .. controls (1458.5,742.16) and (1457.16,743.5) .. (1455.5,743.5) -- (1404.5,743.5) .. controls (1402.84,743.5) and (1401.5,742.16) .. (1401.5,740.5) -- cycle  ;
\draw (1430,708) node  [font=\huge]  {$\accentset{( 0)}{[ \Phi ]}$};
\draw [line width=1.5]    (898.01,581.5) -- (837.88,663.28) ;
\draw [shift={(835.51,666.5)}, rotate = 306.33] [fill={rgb, 255:red, 0; green, 0; blue, 0 }  ][line width=0.08]  [draw opacity=0] (20.36,-9.78) -- (0,0) -- (20.36,9.78) -- (13.52,0) -- cycle    ;
\draw [line width=1.5]    (961.99,581.5) -- (1022.12,663.28) ;
\draw [shift={(1024.49,666.5)}, rotate = 233.67] [fill={rgb, 255:red, 0; green, 0; blue, 0 }  ][line width=0.08]  [draw opacity=0] (20.36,-9.78) -- (0,0) -- (20.36,9.78) -- (13.52,0) -- cycle    ;
\draw [line width=1.5]    (984.5,575.06) -- (1115.66,664.25) ;
\draw [shift={(1118.97,666.5)}, rotate = 214.22] [fill={rgb, 255:red, 0; green, 0; blue, 0 }  ][line width=0.08]  [draw opacity=0] (20.36,-9.78) -- (0,0) -- (20.36,9.78) -- (13.52,0) -- cycle    ;
\draw [line width=1.5]    (835.51,749.5) -- (895.65,831.28) ;
\draw [shift={(898.01,834.5)}, rotate = 233.67] [fill={rgb, 255:red, 0; green, 0; blue, 0 }  ][line width=0.08]  [draw opacity=0] (20.36,-9.78) -- (0,0) -- (20.36,9.78) -- (13.52,0) -- cycle    ;
\draw [line width=1.5]    (1024.49,749.5) -- (964.35,831.28) ;
\draw [shift={(961.99,834.5)}, rotate = 306.33] [fill={rgb, 255:red, 0; green, 0; blue, 0 }  ][line width=0.08]  [draw opacity=0] (20.36,-9.78) -- (0,0) -- (20.36,9.78) -- (13.52,0) -- cycle    ;
\draw [line width=1.5]    (930,921.5) -- (930,1006.5) ;
\draw [shift={(930,1010.5)}, rotate = 270] [fill={rgb, 255:red, 0; green, 0; blue, 0 }  ][line width=0.08]  [draw opacity=0] (20.36,-9.78) -- (0,0) -- (20.36,9.78) -- (13.52,0) -- cycle    ;
\draw [line width=1.5]    (930,581.5) -- (930,666.5) ;
\draw [shift={(930,670.5)}, rotate = 270] [fill={rgb, 255:red, 0; green, 0; blue, 0 }  ][line width=0.08]  [draw opacity=0] (20.36,-9.78) -- (0,0) -- (20.36,9.78) -- (13.52,0) -- cycle    ;
\draw [line width=1.5]    (1210.51,749.5) -- (1272.12,833.28) ;
\draw [shift={(1274.49,836.5)}, rotate = 233.67] [fill={rgb, 255:red, 0; green, 0; blue, 0 }  ][line width=0.08]  [draw opacity=0] (20.36,-9.78) -- (0,0) -- (20.36,9.78) -- (13.52,0) -- cycle    ;
\draw [line width=1.5]    (138,405.5) -- (138,490.5) ;
\draw [shift={(138,494.5)}, rotate = 270] [fill={rgb, 255:red, 0; green, 0; blue, 0 }  ][line width=0.08]  [draw opacity=0] (20.36,-9.78) -- (0,0) -- (20.36,9.78) -- (13.52,0) -- cycle    ;
\draw [line width=1.5]    (106.01,581.5) -- (45.88,663.28) ;
\draw [shift={(43.51,666.5)}, rotate = 306.33] [fill={rgb, 255:red, 0; green, 0; blue, 0 }  ][line width=0.08]  [draw opacity=0] (20.36,-9.78) -- (0,0) -- (20.36,9.78) -- (13.52,0) -- cycle    ;
\draw [line width=1.5]    (169.99,581.5) -- (230.12,663.28) ;
\draw [shift={(232.49,666.5)}, rotate = 233.67] [fill={rgb, 255:red, 0; green, 0; blue, 0 }  ][line width=0.08]  [draw opacity=0] (20.36,-9.78) -- (0,0) -- (20.36,9.78) -- (13.52,0) -- cycle    ;
\draw [line width=1.5]    (192.5,575.06) -- (323.66,664.25) ;
\draw [shift={(326.97,666.5)}, rotate = 214.22] [fill={rgb, 255:red, 0; green, 0; blue, 0 }  ][line width=0.08]  [draw opacity=0] (20.36,-9.78) -- (0,0) -- (20.36,9.78) -- (13.52,0) -- cycle    ;
\draw [line width=1.5]    (43.51,749.5) -- (103.65,831.28) ;
\draw [shift={(106.01,834.5)}, rotate = 233.67] [fill={rgb, 255:red, 0; green, 0; blue, 0 }  ][line width=0.08]  [draw opacity=0] (20.36,-9.78) -- (0,0) -- (20.36,9.78) -- (13.52,0) -- cycle    ;
\draw [line width=1.5]    (232.49,749.5) -- (172.35,831.28) ;
\draw [shift={(169.99,834.5)}, rotate = 306.33] [fill={rgb, 255:red, 0; green, 0; blue, 0 }  ][line width=0.08]  [draw opacity=0] (20.36,-9.78) -- (0,0) -- (20.36,9.78) -- (13.52,0) -- cycle    ;
\draw [line width=1.5]    (138,921.5) -- (138,1006.5) ;
\draw [shift={(138,1010.5)}, rotate = 270] [fill={rgb, 255:red, 0; green, 0; blue, 0 }  ][line width=0.08]  [draw opacity=0] (20.36,-9.78) -- (0,0) -- (20.36,9.78) -- (13.52,0) -- cycle    ;
\draw [line width=1.5]    (418.51,749.5) -- (480.12,833.28) ;
\draw [shift={(482.49,836.5)}, rotate = 233.67] [fill={rgb, 255:red, 0; green, 0; blue, 0 }  ][line width=0.08]  [draw opacity=0] (20.36,-9.78) -- (0,0) -- (20.36,9.78) -- (13.52,0) -- cycle    ;
\draw [line width=1.5]    (138,581.5) -- (138,668.5) ;
\draw [shift={(138,672.5)}, rotate = 270] [fill={rgb, 255:red, 0; green, 0; blue, 0 }  ][line width=0.08]  [draw opacity=0] (20.36,-9.78) -- (0,0) -- (20.36,9.78) -- (13.52,0) -- cycle    ;

\end{tikzpicture}
}
\vspace{1cm}
\caption[]{The two diagrams describe how the boost transformations act on the non-relativistic equations of motion. The diagram on the left shows the case with local dilatation symmetry. This case is characterized by the appearance of the Poisson equation $\accentset{(-2)}{[P_{+}]}$ in the limit of the equations of motion. The Poisson equation is the equation with the lowest weight under dilatations. Furthermore, the scalar field equation is obtained by acting with a boost on $\accentset{(-1)}{[V_{+}]_{Aa}}$. This link is deleted in the case $p=1$ as can be seen by eqs.~\autoref{eq:BoostVp} and \autoref{eq:Combination}, since $\CD$ is proportional to $p-1$. On the right hand side we show the case without dilatation symmetry. It is understood that the equations of motion appearing in the two diagrams are present whenever the underlying dimension/foliation allows for them. In both cases the boost transformations define a reducible indecomposable representation.}  \label{fig:boost}
\end{figure}

\end{landscape}

We have visually summarized  the boost transformations of these two cases in \autoref{fig:boost}

\FloatBarrier

\FloatBarrier

\FloatBarrier
\section*{Conclusions}\addcontentsline{toc}{section}{\protect\numberline{}Conclusions}\label{sec:conclusions}

In this work we performed a systematic study of the conditions under which (generalized)\,\footnote{The generalization is in the sense that the `mass' vector field of Newton-Cartan gravity has been replaced by a $(p+1)$-form gauge field.}  Newton-Cartan gravity in the directions transverse to a $p$-brane in $D$ dimensions could be obtained from a non-relativistic limit of general relativity coupled to a $(p+1)$-form gauge field and a scalar field. In particular,
 we  derived a no-divergence condition \eqref{eq:NoDivCondition4}
\begin{align}
\min\{D-p-4,p-1\}\leqslant 0\,,\label{eq:NoDivCondition2}
\end{align}
whose solutions, for dimensions $3\le D\le 11$, we gave in \autoref{tab:pDCases}.

To derive these results, we used two approaches: a Lagrangian approach where the starting point can be defined by a relativistic Lagrangian and a constrained on-shell approach where the starting point is given by a set of equations without a Lagrangian. The different approaches can be summarized as follows:\\

\begin{enumerate}[start=1,label={\bfseries (\arabic*)}]
\item{\bf Lagrangian Approach.} Within the Lagrangian approach we  distinguished between two different cases (see also \autoref{fig:lagrangian}):

\begin{enumerate}[start=1,label={\bfseries (1\alph*)}]
\item {\bf Theories with an emerging local dilatation symmetry.}
Taking a fine-tuned scalar field coupling, we found in the limit an emerging local dilatation symmetry and a Poisson equation. The particular combination of scalar equations of motions leading to the Poisson equation was chosen such that all the order $\cc{2}$ and $\cc{0}$ terms are canceled. The limit of the Lagrangian produces a pseudo-Lagrangian with a missing equation of motion that is precisely the Poisson equation. 
\vskip .3truecm

\item {\bf Theories without an emerging local dilatation symmetry.} For theories without a fine-tuned scalar field coupling there is no emerging local dilatation symmetry and no Poisson equation. The limit of the action produces a non-relativistic action encoding the same number of equations of motion as the relativistic ones. There is no missing equation.
\end{enumerate}

In both cases the Poisson equation together with the other equations of motion form a reducible but indecomposable representation under boosts. For both cases we gave the complete multiplet structure of this representation.\\

\item {\bf Constrained on-shell approach.} We showed that it is possible to obtain the Poisson equation by imposing one or more constraints by hand (see also \autoref{fig:constrainedonshell}). In this approach the initial set of equations (equations of motion + constraints) does not come from an action. We gave two examples of additional constraints: setting to zero the field-strength of the $(p+1)$-form gauge field and/or of the scalar field.
\end{enumerate}

As a by-product of our calculations we gave in a separate appendix an example of a so-called matter-coupled electric Galilei gravity theory.\\

The physical interpretation of case 1b above with no emergent dilatation symmetry is not so clear. On the one hand we do not have the standard Newton-Cartan gravity but, on the other hand, the obtained theory is a legitimate non-relativistic limit of matter-coupled general relativity. It would be interesting to explore the physical properties of this limiting case in more detail.  \\

Looking at table \eqref{tab:pDCases} we notice the absence of the red branes. In the case of the Lagrangian approach, the obstruction in these cases is due to the fact that the action \eqref{eq:RelAction} that we have been using is not general enough. To see what is lacking, it is instructive to consider two prime examples. First of all, a notable missing brane is the eleven-dimensional membrane with $(D,p) = (11,2)$.  In this case, there is a second divergence coming from the $3$-form kinetic term that is not cancelled by the Einstein-Hilbert term. However, this extra divergence can be controlled in this particular case by performing a so-called Hubbard-Stratonovich transformation that is made possible by the presence of an additional Chern-Simons term in the Lagrangian (with a fine-tuned coefficient that is consistent with 11D supersymmetry!) of the form $C_3\wedge dC_3\wedge dC_3$ where $C_3$ is the 3-form potential of 11D supergravity. This transformation leads to the introduction of a 4-form Lagrange multiplier field that is self-dual in the transverse rotation group SO(8) \cite{Blair:2021waq,Bergshoeff:2024nin}.\,\footnote{A next case where such a self-duality naturally arises, is a $10D$ 5-brane with transverse rotation group SO(4).} Note that there is no scalar field in this example. A second striking missing brane is the $\mathcal{N}=1$ NS-NS 5-brane whose 6-form gauge field is part of the dual formulation of $\mathcal{N}=1, 10D$ supergravity. In this case there are no Chern-Simons terms to save us, but there is a scalar field $\Phi$ that could play the role of the dilaton. However, unlike the scalar field that we have been working with the dilaton has a non-trivial coupling to the 6-form gauge field kinetic term. Therefore, the scalar field cannot be identified with the dilaton. It would be interesting to see whether giving the scalar field the correct dilaton coupling could lead to a finite limit. To obtain a cancellation of leading order terms between the Einstein-Hilbert and the gauge field kinetic term in this case one  should reconsider the basic expansion \eqref{eq:ansatzA} that we have been using. More generally, we expect that several of the red branes can be associated with a finite limit provided we use a combination of adding Chern-Simons terms to the Ansatz \eqref{eq:RelAction} together with using a more general dilaton coupling. 
Although string theory is not the only motivation in this work, we expect that many of the branes of string theory and M-theory can be included in this way. Therefore, a natural direction to inspect is to repeat the classification done in this work using a more general scalar field coupling and including  Chern-Simons terms in \autoref{eq:RelAction}. Turned around, the requirement of a finite non-relativistic limit could lead to strong constraints on the form of the  (bosonic part of) the low-energy limiting supergravity theory underlying string theory, see also in this context \cite{Bergshoeff:2023ogz}. \\

It would be interesting to embed the cases we considered in this work into standard supergravity theories and take a critical supersymmetric limit. One could then consider the conditions in which such a supersymmetric limit could lead to a super Poisson equation with a fermionic Newtino potential partner \cite{Andringa:2013mma} of the  Newton potential. Due to the presence of divergences in the supersymmetry rules this is a non-trivial extension that requires to impose extra constraints on the Newton-Cartan geometry.
In the case of $10D\ \mathcal{N}=1$ supergravity these constraints can be made supersymmetric and this leads to a well-defined Newton-Cartan supergravity theory \cite{Bergshoeff:2021tfn}. However, in the 11D case, even in the presence of the additional Chern-Simons term, the constraints on the geometry can be made supersymmetric but one cannot identify a Poisson equation anymore \cite{Bergshoeff:2024nin}.\\ 

Finally, our treatment could be applied to cases where more than one gauge field is present. A prototypical example is the bosonic sector of heterotic supergravity. The non-relativistic limit of this theory requires a novel expansion unveiling a new type of geometry \cite{Bergshoeff:2023fcf}.
We hope to report on some of the generalizations mentioned here, supersymmetric or not,  in a forthcoming work.

\FloatBarrier

\section*{Acknowledgments}
The work of L.R. has been supported by Next Generation EU through the Maria Zambrano grant from the Spanish Ministry of Universities under the Plan de Recuperacion, Transformacion y Resiliencia. E.A.B.~wishes to thank the physics department of the University of Murcia, where this work was started, for its hospitality. The work of G.G. has been supported by the predoctoral fellowship FPI-UM R-1006-2021-01.

\newpage
\appendix
\section{Notation and Conventions}\label{sec:Notation}
In this work, we have adopted the following notation
\FloatBarrier
\begin{table}[!ht]
\centering
\renewcommand{\arraystretch}{1.5}
\begin{center}
\begin{tabular}{lcl}
\bf {Index}&&\bf{Definition \& Values}\\
$\mu,\nu,\rho,...$&& $D$-dim curved index\\
$\hat{A},\hat{B},\hat{C},...$&&$D$-dim flat index ($\hat{A}=\{A,a\}$)\\
$A,B,C...$&&Longitudinal flat index $A=0,1,...,p$\\
$a ,b ,c ...$&&Transverse flat index $a=p,...,D-1$
\end{tabular}
\end{center}
\caption{}
\end{table}
\FloatBarrier
We use the notation
\begin{multieq}[2]
t_{\mu\nu}{}^{A}&=2\partial_{[\mu}\tau_{\nu]}{}^{A}\,,\\
z_{\mu\nu}{}^{a}&=2\partial_{[\mu}e_{\nu]}{}^{a}\,,\\
\tau_{\mu\nu}&=\tau_{\mu A}\tau_{\nu}{}^{A}\,,\\
\tau^{\mu\nu}&=\tau^{\mu A}\tau^{\nu}{}_{A}\,,\\
h^{\mu\nu}&=e^{\mu a}e^{\nu}{}_{a}\,,\\
h_{\mu\nu}&=e_{\mu a}e_{\nu}{}^{a}\,.
\end{multieq}
When the longitudinal or transverse flat indices take only one value they are suppressed in the definition of the torsions. $p$ in principle can take values $p=-1,...,D-1$. The value $p=-1$ corresponds to the case with no longitudinal space, which is the fully transverse Euclidean case. \\

We convert curved indices into flat as
\begin{multieq}[2]
\tau^{\mu}{}_{A}T_{\mu}&=T_{A}\,,\\
e^{\mu}{}_{a}T_{\mu}&=T_{a}\,,
\end{multieq}
where $T_{\mu}$ is a generic vector. \\

For symmetrization or antisymmetrization of indices we adopt the following conventions
\begin{subequations}
\begin{align*}
T_{[\mu_{1}...\mu_{n}]}&=\frac{1}{n!}\sum_{\sigma\in S_{n}}(-)^{\sigma}\, T_{\mu_{\sigma(1)}...\mu_{\sigma(n)}}\,,\\
T_{(\mu_{1}...\mu_{n})}&=\frac{1}{n!}\sum_{\sigma	\in S_{n}}T_{\mu_{\sigma(1)}...\mu_{\sigma(n)}}\,,
\end{align*}
\end{subequations}
where the sum over $\sigma$  is the sum over the permutation of $n$ elements $S_{n}$.\\

We denote the symmetric traceless part of a tensor with longitudinal flat indices as
\begin{align}
T_{\{AB\}}&=\frac{1}{2}\bigg(T_{AB}+T_{BA}-\frac{2}{p+1}\eta_{AB}T_{C}{}^{C}	\bigg)\,.
\end{align}
\FloatBarrier
\section{Equations of Motion Expansion}\label{sec:appB}
In this section we provide the reader with the detailed expansion of the relativistic equations of motion, except the Poisson combination, discussed in \autoref{sec:Poisson}. The equations of motion expand as
\begin{subequations}
\begin{alignat}{3}
[G]_{Aa}&=\cc{}&&\accentset{(1)}{[G]}_{Aa}+\mathcal{O}(\cc{-1})\,,\\
[G]_{\{AB\}}&=&&\accentset{(0)}{[G]}_{\{AB\}}+\mathcal{O}(\cc{-2})\,,\\
[G]_{ab}&=&&\accentset{(0)}{[G]}_{ab}+\mathcal{O}(\cc{-2})\,,\\
[\Phi]&=&&\accentset{(0)}{[\Phi]}+\mathcal{O}(\cc{-2})\,,\\
[A]_{A_{1}...A_{p}a}&=\cc{}&&\accentset{(1)}{[A]}_{A_{1}...A_{p}a}+\mathcal{O}(\cc{-1})\,,\\
[A]_{A_{1}...A_{p-1}a_{1}a_{2}}&=\cc{2}&&\accentset{(2)}{[A]}_{A_{1}...A_{p-1}a_{1}a_{2}}+\mathcal{O}(\cc{0})\,,\\
[A]_{A_{1}...A_{k}a_{1}...a_{p+1-k}}&=\cc{p+1-k}\quad &&\accentset{(p+1-k)}{[A]}_{A_{1}...A_{k}a_{1}...a_{p+1-k}}+\mathcal{O}(\cc{p-1-k})\,,\quad \text{for}\quad  k\leqslant p-2\,,
\end{alignat}
\end{subequations}
with
\begin{subequations}
\begin{align}
\accentset{(1)}{[A]}_{A_{1}...A_{p}a}&=e^{\mu}{}_{a}\epsilon_{A_{1}...A_{p}B} h^{\nu\rho}\accentset{(nr)}{\nabla}_{\nu}t_{\rho\mu}^{B}-\coefa \partial^{b}\phi t_{ba}{}^{B}\epsilon_{A_{1}...A_{p}B}+\nonumber\\
&-\frac{1}{2}t^{bc}{}_{[A_{1}}f_{A_2 ... A_p]abc}+p(-)^{p} t^{bC}{}_{[A_1}\epsilon_{A_{2}...A_{p}]CB}t_{ba}{}^{B}+\nonumber\\
&-t_{ab}{}^{C}t^{bB}{}_{B}\epsilon_{A_{1}...A_{p}C}\,,\\
\accentset{(-1)}{[A]}_{A_{1}...A_{p}a}&=(-)^{p}e^{\mu_1}{}_{a}\tau^{\mu_2}{}_{A_1}...\tau^{\mu_{p+1}}{}_{A_p} \bigg[\frac{1}{2}(p+2)(p+1)\tau^{\alpha\beta}\accentset{(nr)}{\nabla}_{\alpha}t_{[\beta\mu_1}\tau_{\mu_2}{}^{B_{2}}...\tau_{\mu_{p+1}}{}^{B_{p+1}}\epsilon_{B_{1}...B_{p+1}}+\nonumber\\
&+h^{\alpha\beta}\accentset{(nr)}{\nabla}_{\alpha}f_{\beta\mu_{1}...\mu_{p+1}}\bigg]+\nonumber\\
&+\coefa (-)^{p}\partial^{B}\phi t_{aC}{}^{C}\epsilon_{BA_{1}...A_{p}}-\coefa \partial^{b}\phi f_{bA_{1}...A_{p}a}+\frac{1}{2}t_{a}{}^{bB}f_{bA_{1}...A_{p}B}+t_{B}{}^{bB}f_{aA_{1}...A_{p}b}+\nonumber\\
&+pt^{bB}{}_{[A_{1}}f_{A_{2}...A_{p}]baB}+ (z_{[b}{}^{Bb}+z^{bB}{}_{[b})t_{a]C}{}^{C}\epsilon_{A_{1}...A_{p}B}+\nonumber\\
&+\frac{1}{2}(-)^{p}p\, z_{[A_{1}}{}^{Bb}\epsilon_{A_{2}...A_{p}]BC}t_{bc}{}^{C}-t_{B}{}^{DB}\epsilon_{DA_{1}...A_{p}}t_{aC}{}^{C}-\frac{1}{2}pt^{BD}{}_{[A_{1}}\epsilon_{A_{2}...A_{p}]BD}t_{aC}{}^{C}\,,\\
\nonumber\\
\accentset{(2)}{[A]}_{A_{1}...A_{p-1}a_{1}a_{2}}&=-\frac{1}{2}(p-1)t^{bc}{}_{[A_{1}}f_{A_{2}...A_{p-1}]a_{1}a_{2}bc}\,,\\
\accentset{(0)}{[A]}_{A_{1}...A_{p-1}a_{1}a_{2}}&=e^{\mu_{1}}{}_{a_{1}}e^{\mu_{2}}{}_{a_{2}}\tau^{\mu_{3}}{}_{A_{1}}...\tau^{\mu_{p+1}}{}_{A_{p-1}}h^{\alpha\beta}\accentset{(nr)}{\nabla}_{\alpha}f_{\beta \mu_{1}...\mu_{p+1}}+\nonumber\\
&-\coefa \partial^{b}\phi f_{a_{1}a_{2}bA_{1}...A_{p-1}}+\coefa\partial^{B}\phi t_{a_{1}a_{2}}{}^{C}\epsilon_{A_{1}...A_{p-1}BC}+\nonumber\\
&-\frac{3}{2}(z_{[b}{}^{Cb}+z^{bC}{}_{[b})t_{a_{1}a_{2}]}{}^{B}\epsilon_{A_{1}...A_{p-1}BC}-t_{B}{}^{bB}f_{a_{1}a_{2}bA_{1}...A_{p-1}}+\nonumber\\
&-\frac{1}{2}(-)^{p}(p-1)t^{BC}{}_{[A_{1}}\epsilon_{A_{2}...A_{p-1}]BCD}t_{a_{1}a_{2}}{}^{D}+t_{B}{}^{CB}\epsilon_{A_{1}...A_{p-1}CD}t_{a_{1}a_{2}{}}{}^{D}\,,\\
\nonumber\\
\accentset{(p+1-k)}{[A]}_{A_{1}...A_{k}a_{1}...a_{p+1-k}}&=\frac{1}{2}(-)^{p+1}k t^{bc}{}_{[A_{1}}f_{A_{2}...A_{k}]a_{1}...a_{p+1-k}bc}\,,\quad \text{for}\quad  k\leqslant p-2\,,\\
\accentset{(p-1-k)}{[A]}_{A_{1}...A_{k}a_{1}...a_{p+1-k}}&=(-)^{kp}e^{\mu_1}{}_{a_1}...e^{\mu_{p+1-k}}{}_{a_{p+1-k}}\tau^{\mu_{p+2-k}}{}_{A_1}...\tau^{\mu_{p+1}}{}_{A_{k}}h^{\alpha\beta}\accentset{(nr)}{\nabla}_{\alpha}f_{\beta\mu_{1}...\mu_{p+1}}+\nonumber\\
&-\coefa\partial^{b}\phi f_{bA_{1}...A_{k}a_{1}...a_{p+1-k}}+(-)^{k}kt^{bB}{}_{[A_{1}}f_{A_{2}...A_{k}]ba_{1}...a_{p+1-k}B}+\nonumber\\
&+(k+1)t_{B}{}^{bB}f_{A_{1}...A_{k}a_{1}...a_{p+1-k}b}\,,\quad \text{for}\quad  k\leqslant p-2\,,\\
\nonumber\\
\accentset{(1)}{[G]}_{Aa}&=-\frac{1}{4(p-1)!}t^{bcD}\epsilon_{DAB_{1}...B_{p-1}}f_{abc}{}^{B_{1}...B_{p-1}}-\frac{\coefa}{2}t_{abA}\partial^{b}\phi+\nonumber\\
&+\frac{1}{2}e^{\nu}{}_{a}h^{\mu\rho}\accentset{(nr)}{\nabla}_{\mu}t_{\nu\rho A}-\frac{1}{2}t_{A}{}^{bB}t_{abB}\,,\\
\nonumber\\
\accentset{(-1)}{[G]}_{Aa}&=\frac{1}{2p!}t^{bD}{}_{D}\epsilon_{AB_{1}...B_{p}}f_{ab}{}^{B_{1}...B_{p}}-\frac{1}{2p!}f_{A}{}^{bB_{1}...B_{p}}t_{ab}{}^{D}\epsilon_{DB_{1}...B_{p}}+\nonumber\\
&-\frac{1}{4(p-1)!}f_{A}{}^{B_{1}...B_{p-1}bc}f_{abcB_{1}...B_{p-1}}+\accentset{(nr)}{\rm R}_{Aa}+\nonumber\\
&+\frac{1}{2}\tau^{\mu\rho}e^{\nu}{}_{a}\accentset{(nr)}{\nabla}_{\mu}t_{\nu\rho A}-e^{\mu}{}_{a}\tau^{\nu}{}_{A}e^{\rho}{}_{B}\accentset{(nr)}{\nabla}_{\mu}t_{\nu\rho}{}^{B}-\frac{1}{2}\tau^{\mu}{}_{B}e^{\nu}{}_{a}\tau^{\rho}{}_{A}\accentset{(nr)}{\nabla}_{\mu}t_{\nu\rho}{}^{B}+\nonumber\\
&-\frac{1}{2}t_{A}{}^{BC}(t_{aBC}+t_{aCB})+\frac{1}{2}t_{Aa}{}^{B}(t_{BC}{}^{C}+z_{Bb}{}^{b})+\frac{1}{4}t_{A}{}^{bB}(z_{aBb}+z_{bBa})+\nonumber\\
&+\frac{1}{4}t^{Bb}{}_{A}(z_{Bba}+z_{Bab})+\frac{1}{2}t_{aBA}(t^{BC}{}_{C}+z^{Bb}{}_{b})-\frac{1}{4}t_{a}{}^{bB}z_{ABb}+\tau^{\mu}{}_{A}e^{\nu}{}_{a}\accentset{(nr)}{\nabla}_{\mu}\partial_{\mu}\phi+\nonumber\\
&-t_{a(BA)}\partial^{B}\phi-(\coefa^2-\coefb)\partial_{A}\phi\partial_{a}\phi\,,\\
\nonumber\\
\accentset{(0)}{[G]}_{\{AB\}}&=-\accentset{(nr)}{\nabla}_{\mu}t_{\nu\rho \{A}\tau^{\rho}{}_{B\}}h^{\mu\nu}-t^{aD}{}_{D}t_{a\{AB\}}-\frac{1}{2}t_{AaC}t_{B}{}^{aD}+\frac{1}{2}t_{DaA}t^{Da}{}_{B}+\coefa t^{a}{}_{\{AB\}}\partial_{a}\phi+\nonumber\\
&+\frac{1}{2(p-1)!}t_{abD}\epsilon_{\{A}{}^{DC_{1}...C_{p-1}}f_{B\}C_{1}...C_{p-1}ab}-\frac{1}{12(p-2)!}f_{\{A}{}^{C_{1}...C_{p-1}abc}f_{B\}C_{1}...C_{p-1}abc}\,,\\
\nonumber\\
\accentset{(0)}{[G]}_{ab}&=\accentset{(nr)}{\rm R}_{(ab)}-e^{\mu}{}_{(a}e^{\nu}{}_{b)}\tau^{\rho}{}_{A}\accentset{(nr)}{\nabla}_{\mu}t_{\nu\rho}{}^{A}-\frac{1}{2}t_{a}{}^{AB}(t_{bAB}+t_{bBA}-\eta
_{AB}t_{bC}{}^{C})+\nonumber\\
&+\coefa \bigg[e^{\mu}{}_{a}e^{\nu}{}_{b}\accentset{(nr)}{\nabla}_{\mu}\partial_{\nu}\phi-\frac{1}{2}t_{ab}{}^{A}\partial_{A}\phi\bigg]+\nonumber\\
&+\frac{1}{p!}t_{c(a}{}^{B}f_{b)}{}^{cA_{1}...A_{p}}\epsilon_{BA_{1}...A_{p}}-\frac{1}{4(p-1)!}f_{a(p-1|2)}f_{b}{}^{(p-1|2)}+\nonumber\\
&-(\coefa^{2}-\coefb)\bigg[\partial_{a}\phi\partial_{b}\phi-\delta_{ab}\partial_{c}\phi\partial^{c}\phi+\frac{1}{\coefa}\delta_{ab}\Big(h^{\mu\nu}\accentset{(nr)}{\nabla}_{\mu}\partial_{\nu}\phi-t_{A}{}^{cA}\partial_{c}\phi\Big)\bigg]\,,\\
\nonumber\\
\accentset{(0)}{[\Phi]}&=2h^{\mu\nu}\accentset{(nr)}{\nabla}_{\mu}\partial_{\nu}\phi-2t_{A}{}^{aA}\partial_{a}\phi-\coefa\partial_{a}\phi\partial^{a}\phi
+\nonumber\\
&+\frac{\coefa}{\coefb}\bigg[h^{\mu\nu}\accentset{(nr)}{{\rm R}}_{\mu\nu}-2h^{\mu\nu}\tau^{\rho}{}_{A}\accentset{(nr)}{\nabla}_{\mu}t_{\nu \rho}{}^{A} e_{\mu}\,^{a} e_{\nu a} \tau_{\rho A}-t^{aAB}t_{a(AB)}-\frac{1}{2}t_{aA}{}^{A}t^{aB}{}_{B}+\nonumber\\
&-\frac{1}{12(p-1)!}f_{a_{1}a_{2}a_{3}A_{1}...A_{p-1}}f^{a_{1}a_{2}a_{3}A_{1}...A_{p-1}}\textcolor{black}{-\frac{1}{2\, p!}f^{abA_{1}...A_{p}}t_{ab}{}^{B}\epsilon_{BA_{1}...A_{p}}}\bigg]\,,\\
\end{align}
\end{subequations}
where
\begin{align}
f_{a(p-1|2)}f_{b}{}^{(p-1|2)}&=f_{ac_{1}c_{2}A_{1}...A_{p-1}}f_{b}{}^{c_{1}c_{2}A_{1}...A_{p-1}}\,.
\end{align}
Some remarks are in order. The equations $[A]_{A_{1}...A_{p-1}a_{1}a_{2}}$ has the following  leading order term
\begin{align}
\accentset{(2)}{[A]}_{A_{1}...A_{p-1}a_{1}a_{2}}&=-\frac{1}{2}(p-1)t^{bc}{}_{[A_{1}}f_{A_{2}...A_{p-1}]a_{1}a_{2}bc}\,.
\end{align}
The fact that this equation has a leading part at order $\cc{2}$ could be problematic since it could induce a divergent term in the boost transformation of the equations $[V_{\pm}]_{Aa}$. However, we should consider this term in the presence of the no-divergence condition, that we report here for simplicity,
\begin{align}
\min\{p-1, D-p-4\}\leqslant 0.
\end{align}
If $p-1\leqslant 0$ then the leading term above cannot be there because the differential form should be at least a 4-form. If $D-p-4\leqslant 0$ this implies that the transverse space is at most three dimensional. The term above, to be non-trivial, requires at least a 4-dim transverse space. This implies that in all the theories satisfying the no-divergence condition this term is simply not there and the leading order term is of order zero in $\cc{}$. The same applies to  $\accentset{(p+1-k)}{[A]}_{A_{1}...A_{k}a_{1}...a_{p+1-k}}$.

\section{Matter Coupled Electric Galilei Gravity}\label{sec:ElectricGalilei2}

In this section we do not cancel the leading order term coming from the expansion in powers of $\cc{}$ of the Einstein-Hilbert term against a gauge field kinetic term like we do when taking a Newton-Cartan limit. Instead, we take a Galilei limit and consider the same term coming from the Einstein-Hilbert term as part of the finite action. The gauge field kinetic term leads in this case to additional matter couplings which we derive in this appendix.

We use a slight modified version of the Ansatz\autoref{eq:ansatzA},
\begin{subequations}
\begin{align}
E_{\mu}{}^{A}&=\cc{\alpha}\tau_{\mu}{}^{A}\,,\label{eq:ansatztau2}\\
E_{\mu}{}^{a}&=\cc{\beta}e_{\mu}{}^{a}\,,\\
E^{\mu}{}_{A}&=\cc{-\alpha}\tau^{\mu}{}_{A}\,,\\
E^{\mu}{}_{a}&=\cc{-\beta}e^{\mu}{}_{a}\,,\\
\Phi&=\phi +\frac{p+1}{\coefa}\ln \cc{}\,,\\
A_{\mu_{1}...\mu_{p+1}}&=\eta\,\cc{\xi}\tau_{\mu_{1}}{}^{A_{1}}...\tau_{\mu_{p+1}}{}^{A_{p+1}}\epsilon_{A_{1}...A_{p+1}}+ \cc{\gamma} a_{\mu_{1}...\mu_{p+1}}\,,\label{eq:ansatzA2}
\end{align}
\end{subequations}
where $\eta$ is a constant. The finiteness of the boost transformations of the Vielbein under the limit $\cc{}\rightarrow \infty$ still imposes the same conditions as in \autoref{eq:nodivboost} , that we report here for convenience
\begin{multieqref}[2]{eq:nodivboost2}
\delta&=\beta-\alpha\,,\\
\alpha-\beta&>0\,.\
\end{multieqref}
This implies that the expansion \autoref{eq:Rexpansion} is not modified.
The boost transformation of the gauge field \autoref{eq:aboost},
\begin{align}
\delta a_{\mu_{1}...\mu_{p+1}}&=-(p+1)\eta\, \cc{\xi-2\alpha+2\beta-\gamma}\lambda^{A_{1}}{}_{b}e_{[\mu_{1}}{}^{b}\tau_{\mu_{2}}{}^{A_{2}}...\tau_{\mu_{p+1}]}{}^{A_{p+1}}\epsilon_{A_{1}...A_{p+1}}\,,
\end{align}
requires
\begin{align}
\gamma\geqslant \xi-2(\alpha-\beta)\,.\label{eq:nodivboost3}
\end{align}
When this bound is saturated the non-relativistic gauge field transforms under boosts, otherwise it is inert. Since we are interested to the highest power in the expansion we give here only the contributions to the expansion of the Maxwell term,   for the three types of terms defined in \autoref{eq:F2terms},  with the highest $\cc{}$ power:
\begin{align}
F_{\mu_{1}...\mu_{p+2}}F^{\mu_{1}...\mu_{p+2}}&=
-\frac{1}{2}(p+2)!\cc{2\xi-2p\alpha-2\beta}\,   \eta^2\, t_{abA}t^{abA}+\nonumber\\
&+(p+2)(p+1)\cc{\xi+\gamma-2p\alpha-2\beta}t_{ab}{}^{B}\epsilon_{BA_{1}...A_{p}}f^{abA_{1}...A_{p}}+\nonumber\\
&+\cc{2\gamma-2\alpha k_{0}-2\beta(p+2-k_{0})}\binom{p+2}{k_{0}} f_{A_{1}...A_{k_{0}}a_{1}...a_{p+2-k_{0}}}f^{A_{1}...A_{k_{0}}a_{1}...a_{p+2-k_{0}}}+...\,,\label{eq:F2expansionleading}
\end{align}
where we have defined
\begin{align}
k_{0}=\max\{0,2p+3-D\}\,.
\end{align}
The leading order term coming from the Einstein-Hilbert Lagrangian is at order $\cc{2\alpha-4\beta}$. Although it is possible to be more general, we will limit our analysis   to the cases where the terms in \autoref{eq:F2expansionleading} have at most the same order as the one  coming from the Einstein-Hilbert term, i.e. we impose for each of the three terms:
\begin{subequations}
\begin{align}
&I_{1}:&\xi&\leqslant (p+1)\alpha\,,\\
&I_{2}:&\gamma+\xi&\leqslant (2p+1)\alpha\,,\\
&I_{3}:&\gamma &\leqslant \alpha+p\beta +(\alpha-\beta)k_{0}\,.
\end{align}\label{eq:diseq3}
\end{subequations}
There are eight cases to consider depending on whether each of the three terms in \autoref{eq:F2expansionleading} has exactly the order $\cc{2\alpha-4\beta}$ or is sub-leading with respect to it, i.e.~for each of the inequalities in \autoref{eq:diseq3} we have two cases, either the bound is saturated or the inequality is a strict inequality. In the first case the corresponding term will appear in the action after sending $\cc{}$ to infinity, otherwise, being sub-leading, it will vanish. Not every case will be possible for any $p$ and $D$ since the inequalities \autoref{eq:diseq3} together with \autoref{eq:nodivboost2} and \autoref{eq:nodivboost3} can impose restrictions on $p$ and $D$. The final finite action can be parametrized as follows
\begin{align}
\mathcal{S}_{\rm electric\ Galilei}&=\int\,d^{D}x\, E\, {\rm e}^{-\coefa\phi}\bigg[\frac{1}{4}(-1+\eta^{2}\CE)\, t_{abA}t^{abA}+\nonumber\\
&-\frac{\CF}{2p!}\, t_{ab}{}^{B}\epsilon_{BA_{1}...A_{p}}f^{abA_{1}...A_{p}}+\nonumber\\
&-\frac{\CG}{2(p+2)!}\,\binom{p+2}{k_{0}} f_{A_{1}...A_{k_{0}}a_{1}...a_{p+2-k_{0}}}f^{A_{1}...A_{k_{0}}a_{1}...a_{p+2-k_{0}}}\bigg]\,, \label{eq:electricaction2}
\end{align}
where the constants $\CE, \CF$ and $\CG$  capture all the different cases. They will be $+1$ if the term they multiply is at the leading order, 0 if it is sub-leading. As an example let us examine the case where all the bounds in \autoref{eq:diseq3} are saturated, corresponding to the fact that all the three types of terms coming from the Maxwell term will be in the action, i.e. $\CE=\CF=\CG=1$. The conditions are
\begin{subequations}
\begin{align}
\alpha-\beta &>0\,,\\
\gamma &\geqslant \xi-2(\alpha-\beta)\,,\\
\xi&=(p+1)\alpha\,,\\
\gamma+\xi&=(2p+1)\alpha\,,\\
\gamma &= \alpha+p\beta +(\alpha-\beta)k_{0}\,.
\end{align}
\end{subequations}
We find the following solution
\begin{align}
\xi&=(p+1)\alpha\,,&\gamma&=(p+1)\alpha\,,
\end{align}
together with the condition
\begin{align}
k_{0}&=\max\{0,2p+3-D\}=p\,,
\end{align}
that is equivalent to $p=D-3$. \\

It is interesting to see how the boost invariance is realized in the action. Regardless of the choice $\CE=\CF=\CG=1$, we know that $t_{ab}{}^{A}$ is invariant under boost transformations. However for $f_{A_{1}...A_{k}a_{1}...a_{p+2-k}}$ the transformation rule is related to the values of $\gamma$ and thus depends on the specific case we are considering. For the case we have just considered the inequality \autoref{eq:nodivboost3} is not saturated thus the gauge field does not transform. This implies that the only contribution to the transformation comes from the inverse longitudinal Vielbein ,
\begin{align}
\delta f_{A_{1}...A_{k}a_{1}...a_{p+2-k}}&=(-)^{p+1}k\lambda_{[A_{1}}{}^{b}f_{A_{2}...A_{k}]a_{1}...a_{p+2-k}b}. \label{eq:fk0boost}
\end{align}
It then seems that the action \autoref{eq:electricaction2} is not invariant under boost. However in this case $k_{0}=p$ implying
\begin{align}
f_{A_{1}...A_{k_{0}}a_{1}...a_{p+2-k_{0}}}&=f_{A_{1}...A_{p}a_{1}a_{2}}\,.
\end{align}
This is the only component of $f_{\mu_{1}...\mu_{p+2}}$ appearing in the action for $\CE=\CF=\CG=1$ and since $f_{A_{1}...A_{k_{0}}a_{1}...a_{p+2-k_{0}}}$ is the component of the field strength with the maximum number of transverse indices the right hand side of \autoref{eq:fk0boost} is trivial. This implies that the action is invariant under boost transformations. A similar mechanisms hold in all the other cases. \\

We have summarized all the possibilities choices for $\CE, \CF$ and $\CG$ in  \autoref{tab:electricGalilei}.
\FloatBarrier
\begin{table}[!ht]
 \rowcolors{2}{gray!25}{white}
\centering
\renewcommand{\arraystretch}{1.5}
\begin{center}
\begin{tabular}{|c|c|c|c|c|c|}
\rowcolor{gray!50}
\hline
 \boldmath\CE &  \CF &\CG & \boldmath $\xi$  &\boldmath $\gamma$& \bf Conditions\\
\hline
1&1&1&$(p+1)\alpha$&$(p+1)\alpha$&$p=D-3\,, D\geqslant 3$\\
\hline
1&1&0&$(p+1)\alpha$&$(p+1)\alpha$&$p=D-2\,, D\geqslant 3$\\
\hline
1&0&1&\begin{tabular}{c}
$2\alpha$\\
$(p+1)\alpha$
\end{tabular}&\begin{tabular}{c}
$\alpha+\beta$\\
$(D-4)\alpha+\beta$
\end{tabular}&\begin{tabular}{c}
$p=1\,, D\geqslant 5$\\
$p=D-4\,, D\geqslant 6$
\end{tabular}\\
\hline
1&0&0&$(p+1)\alpha$&\begin{tabular}{c}
$2\beta-\alpha\leqslant \gamma < \alpha$\\
$2\beta\leqslant \gamma < 2\alpha$\\
$2\beta\leqslant \gamma < \alpha+\beta$\\
$2\beta+\alpha(p-1)\leqslant \gamma < p\alpha+\beta$\\
$2\beta+\alpha(p-1)\leqslant \gamma < (p+1)\alpha$\\
$2\beta+\alpha(p-1)\leqslant \gamma < (p+1)\alpha$\\
\end{tabular}&
\begin{tabular}{c}
$p=0\,, D\geqslant 3$\\
$p=1\,, D= 3,4$\\
$p=1\,, D\geqslant 5$\\
$p=D-4\,, D \geqslant 6$\\
$p=D-2\,, D \geqslant 4$\\
$p=D-3\,, D \geqslant 5$\\
\end{tabular}\\
0&1&1&$\alpha p+\beta$& $(p+2)\alpha-\beta$ &$p=D-2\,, D\geqslant 3$\\
0&1&0&$\alpha p + \beta <\xi< (p+1)\alpha$&$\xi-2\alpha(p+1)$&$p=D-2\,, D\geqslant 3$\\
0&0&1&\begin{tabular}{c}

$\xi < \alpha $\\
$\xi < 2\alpha$\\
$\xi < \alpha p +\beta$\\
$\xi < (p+1)\alpha$\\
$\xi < (p+1)\alpha$\\
\end{tabular}
&\begin{tabular}{c}
$\alpha$\\
$\alpha +\beta$\\
$(p+2)\alpha -\beta$\\
$(p+1)\alpha$\\
$p\alpha+\beta$\\
\end{tabular}&
\begin{tabular}{c}
$p=0\,, D\geqslant 3$\\
$p=1\,, D\geqslant 5$\\
$p=D-2\,, D\geqslant 3$\\
$p=D-3\,, D\geqslant 3$\\
$p=D-4\,, D\geqslant 5$\\
\end{tabular}\\
0&0&0&$\cdots$&$\cdots$&$\cdots$\\
\hline
\end{tabular}
\end{center}
\caption{In this table we summarize the different possible limits captured by the action \autoref{eq:electricaction2}. In the first three columns we list the values $\CE, \CF$ and $\CG$. Value 1 means that the corresponding term in \autoref{eq:electricaction2} is leading, value 0 that it is sub-leading. The other columns describe how to choose the parameters $\xi$ and $\gamma$ occurring in the expansion of the gauge field and if there are restrictions on $p,D$ for the given configuration. We do not list the details about the case $\CE=\CF=\CG=0$ since this case is equivalent to the absence of the Maxwell contribution. }\label{tab:electricGalilei}
\end{table}

\FloatBarrier

\bibliography{bibliography}{}
\end{document}